\documentclass[article]{IEEEtran}

\usepackage{amsmath,amssymb,latexsym,cite,etex}
\usepackage{epsf, pstricks,pgf,xcolor}
\usepackage{epsfig}
\usepackage{bbm} 
\usepackage{stmaryrd}
\usepackage{soul}
\usepackage{pgfplots}
\usepackage{graphicx}
\pgfplotsset{compat=newest}
\pgfplotsset{plot coordinates/math parser=false}

\newrgbcolor{ColorLink}{0 .4 .6}
\newrgbcolor{ColorCite}{0 .6 .4}

\usepackage[plainpages=false,pdfpagelabels,colorlinks=true,linkcolor=ColorLink,citecolor=ColorCite]{hyperref}

\allowdisplaybreaks

\newtheorem{theorem}{Theorem}
\newtheorem{corollary}{Corollary}
\newtheorem{lemma}{Lemma}

\newtheorem{example}{Example}

\newtheorem{definition}{Definition}

\newtheorem{remark}{Remark}

\def\etal{\textit{et al.~}}

\newcommand{\CC}{\mathbb{C}}
\newcommand{\RR}{\mathbb{R}}
\newcommand{\ZZ}{\mathbb{Z}}
\newcommand{\NN}{\mathbb{N}}

\newcommand{\av}{{\mathbf a}}
\newcommand{\bv}{{\mathbf b}}
\newcommand{\cv}{{\mathbf c}}
\newcommand{\dv}{{\mathbf d}}

\newcommand{\gv}{{\mathbf g}}
\newcommand{\hv}{{\mathbf h}}

\newcommand{\lv}{{\mathbf l}}

\newcommand{\rv}{{\mathbf r}}
\newcommand{\sv}{{\mathbf s}}
\newcommand{\tv}{{\mathbf t}}
\newcommand{\uv}{{\mathbf u}}
\newcommand{\wv}{{\mathbf w}}
\newcommand{\vv}{{\mathbf v}}
\newcommand{\xv}{{\mathbf x}}
\newcommand{\yv}{{\mathbf y}}
\newcommand{\zv}{{\mathbf z}}
\newcommand{\zerov}{{\mathbf 0}}

\newcommand{\deltav}{\boldsymbol{\delta}}

\newcommand{\lambdav}{\boldsymbol{\lambda}}

\newcommand{\nuv}{\boldsymbol{\nu}}
\newcommand{\muv}{\boldsymbol{\mu}}

\newcommand{\chiv}{\boldsymbol{\chi}}

\newcommand{\atv}{\mathbf{\tilde{a}}}

\newcommand{\wtv}{\mathbf{\tilde{w}}}

\newcommand{\ytv}{\mathbf{\tilde{y}}}

\newcommand{\lambdatv}{\boldsymbol{\tilde{\lambda}}}

\newcommand{\shv}{\mathbf{\hat{s}}}
\newcommand{\uhv}{\mathbf{\hat{u}}}

\newcommand{\muhv}{\boldsymbol{\hat{\mu}}}
\newcommand{\nuhv}{\boldsymbol{\hat{\nu}}}
\newcommand{\chihv}{\boldsymbol{\hat{\chi}}}

\newcommand{\Am}{{\mathbf A}}
\newcommand{\Atm}{{\mathbf {\tilde{A}}}}

\newcommand{\Fm}{{\mathbf F}}
\newcommand{\Gm}{{\mathbf G}}

\newcommand{\Hm}{{\mathbf H}}
\newcommand{\Id}{{\mathbf I}}

\newcommand{\Lm}{{\mathbf L}}
\newcommand{\Mm}{{\mathbf M}}
\newcommand{\Nm}{{\mathbf N}}
\newcommand{\Om}{{\mathbf O}}
\newcommand{\Pm}{{\mathbf P}}
\newcommand{\Qm}{{\mathbf Q}}

\newcommand{\Tm}{{\mathbf T}}
\newcommand{\Um}{{\mathbf U}}

\newcommand{\Xm}{{\mathbf X}}
\newcommand{\Ym}{{\mathbf Y}}
\newcommand{\Zm}{{\mathbf Z}}

\newcommand{\Shm}{{\mathbf{\hat{S}}}}

\newcommand{\Abm}{{\mathbf {\bar{A}}}}
\newcommand{\Lbm}{{\mathbf {\bar{L}}}}

\newcommand{\Bc}{{\mathcal B}}
\newcommand{\Cc}{{\mathcal C}}
\newcommand{\Dc}{{\mathcal D}}

\newcommand{\Ic}{{\mathcal I}}

\newcommand{\Lc}{{\mathcal L}}
\newcommand{\Mc}{{\mathcal M}}
\newcommand{\Nc}{{\mathcal N}}

\newcommand{\Rc}{{\mathcal R}}
\newcommand{\Sc}{{\mathcal S}}

\newcommand{\Vc}{{\mathcal V}}

\newcommand{\diag}{{\mathrm{diag}}}
\renewcommand{\det}{{\mathrm{det}}}

\newcommand{\conv}{{\mathrm{conv}}}

\newcommand{\rowspan}{\mathrm{rowspan}}
\newcommand{\rank}{\mathrm{rank}}

\newcommand{\T}{^{\mathsf T}}

\newcommand{\llb}{\llbracket}
\newcommand{\rrb}{\rrbracket}
\newcommand{\ones}{\mathbf{1}}
\newcommand{\zeros}{\mathbf{0}}
\newcommand{\ant}{{N_{\text{r}}}}
\newcommand{\ex}{\mathbb{E}}
\newcommand{\pr}{\mathbb{P}}
\newcommand{\opt}{{\text{opt}}}

\newcommand{\tildebf}[1]{\mathbf{\tilde{#1}}}

\def\L{\Lambda}
\def\f{\text{\textnormal{F}}}
\def\c{\text{\textnormal{C}}}
\def\Vol{\mathrm{Vol}}
\def\eff{\text{\textnormal{eff}}}
\def\comp{\text{\textnormal{comp}}}
\def\para{\text{\textnormal{para}}}
\def\succ{\text{\textnormal{succ}}}
\def\prim{\text{\textnormal{prim}}}
\def\opt{\text{\textnormal{opt}}}
\def\modl{\hspace{-0.1in}\mod\Lambda}

\def\modp{\hspace{-0.1in}\mod{p}}

\def\kcl{k_{\c,\ell}}
\def\kfl{k_{\f,\ell}}
\def\kcmin{k_{\c}}
\def\kfmax{k_{\f}}

\def\argmax{\operatornamewithlimits{arg\,max}}
\def\argmin{\operatornamewithlimits{arg\,min}}
\newcommand{\define}{\triangleq}

\newrgbcolor{BoxRed}{1 .6 .6}
\newrgbcolor{BoxBlue}{.8 .8 1}
\newrgbcolor{BoxGreen}{.8 1 .8}
\newrgbcolor{BoxOrange}{1 .65 0}
\newrgbcolor{BoxPurple}{.87 .79 .85}
\newrgbcolor{LineGray}{.6 .6 .6}
\newrgbcolor{AxisGray}{.3 .3 .3}
\newrgbcolor{LineBlue}{.2 .6 .9}
\newrgbcolor{LineRed}{.9 .3 .3}
\newrgbcolor{LineGreen}{.3 .8 .3}

\newcommand{\remove}[1]{}

\newcommand{\defsym}{\hfill $\lozenge$}
\newcommand{\thmsym}{\hfill $\square$}

\newcommand\scalemath[2]{\scalebox{#1}{\mbox{\ensuremath{\displaystyle #2}}}}

\begin{document}

\title{Expanding the Compute-and-Forward Framework: Unequal Powers, Signal Levels,\\ and Multiple Linear Combinations}
\author{\thanks{B. Nazer was supported by NSF grant CCF-1253918. This work was presented at the 51st Allerton Conference on Communications, Control, and Computing in 2013 and the 10th Information Theory and Applications Workshop in 2015.}Bobak Nazer\thanks{B. Nazer is with the Department of Electrical and Computer Engineering, Boston University, Boston, MA. Email: \texttt{bobak@bu.edu}.}, \IEEEmembership{Member, IEEE,} Viveck Cadambe\thanks{V. Cadambe is with the Department of Electrical Engineering, Pennsylvania State University, University Park, PA. Email: \texttt{viveck@engr.psu.edu}.}, \IEEEmembership{Member, IEEE,} \\Vasilis Ntranos\thanks{V. Ntranos is with the Department of Electrical Engineering, University of Southern California, Los Angeles, CA. Email: \texttt{ntranos@usc.edu}.}, \IEEEmembership{Member, IEEE,} and Giuseppe Caire\thanks{G. Caire is with the Department of Telecommunication Systems, Technical Universit\"at Berlin, Berlin, Germany and the Department of Electrical Engineering, University of Southern California, Los Angeles, CA. Email: \texttt{caire@tu-berlin.de}.}, \IEEEmembership{Fellow, IEEE}}

\markboth{IEEE Trans Info Theory, to appear}{~}

\maketitle

\begin{abstract}
The compute-and-forward framework permits each receiver in a Gaussian network to directly decode a linear combination of the transmitted messages. The resulting linear combinations can then be employed as an end-to-end communication strategy for relaying, interference alignment, and other applications. Recent efforts have demonstrated the advantages of employing unequal powers at the transmitters and decoding more than one linear combination at each receiver. However, neither of these techniques fit naturally within the original formulation of compute-and-forward. This paper proposes an expanded compute-and-forward framework that incorporates both of these possibilities and permits an intuitive interpretation in terms of signal levels. Within this framework, recent achievability and optimality results are unified and generalized.
\end{abstract}


\section{Introduction} \label{s:intro}

Consider a Gaussian wireless network consisting of multiple transmitters and receivers. In this context, the compute-and-forward framework of \cite{ng11IT} enables the receivers to decode linear combinations of the messages, often at much higher rates than what would be possible for decoding the individual messages. This strategy can be used as a building block for relaying strategies~\cite{ng11PIEEE}, MIMO integer-forcing transceiver architectures~\cite{zneg14,hc12,hc13,shv13,oen13,oe15}, or interference alignment schemes~\cite{oen14,ncnc13ISIT,fn13ISIT}.

The coding scheme underlying this compute-and-forward framework maps the messages, which are viewed as elements of a vector space over a prime-sized finite field, to nested lattice codewords. The receivers are then able to decode integer-linear combinations of the codewords with coefficients chosen to approximate the real-valued channel coefficients (with better approximations yielding higher rates). Finally, these integer-linear combinations of lattice codewords are mapped back to the vector space over the finite field to yield linear combinations of the original messages. In other words, compute-and-forward creates a direct connection between network coding over a finite field and signaling over a Gaussian channel. 

We now recall two simple properties of Gaussian networks with multiple transmitters and receivers. First, the amount of available power may vary across transmitters. Second, the noise variance may vary across receivers. Thus, we would like our coding scheme to be versatile enough to allocate power unequally across transmitters as well as space codewords far enough apart to tolerate the noise at the targeted receivers. For classical random coding strategies that aim to deliver subsets of the messages to the receivers, these two forms of versatility can be viewed as simply the flexibility to adjust the targeted signal-to-noise ratio (SNR) for each codeword. However, in the compute-and-forward setting, the receivers want linear combinations of the messages, and the effect of noise variance and power on the nested lattice codebooks is more nuanced than in the classical random coding setting. In particular, the codeword spacing for a given message is determined by the maximum noise variance across all receivers whose desired linear combinations involve the message. This codeword spacing corresponds to the density of the fine lattice from which the codewords are drawn. Additionally, the power level for a given message is determined by the power constraint of the associated transmitter. This power level corresponds to the second moment of the coarse lattice used for shaping. 

The goal of this paper is to expand the compute-and-forward to include these two forms of versatility while retaining the connection between the finite field messages and the lattice codewords. Prior work has focused on either varying the noise tolerances (i.e., the codeword spacings) or the power levels but not both. For instance, the framework in~\cite{ng11IT} permits the codewords to tune their noise tolerances but requires that the codewords have the same power level. In~\cite{ncl11}, Nam \etal proposed a nested lattice technique that permits unequal power levels for multiple transmitters that communicate the sum of their codewords to a single receiver over a symmetric channel (i.e., all channel gains are equal to one). However, this technique does not establish a connection between messages drawn from a finite field and lattice codewords. Part of the motivation for this paper is to unify the techniques from~\cite{ng11IT} and~\cite{ncl11} into a single framework. 

The primary contribution of this paper is an \textit{expanded compute-and-forward framework} that permits both unequal powers and noise tolerances across transmitters while providing a mapping between finite field messages and lattice codewords. Interestingly, this framework allows us to interpret both the power constraint and noise tolerance associated to each message in terms of ``signal levels,'' in a manner reminiscent of the deterministic model of Avestimehr \etal~\cite{adt11}. Specifically, each transmitter's message is a vector from $\mathbb{F}_p^k$ where $p$ is prime. The power level of the transmitter determines a ``ceiling'' above which the message vector must be zero. Similarly, the noise tolerance of the transmitter determines a ``floor'' below which the message vector must be zero. The information symbols of the transmitter are placed between these constraints. 

Recent work has studied the problem of recovering multiple linear combinations at a single receiver. In particular, Feng \etal\cite{fsk13} linked this problem to the shortest independent vector problem~\cite{miccianciogoldwasser} and a sequence of papers has demonstrated its value for integer-forcing MIMO decoding~\cite{zneg14,shv13,oen13,oe15} as well as for integer-forcing interference alignment~\cite{oen14,ncnc13ISIT,fn13ISIT}. However, the original compute-and-forward framework does not capture some of the subtleties that arise when decoding multiple linear combinations. For instance, as shown by Ordentlich \etal\cite{oen14}, after one or more linear combinations have been decoded, they can be used as side information to eliminate some of the codewords from subsequent linear combinations. This \textit{algebraic successive cancellation} technique eliminates some of the rate constraints placed on codewords, i.e., it enlarges the rate region. Also, as shown in~\cite{oen14}, this technique can be used to approach the multiple-access sum capacity within a constant gap. Additionally, recent work by the first author~\cite{nazer12} as well as Ordentlich \etal\cite{oen13} revealed that decoded linear combinations can be used to infer the corresponding integer-linear combination of channel inputs, which can in turn be used to reduce the effective noise encountered in subsequent decoding steps. As argued in~\cite{oen13}, this \textit{successive computation} technique can reach the exact multiple-access sum capacity.

Our expanded compute-and-forward framework is designed with multiple linear combinations in mind. Specifically, we use a \textit{computation rate region} to capture the dependencies between rate constraints. Our achievability results broaden the algebraic successive cancellation and successive computation techniques to permit unequal powers as well as scenarios where the number of messages exceeds the number of desired linear combinations. We capture the prior results of~\cite{zneg14,oen14,nazer12,oen13} as special cases and shed additional light on the structure of optimal integer matrices for successive decoding. 

Beyond unifying existing results, this expanded framework is meant to serve as a foundation for ongoing and future applications of compute-and-forward and integer-forcing techniques. For example, the initial motivation behind developing this framework was the authors' exploration of integer-forcing for interference alignment~\cite{ncnc13ISIT}. Subsequently, He \etal\cite{hns14ISIT} have used this framework to propose a notion of uplink-downlink duality for integer-forcing and Lim \etal\cite{lfng15} used it to propose a discrete memoryless version of compute-and-forward.

\subsection{Related Work}

The main concept underlying compute-and-forward is that the superposition property of the wireless medium can  be exploited for network coding~\cite{acly00,lyc03,km03}. This phenomenon was independently and concurrently discovered by~\cite{py06ICC,ng06,zll06}, with the latter coining the phrase \textit{physical-layer network coding.} Subsequent efforts~\cite{ng07allerton,wnps10,ncl11} developed lattice coding strategies for communicating the sum of messages to a single receiver. This lead to the compute-and-forward framework~\cite{ng11IT} for multiple receivers that recover linear combinations of the messages (albeit with equal power constraints, unlike the single receiver framework of~\cite{ncl11}). 

As shown by Feng \etal~\cite{fsk13}, any compute-and-forward scheme based on nested lattice codes can be connected to network coding over a finite commutative ring. From this algebraic perspective, the compute-and-forward framework of~\cite{ng11IT} can be viewed as a special case that connects nested lattice codes generated via Construction A to network coding over a prime-sized finite field. Another important special case is the recent work of Tunali \etal\cite{thbn15} that develops a compute-and-forward scheme based on nested lattices over Eisenstein integers. For complex-valued channels, this scheme can offer higher computation rates on average (e.g., for Rayleigh fading) since the Eisenstein integers are a better covering of the complex plane than the Gaussian integers employed by \cite{ng11IT}. Several recent papers have also used the algebraic perspective of~\cite{fsk13} to propose practical codes and constellations for compute-and-forward~\cite{bl12,fm13,tnp13,hnt14}.

The line of work on compute-and-forward is part of a broader program aimed at uncovering the role of algebraic structure in network information theory, inspired by the paper of K\"orner and Marton~\cite{km79}. For instance, there are advantages for using algebraic structure in coding schemes for dirty multiple-access~\cite{pz09,pzek11,wang12}, distributed source coding~\cite{kp09,kp11,wagner11,mt10ISIT,yx14}, relaying~\cite{ng11PIEEE,nw12,sd13,hc13,nnw13,rgwg14}, interference alignment~\cite{bpt10,mgmk14,nm13,oen14,sa14,ncnc13ISIT,psp16}, and physical-layer secrecy~\cite{hy14,bmk10,vkt15,xu14}. Many of these works benefit from the development of lattice codes that are good for source and channel coding as well as binning~\cite{zf96,loeliger97,ftc00,zse02,elz05,oe16}. We refer readers to the textbook of Zamir~\cite{zamir} for a full history of these developments, an in-depth look at lattice constructions, achievable rates, and applications as well as a chapter~\cite{nz14} on the use of lattice codes in network information theory.

In recent, independent work, Zhu and Gastpar~\cite{zg14} proposed a compute-and-forward scheme for unequal powers based on the scheme of~\cite{ncl11}. They also showed how to use this scheme to reach the two-user multiple-access sum capacity (if the channel strength lies above a small constant). However, their scheme does not retain the connection to a finite field. 

The original motivation for the compute-and-forward strategy was the possibility of relaying in a multi-hop network while avoiding the harmful effects of interference between users (by decoding linear combinations) as well as noise accumulation (by decoding at every relay). Several works have investigated and improved upon the performance of the original compute-and-forward framework in the context of multi-hop relaying~\cite{nw12,nnw13,hc14,hc15,hc13ISIT,sa14}. Our expanded framework can improve performance further by permitting relays to employ unequal powers and decode multiple linear combinations when appropriate.\footnote{Following the conference publication of this work, Tan \etal\cite{tylk14} noted that our proposed message representation may be inefficient for multi-hop relaying if each relay naively treats a linear combination over $\ZZ_p^k$ as $k$ information symbols for the next hop. They proposed a lattice-based solution to this issue.  It is also possible to resolve this issue directly over the message representation by having each relay only use some of its $k$ received symbols as information symbols for the next hop. See~\cite[Section III.F]{ncncarxiv} for details.}

\subsection{Paper Organization}

We have strived to present our results, some of which are rather technical, in an accessible fashion. To this end, we begin in Section~\ref{s:probstate} with an informal overview of our framework to build intuition, before giving a formal problem statement. We then state our main results in Section~\ref{s:mainresults} without using any lattice definitions or properties. Afterwards, we introduce our nested lattice code construction in Section~\ref{s:lattices} and proceed to prove our main achievability theorems in Sections~\ref{s:proofcomppara} and~\ref{s:proofcompsucc}. 

\section{Problem Statement} \label{s:probstate}
In this section, we provide a problem statement for the expanded compute-and-forward framework. As mentioned earlier, our message structure can be interpreted in terms of signal levels that resemble the deterministic model of Avestimehr~\etal \cite{adt11}.\footnote{Unlike \cite{adt11}, we do not propose a deterministic model for analyzing communication networks. Instead, here, we use a deterministic model as an expository tool to explain the decoding requirements of our problem statement.} Unlike the original compute-and-forward problem of~\cite{ng11IT}, we will not aim to directly decode linear combinations of the messages. Instead, we associate each message realization with a coset and aim to decode linear combinations of vectors that belong to the same cosets as the transmitted messages. We describe a simple method of understanding this class of linear combinations through the use of ``don't care'' entries. As we will argue, if the coefficients of the linear combinations would suffice to recover a given subset of the messages in the original compute-and-forward framework~\cite{ng11IT}, they also suffice to recover these messages in the expanded compute-and-forward framework. For example, if the receiver obtains a full-rank set of linear combinations, all of the transmitted messages can be recovered successfully.  

For the sake of conciseness, we will focus on real-valued channel models with additive white Gaussian noise (AWGN). Our coding theorems can be applied to complex-valued channel models either via a real-valued decomposition of the channel~\cite{ng11IT,zneg14} or by building nested lattice codebooks directly over the complex field using either Gaussian or Eisenstein\footnote{For the case of Eisenstein integers, our lattice achievability proof from Theorem~\ref{t:mainlattice} will need to be generalized following the approach of Huang \etal\cite{thbn15}.} integers~\cite{fsk13,tnbh12}. While our framework is intended for AWGN networks with any number of sources, relays, and destinations, we find it clearer to first state our main results from the perspective of a single receiver that wishes to decode one or more linear combinations. In Section~\ref{s:multiplerx}, we will show how to apply our coding theorems to scenarios with multiple receivers. \remove{In Section~\ref{s:multihop}, we will discuss how to apply this framework to multi-hop networks.} Below, we state our notational conventions, essential definitions for our channel model, a high-level overview of the compute-and-forward problem, and a formal problem statement. 

\subsection{Notation} \label{s:notation}

We will employ the following notation. Lowercase, bold font (e.g., $\mathbf{x})$ will be used to denote column vectors and uppercase, bold font ({e.g.}, $\mathbf{H})$ will be used to denote matrices. For any matrix $\Hm$, we denote the transpose by $\Hm\T$, the span of its rows by $\rowspan(\Hm)$, and its rank by $\rank(\Hm)$. The notation $\| \xv \|$ denotes the Euclidean norm of the vector $\xv$ while $\lambda_{\text{min}}(\Hm)$ and $\lambda_{\text{max}}(\Hm)$ denote the minimum and maximum singular values of $\Hm$, respectively. We will denote the all-zeros column vector of length $k$ by $\zeros_k$, the $k \times n$ all-zeros matrix by $\Om_{k \times n}$, the all-ones column vector of length $k$ by $\ones_k$, and the $k \times k$ identity matrix by $\Id_k$. We will sometimes drop the subscript when the size can be inferred from the context. The $\log$ operation will always be taken with respect to base $2$ and we define $\log^+(x) = \max(0,\log(x))$.  

Our framework will make frequent use of operations over both the real field $\RR$ and the finite field consisting of the integers modulo $p$,  $\ZZ_p = \{0,1,\ldots,p-1\}$, where $p$ is prime.\footnote{Historically, the set of integers modulo $p$ has been denoted by $\mathbb{Z} / p \mathbb{Z}$, $\mathbb{Z} / (p)$, or $\mathbb{Z}_p$. Some mathematicians prefer to avoid the notation $\mathbb{Z}_p$ because it can be confused with the set of $p$-adic integers if $p$ is a prime number. Here, we will use the notation $\mathbb{Z}_p$ for the sake of conciseness, especially since we will frequently refer to vector spaces of the form $\mathbb{Z}_p^k$ (and have no need to refer to $p$-adic integers).} Addition and summation over $\RR$ will be denoted by $+$ and $\sum$, respectively. Similarly, addition and summation over the finite field $\ZZ_p$ where $p$ is prime will be denoted by $\oplus$ and $\bigoplus$, respectively. We will write the modulo-$p$ reduction of an integer $a \in \ZZ$ as $[a]\bmod{p} = r$ where $r \in \ZZ_p$ is the unique element satisfying $a = qp + r$ for some integer $q$. It will also be convenient to write the elementwise modulo-$p$ reduction of an integer vector $\av \in \ZZ^L$ and an integer matrix $\Am \in \ZZ^{M \times L}$ as $[\av] \bmod{p}$ and $[\Am] \bmod{p}$, respectively. Recall that addition and multiplication over $\ZZ_p$ are equivalent to addition and multiplication over $\RR$ followed by a modulo-$p$ reduction, i.e., 
$$q_1 w_1 \oplus q_2 w_2 = [ q_1 w_1 + q_2 w_2] \bmod{p} $$ where $q_1,q_2,w_1,w_2 \in \ZZ_p$. Note that on the right-hand side of the equation above, we have implicitly viewed $q_1,q_2,w_1,w_2$ as the corresponding elements of $\ZZ$ (under the natural mapping) in order to evaluate the real addition and multiplication. This will be the case throughout the paper: whenever elements of $\ZZ_p$ appear as part of operations over the reals, they will be implicitly viewed as the corresponding elements of $\ZZ$.

\subsection{Channel Model} \label{s:channelmodel}

Consider $L$ single-antenna transmitters that communicate to a receiver over a Gaussian multiple-access channel. See Figure~\ref{f:cfprobstatesinglerx} for an illustration. Each transmitter (indexed by $\ell = 1,2,\ldots,L$) produces a length-$n$ \textit{channel input} $\xv_\ell \in \RR^n$ subject to the \textit{power constraint}\footnote{In~\cite[Appendix C]{ng11IT}, it is argued that, for symmetric compute-and-forward, the expected power constraint $\mathbb{E} \big[\| \xv_\ell \|^2 \big]  \leq n P$ can be replaced with a hard power constraint $\| \xv \|^2 \leq nP$ without affecting the achievable rates. A similar argument should apply in our setting by first refining the nested lattice existence proof in~\cite[Theorem 2]{oe16} to show that the coarse lattices are also good for covering~\cite{elz05}. Alternatively, each encoder can throw out a constant fraction of its codebook to obtain a subcodebook that satisfies a hard power constraint while still maintaining the same achievable rate asymptotically.}
 
\begin{equation}
\ex \| \xv_\ell \|^2  \leq n P_\ell \label{e:powerconstraint}
\end{equation} where $P_\ell \geq 0$.

\begin{figure}[h]
\psset{unit=.77mm}
\begin{center}
\begin{pspicture}(7,2)(120,64)

\rput(14.5,55){$\wv_1$} \psline{->}(18,55)(23,55) \psframe(23,50)(33,60)
\rput(28,55){$\mathcal{E}_1$} \rput(40,58.75){$\xv_1\T$}
\psline[linecolor=black]{->}(33,55)(55,55)
\psframe(55,50)(65,60) \rput(60,55){$\hv_1$}
\psline{->}(65,55)(77,55)(84,42)

\rput(14.5,40){$\wv_2$} \psline{->}(18,40)(23,40) \psframe(23,35)(33,45)
\rput(28,40){$\mathcal{E}_2$} \rput(40,43.75){$\xv_2\T$}
\psline[linecolor=black]{->}(33,40)(55,40)
\psframe(55,35)(65,45) \rput(60,40){$\hv_2$}
\psline{->}(65,40)(83,40)

\rput(15,15){$\wv_L$} \psline{->}(18.5,15)(23,15) \psframe(23,10)(33,20)
\rput(28,15){$\mathcal{E}_L$} \rput(40,18.75){$\xv_L\T$}
\psline[linecolor=black]{->}(33,15)(55,15)
\psframe(55,10)(65,20) \rput(60,15){$\hv_L$}
\psline{->}(65,15)(77,15)(84,38)

\rput(28,29){\large{$\vdots$}}
\rput(60,29){\large{$\vdots$}}

\psframe[linewidth=2pt,linestyle=dashed,linecolor=LineBlue](45,-3)(74,64)
\rput(56.5,6.5){Channel}
\rput(56.5,2){Matrix}
\rput(68.5,4){\large{$\mathbf{H}$}}
\pscircle(85.5,40){2.5} \psline{-}(84.25,40)(86.75,40)
\psline{-}(85.5,38.75)(85.5,41.25) \psline{<-}(85.5,42.5)(85.5,47.5) \rput(85.5,50){$\Zm$}

\psline{->}(88,40)(98,40) \rput(92.5,43){$\Ym$}
\psframe(98,35)(108,45) \rput(103,40){$\mathcal{D}$}
\psline{->}(108,40)(113,40)
\rput(116,40){$\begin{array}{c}\uhv_1 \\ \uhv_2 \\ \vdots \\ \uhv_L\end{array}$}

\rput(102,20){$\uv_m = {\displaystyle \bigoplus_{\ell=1}^L}\  q_{m,\ell}  \wtv_{\ell}$}
\rput(96,8){$\wtv_\ell \in \llb \wv_\ell \rrb$}

\end{pspicture}
\end{center}
\caption{Block diagram for the compute-and-forward problem with a single receiver. Each transmitter has a message $\wv_\ell$ whose elements are taken from $\ZZ_p$. This message is embedded into $\ZZ_p^k$ (by zero-padding), mapped into a codeword $\xv_\ell \in \RR^n$, and sent over the channel. The receiver observes a noisy linear combination of these codewords, $\Ym = \sum_\ell \hv_\ell \xv_\ell\T + \Zm$ and attempts to recover the linear combinations $\uv_1,\uv_2,\ldots,\uv_L$ of the coset representatives of the original messages.}
\label{f:cfprobstatesinglerx}
\end{figure}

The receiver has $N_{\text{r}}$ antennas and observes an $N_{\text{r}} \times n$ dimensional \textit{channel output} $\mathbf{Y}$ that is a noisy linear combination of the inputs:
\begin{equation*}
\mathbf{Y} = \sum_{\ell=1}^L \mathbf{h}_{\ell} \xv_\ell^{\mathsf{T}} + \mathbf{Z} \label{e:channeloutput}
\end{equation*} where $\mathbf{h}_{\ell} \in \RR^{N_{\text{r}}}$ is the \textit{channel vector} between the $\ell$th transmitter and the receiver and $\mathbf{Z}\in\RR^{N_{\text{r}} \times n} $ is elementwise i.i.d.~$\Nc(0,1)$. It will often be convenient to group the channel vectors into a \textit{channel matrix}
\begin{equation*}
\mathbf{H} \triangleq \begin{bmatrix}\mathbf{h}_{1} & \mathbf{h}_{2} & \cdots & \mathbf{h}_{L}  \end{bmatrix} \ ,
\end{equation*} and concisely write the channel output as
\begin{equation}
\Ym = \Hm \Xm + \Zm \label{e:channeloutputmatrix}
\end{equation} where $\Xm \triangleq \big[ \xv_1~ \cdots ~ \xv_L \big]\T$ is the matrix of channel inputs. We will assume throughout that the channel matrix $\mathbf{H}$ is known to the receiver and unknown to the transmitters. However, the transmitters may assume that the maximum singular value $\lambda_{\text{max}}(\Hm)$ of the channel matrix is upper bounded by a constant.\footnote{This assumption will be used to make a connection between the solvability of the linear combinations over $\ZZ_p$ and the rank of the integer matrix $\Am$ over $\RR$. If we further assume that the channel matrix is generated randomly and the receiver is able to tolerate some probability of outage, then this condition can by replaced by the milder condition that $\pr\big(\lambda_{\text{max}}(\Hm) \geq \gamma\big) \rightarrow 0$ as $\gamma \rightarrow \infty$. See~\cite[Remark 10]{ng11IT} for further details.}

\subsection{High-Level Overview} \label{s:highlevel}

We now provide a high-level, informal overview of our compute-and-forward framework, which will help build intuition for the formal problem statement to follow. We begin by summarizing the original compute-and-forward framework from~\cite{ng11IT} and its MIMO generalization from~\cite{zneg14}. We then discuss how to incorporate unequal power constraints and recovering multiple linear combinations into an expanded framework. 

\noindent\textbf{Compute-and-Forward with Equal Powers:} The $\ell$th transmitter's message is a length-$k_\ell$ vector $\wv_{\ell}$ whose elements are from $\ZZ_p$ where $p$ is prime. Each message is zero-padded to a common length 
\begin{align*}
\mathbf{\bar{w}}_\ell = \begin{bmatrix} \wv_{\ell} \\ \mathbf{0}_{k - k_\ell} \end{bmatrix}
\end{align*} where $k = \max_\ell k_\ell$, and mapped to a length-$n$ real-valued codeword $\xv_\ell$ that satisfies the symmetric power constraint $\mathbb{E}\big[\| \xv_\ell \|^2\big] \leq nP$. The rate $R_\ell$ associated with a transmitter is the number of bits in its message normalized by the length of the codeword, $R_\ell = (k_\ell / n) \log{p}$. Given coefficients $q_1,q_2, \ldots, q_L \in \ZZ_p$, the receiver's goal is to recover a linear combination $\uv$ of the (zero-padded) messages,
\begin{equation*}
\uv = \bigoplus_{\ell = 1}^L q_\ell \mathbf{\bar{w}}_\ell.
\end{equation*} 

As argued in~\cite{ng11IT}, the main idea underlying compute-and-forward is to establish a connection between linear combinations of the messages and \textit{integer-linear combinations of the codewords}, in order to exploit the noisy linear combination taken by the channel. For instance, after applying an equalization vector $\bv \in \RR^{N_{\text{r}}}$, the channel output can be expressed as an integer-linear combination of the codewords\footnote{To be precise, our coding scheme employs dithered lattice codewords as the channel inputs $\xv_\ell$. However, the dithers can be removed at the receiver prior to decoding, and are thus ignored in this high-level overview.} with coefficients $a_1,a_2,\ldots,a_L \in \ZZ$ plus effective noise,
\begin{equation}
\bv\T \Ym  = \sum_{\ell = 1}^L a_\ell \xv\T_\ell  + \underbrace{\sum_{\ell = 1}^L \big( \bv\T \hv_\ell - a_\ell \big) \xv_\ell\T + \bv\T \Zm}_{\text{effective noise}} \ . \label{e:symcomp_effecnoise}
\end{equation} Each integer-linear combination of codewords is associated with a linear combination of the messages with coefficients $q_\ell = [a_\ell] \ \modp$. The performance of a compute-and-forward scheme is given by a computation rate region, which is specified by a function $R_{\comp}(\Hm, \av)$ that maps each channel matrix $\Hm$ and integer coefficient vector $\av = [a_1~a_2~\cdots~a_L]\T$ to a rate. Specifically, if the rates associated to messages with non-zero coefficients are less than the computation rate 
\begin{equation*}
\max_{\ell: a_\ell \neq 0} R_\ell < R_{\comp}(\Hm, \av)  \ ,
\end{equation*} then the linear combination with coefficients $q_\ell = [a_\ell] \ \modp$ is decodable with vanishing probability of error (with respect to the blocklength $n$). Operationally, this means that the scheme works in the absence of channel state information at the transmitter (CSIT) and that the receiver is free to choose which linear combination to decode, among those satisfying the computation rate constraint. Owing to this form of universality, compute-and-forward is applicable to scenarios with multiple receivers, each facing a different channel matrix and aiming to decode its own linear combination.

It shown in~\cite[Theorem 1]{ng11IT} and~\cite[Theorem 3]{zneg14} that the computation rate region described by 
\begin{equation}
	R_{\comp}(\Hm, \av) =  \frac{1}{2} \log^+\Bigg( \frac{P}{ \av\T \big( P^{-1} \Id   + \Hm\T \Hm\big)^{-1} \av} \Bigg) \label{e:symcomp_rate}
\end{equation} is achievable. The achievability proof utilizes nested lattice codebooks, which guarantees that any integer-linear combination of codewords is itself a codeword and thus afforded protection from noise. In particular, each transmitter's codebook is constructed using a fine lattice with effective noise tolerance $\sigma^2_{\eff,\ell}$ and a common coarse lattice that enforces the power constraint $P$. These nested lattices are chosen such that the $\ell$th transmitter's rate $R_\ell$ converges to $\frac{1}{2} \log^{+}(P/ \sigma^2_{\eff,\ell})$ asymptotically in the blocklength $n$. The nested lattice construction sends the field size $p$ to infinity with the blocklength $n$, in order to produce Gaussian-like channel inputs and obtain closed-form rate expressions.

\begin{remark}
If the field size is held fixed, we encounter similar issues as seen when evaluating the capacity of a point-to-point Gaussian channel under a finite input alphabet, i.e., we do not obtain closed-form rate expressions. For practical point-to-point codes, a common approach is to pick a finite constellation size based on the SNR~\cite{fc07} and accept a small rate loss. A similar approach enables practical codes for compute-and-forward~\cite{fsk13,hnt14}.\defsym
\end{remark}

\begin{remark}
Since the field size $p$ changes with the blocklength $n$, it does not make sense to specify the desired linear combinations via fixed coefficients $q_{\ell} \in \ZZ_p$. Instead, we fix desired integer coefficients $a_{\ell}$ and specify the desired linear combinations as those with coefficients satisfying $q_{m,\ell} = [a_{m,\ell}] \bmod{p}$.\defsym
\end{remark}

The effective noise tolerance of the codeword associated to $\sum_{\ell} a_\ell \xv_\ell$ is determined by the minimum noise tolerance over all participating fine lattices, $\min_{\ell: a_\ell \neq 0} \sigma^2_{\eff,\ell}$. Roughly speaking, an integer-linear combination is decodable if the variance of the effective noise in~\eqref{e:symcomp_effecnoise} is less than its effective noise tolerance. It can be shown that the denominator in~\eqref{e:symcomp_rate} corresponds to the variance of the effective noise when $\bv$ is chosen as the minimum mean-squared error (MMSE) projection.

\noindent\textbf{Compute-and-Forward with Unequal Powers:} In this paper, we expand the original compute-and-forward framework~\cite{ng11IT} in two aspects. First, we allow for an unequal power allocation across transmitters. Second, we explicitly consider the scenario where the receiver may wish to recover more than one linear combination. Decoding more than one linear combination appears in many contexts, such as recovering the $L$ transmitted messages in an integer-forcing MIMO receiver~\cite{zneg14}, relaying in a network where there are more transmitters than relays, and integer-forcing interference alignment~\cite{ncnc13ISIT}. In order to incorporate these two generalizations, we will expand the definition of the computation rate region. Here, we provide an intuitive description of our modifications before presenting a formal problem statement. 

We first describe our modification to the message structure. To each transmitter, we associate a power constraint $P_\ell$ and, as before, an effective noise tolerance $\sigma^2_{\eff,\ell}$. In the equal power setting, the rates varied across transmitters due only to the change in the effective noise tolerance. To cope with the fact that the messages have different lengths, they are zero-padded prior to taking linear combinations. Here, the rates will vary due to both changes in power and effective noise tolerance, for which zero-padding will not suffice. Instead, we take inspiration from the idea of \emph{signal levels} as introduced in~\cite{adt11}.

The length of the $\ell$th transmitter's message is a length-$(\kfl - \kcl)$ vector $\wv_\ell$ whose elements are drawn from $\ZZ_p$ where $p$ is prime and the parameters $k_{\c,\ell}, k_{\f,\ell} \in \NN$ will be determined by the power constraint $P_\ell$ and effective noise tolerance $\sigma_{\eff,\ell}^2$, respectively. Define $\kcmin = \min_\ell k_{\c,\ell}$ and $\kfmax = \max_\ell k_{\f,\ell}$. The total number of available signal levels is $k =  \kfmax - \kcmin$. Each message is embedded into $\ZZ_p^k$ using $\kcl - \kcmin$ leading zeros and $\kfmax - \kfl$ trailing zeros, 
\begin{equation}
\begin{bmatrix} 
\mathbf{0}_{\kcl - \kcmin} \\ \wv_{\ell} \\ \mathbf{0}_{\kfmax - \kfl}   
\end{bmatrix}  \ , 
\end{equation} and mapped to a length-$n$ real-valued codeword $\xv_\ell$ that satisfies the power constraint $\ex\big[\| \xv_\ell \|^2\big] \leq n P_\ell$. The rate $R_\ell$ associated with this transmitter is the number of message bits normalized by the codeword length, $R_\ell = ((k_{\f,\ell}- k_{\c,\ell})/n) \log{p}$.

\begin{figure}[h]
\psset{unit=0.8mm}
\begin{center}
\begin{pspicture}(0,-3)(100,32)

\rput(20,0){
\rput(-1,16){
$\wv_1\begin{cases}
~ & ~ \\
~ & ~ \\
~ & ~
\end{cases}$
}
\psframe[linewidth=1.25pt](0,0)(7.5,32)
\rput(4,-5){$\llb \wv_1 \rrb$}
\rput(3.75,1){
\pscircle[fillstyle=solid,fillcolor=black](0,2.5){2}
\pscircle[fillstyle=solid,fillcolor=LineBlue](0,7.5){2}
\pscircle[fillstyle=solid,fillcolor=LineBlue](0,12.5){2}
\pscircle[fillstyle=solid,fillcolor=LineBlue](0,17.5){2}
\pscircle[fillstyle=solid,fillcolor=LineBlue](0,22.5){2}
\pscircle(0,27.5){2}
\rput(0,27.5){\Large{\textcolor{LineBlue}{${*}$}}}
}
}

\rput(50,0){
\rput(-1,23.5){
$\wv_2\begin{cases}
~ & ~ \\
~ & ~ 
\end{cases}$
}
\psframe[linewidth=1.25pt](0,0)(7.5,32)
\rput(4,-5){$\llb \wv_2 \rrb$}
\rput(3.75,1){
\pscircle[fillstyle=solid,fillcolor=black](0,2.5){2}
\pscircle[fillstyle=solid,fillcolor=black](0,7.5){2}
\pscircle[fillstyle=solid,fillcolor=black](0,12.5){2}
\pscircle[fillstyle=solid,fillcolor=LineBlue](0,17.5){2}
\pscircle[fillstyle=solid,fillcolor=LineBlue](0,22.5){2}
\pscircle[fillstyle=solid,fillcolor=LineBlue](0,27.5){2}

}
}

\rput(80,0){
\rput(-1,11){
$\wv_3\begin{cases}
~ & ~ \\
~ & ~ \\
~ & ~
\end{cases}$
}
\psframe[linewidth=1.25pt](0,0)(7.5,32)
\rput(4,-5){$\llb\wv_3\rrb$}
\rput(3.75,1){
\pscircle[fillstyle=solid,fillcolor=LineBlue](0,2.5){2}
\pscircle[fillstyle=solid,fillcolor=LineBlue](0,7.5){2}
\pscircle[fillstyle=solid,fillcolor=LineBlue](0,12.5){2}
\pscircle[fillstyle=solid,fillcolor=LineBlue](0,17.5){2}
\pscircle(0,22.5){2}
\rput(0,22.5){\Large{\textcolor{LineBlue}{$*$}}}
\pscircle(0,27.5){2}
\rput(0,27.5){\Large{\textcolor{LineBlue}{$*$}}}
}
}
\end{pspicture}
\end{center}
\caption{Illustration of message cosets for $L = 3$ transmitters. The parameters are $k_{\f,1} = 7,\ k_{\f,2} = 5,\ k_{\f,3} = 8,\ k_{\c,1} = 3,\ k_{\c,2} = 2, \text{~and~}k_{\c,3} = 4$. Therefore, $\kfmax = 8, \ \kcmin = 2, \text{~and~} k = 6$. We use the symbol $*$ to stand for a ``don't care'' entry that can take any value in $\ZZ_p$, lightly shaded (blue) circles to denote information symbols (i.e., elements of $\wv_\ell$) that can take values in $\ZZ_p$, and black circles to denote zeros. Based on the parameters, the coset $\llb \wv_1 \rrb$ associated with the first transmitter has $1$ ``don't care'' entry, then $4$ information symbols, and then $1$ zero. Similarly, the coset $\llb \wv_2 \rrb$ associated with the second transmitter has $3$ information symbols followed by $3$ zeros. Finally, the coset $\llb \wv_3 \rrb$ associated with the third transmitter has $2$ ``don't care'' entries followed by $4$ information symbols. The receiver's goal is to recover a linear combination of vectors drawn from these cosets.}
\label{f:signallevels}
\end{figure}
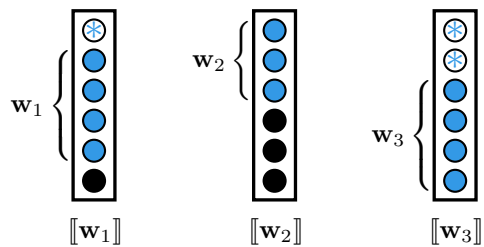

As mentioned above, the receiver may wish to decode more than one linear combination.  We compactly represent the receiver's demands through $L$ desired linear combinations $\uv_1,\uv_2,\ldots,\uv_L$ of the form
\begin{equation*}
\uv_m = \bigoplus_{\ell = 1}^L q_{m,\ell} \wtv_\ell \qquad \qquad \label{e:asym_lincomb}
\end{equation*} where $q_{m,\ell} \in \ZZ_p$ and $\wtv_\ell$ is an element of a certain coset $\llb \wv_\ell \rrb$ with respect to the message $\wv_\ell$.\footnote{To be precise, the linear combinations are affine varieties (i.e., translates of a vector subspace). However, we will simply refer to these as linear combinations throughout the paper.} Specifically, the coset consists of all vectors in $\ZZ_p^k$ for which the first $k_{\c,\ell} - \kcmin$ elements can take any values in $\ZZ_p$ (and can be viewed as ``don't care'' entries), the next $k_{\f,\ell} - k_{\c,\ell}$ elements contain the message $\wv_{\ell}$, and the remaining $\kfmax - k_{\f,\ell}$ entries are equal to zero. 

It is convenient to group the coefficients into a matrix $\Qm = \{q_{m,\ell} \}$. If the receiver wants fewer than $L$ linear combinations, then it can set the entries of the unneeded rows of $\Qm$ to zero. We have illustrated an example message structure in Figure~\ref{f:signallevels}.  Note that this framework includes the original problem statement as a special case by setting $k_{\c,1} = k_{\c, 2} = \ldots = k_{\c,L}$. 

We have relaxed the decoding requirements by allowing the receiver to decode linear combinations of vectors drawn from the same cosets as the message vectors. However, this does not affect the algebraic conditions for recovering messages from their linear combinations. Specifically, if the coefficient matrix $\Qm$ enables the receiver to recover $\wtv_\ell$, it can also immediately recover $\wv_{\ell}$. For example, if $\Qm$ is full rank, the receiver can recover all $L$ transmitted messages.

Our coding scheme will employ nested lattice codes in order to link linear combinations of the messages to integer-linear combinations of the codewords, just as in the symmetric case. We will select $k_{\c,\ell}$ using the transmitter's power constraint $P_\ell$ and $k_{\f,\ell}$ using the effective noise tolerance $\sigma_{\eff,\ell}^2$ so that the $\ell$th transmitter's rate $R_\ell$ converges to $\frac{1}{2} \log^{+}(P_\ell/ \sigma^2_{\eff,\ell})$ asymptotically in the blocklength $n$. As before, the field size $p$ tends to infinity with the blocklength $n$. Thus, we select desired  integer coefficients $a_{m,\ell} \in \ZZ$ and specify the desired linear combinations as those with coefficients satisfying $q_{m,\ell} = [a_{m,\ell}] \bmod{p}$.

The channel output can be written as an integer-linear combination of the codewords plus effective noise as in~\eqref{e:symcomp_effecnoise}. We consider $L$ such integer-linear combinations and collect the desired coefficients into an integer coefficient matrix $\Am = \{a_{m,\ell}\}$, which in turn specifies the coefficient matrix as $\Qm = [\Am]\! \mod{p}$ where the modulo operation is taken elementwise. 

\begin{figure}[h]
\psset{unit=.7mm}
\begin{center}
\begin{pspicture}(0,5)(95,95)

\psline[linewidth=2pt,linecolor=LineBlue](45,5)(45,45)(5,45)
\psline[linewidth=2pt,linecolor=LineRed](75,5)(75,60)(5,60)
\psline[linewidth=2pt,linecolor=LineGreen!70!black](5,75)(60,75)(60,5)

\pscircle[fillstyle=solid,fillcolor=black](54,30){2}

\psline[linewidth=1pt]{->}(5,5)(90,5)
\psline[linewidth=1pt]{->}(5,5)(5,90)
\rput(85,-1){\Large{$R_1$}}
\rput(-1,85){\Large{$R_2$}}

\rput(26,40){\textcolor{LineBlue}{$\Rc_{\comp}(\Hm_1,\Am_1)$}}
\rput(33,55){\textcolor{LineRed}{$\Rc_{\comp}(\Hm_2,\Am_2)$}}
\rput(40,70){\textcolor{LineGreen!70!black}{$\Rc_{\comp}(\Hm_3,\Am_3)$}}

\end{pspicture}
\end{center}
\caption{Sample evaluations of the computation rate region for $L = 2$ transmitters. The dot denotes the rate tuple for the two transmitters. Since this rate tuple falls inside the (red) rate region $\Rc_{\comp}(\Hm_2,\Am_2)$, the receiver can recover the linear combinations with integer coefficient matrix $\Am_2$ under channel matrix $\Hm_2$. Similarly, the linear combinations with integer coefficient matrix $\Am_3$ can be recovered under channel matrix $\Hm_3$ since the rate tuple falls inside the (green) rate region $\Rc_{\comp}(\Hm_3,\Am_3)$. On the other hand, since the rate tuple falls outside the (blue) rate region $\Rc_{\comp}(\Hm_1,\Am_1)$, the receiver is not required to recover the linear combinations with integer coefficient matrix $\Am_1$ under channel matrix $\Hm_1$.}\label{f:comprateregion}
\end{figure}
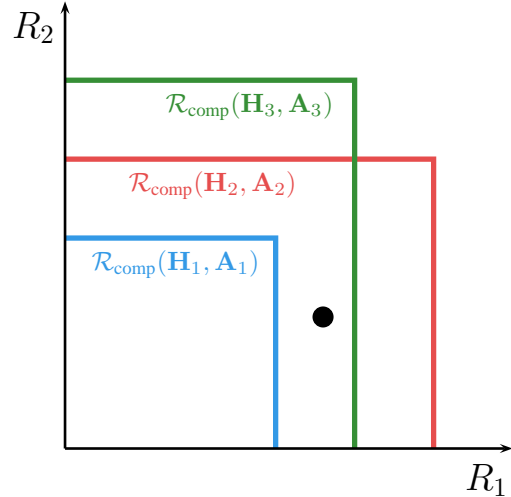
 
We now describe our modification to the computation rate region definition from~\cite{ng11IT}. Unlike the equal power setting, the rate region cannot be described using a single computation rate. For example, even if we are interested in only recovering a single linear combination at the receiver, the computation rate for each transmitter will still be determined by its own power combined with the effective noise for the linear combination. As another example, consider an equal power setting where the receiver wishes to decode more than one linear combination. Once it has decoded a single linear combination, it can use it as side information to help decode the next, meaning that the rate constraints for the linear combinations should be considered jointly. To capture such phenomena, we characterize the performance of a compute-and-forward scheme via a set-valued computation rate function $\Rc_{\comp}(\Hm,\Am)$ that maps each channel matrix $\Hm$ and integer coefficient matrix $\Am$ to a subset of $\RR_+^L$. This subset consists of all rate tuples that are achievable for the specified $\Hm$ and $\Am$ under the chosen coding scheme. That is, if the rate tuple associated to the messages falls inside the computation rate region,
\begin{equation*}
(R_1,R_2,\ldots,R_L) \in \Rc_{\comp}(\Hm,\Am) \ , 
\end{equation*} then the linear combinations with coefficient matrix $\Qm = [\Am]\bmod{p}$ are decodable with vanishing probability of error (with respect to the blocklength $n$). See Figure~\ref{f:comprateregion} for an illustration explaining the computation rate region.

\subsection{Formal Problem Statement} \label{s:formalprobstate}
We now provide a formal problem statement. A coding scheme is parametrized by the following:
\begin{itemize}
\item A positive integer $n$ denoting coding blocklength,
\item a positive prime number $p$ denoting the size of the finite field $\ZZ_p$ over which the linear combinations are taken, and
\item non-negative integers $k_{\c, \ell}, k_{\f,\ell}$ for $\ell \in\{1,2,\ldots, L\}$ satisfying $k_{\c,\ell} \leq k_{\f, \ell}\leq n$, where $k_{\f,\ell} - k_{\c,\ell}$ can be interpreted as the number of available signal levels at the $\ell$th transmitter. 
\end{itemize}

\begin{definition}[Messages]
For $\ell = 1,2,\ldots,L$, the $\ell$th transmitter has a \emph{message} $\wv_{\ell}$ that is drawn independently and uniformly over $\mathbb{Z}_{p}^{k_{\f,\ell}-k_{\c,\ell}}$. The \emph{rate} of the $\ell$th message (in bits per channel use) is 
\begin{equation*}
\frac{k_{\f,\ell}-k_{\c,\ell}}{n} \log{p} \ . 
\end{equation*} \defsym
\end{definition}

\begin{definition}[Encoders]
For $\ell = 1,2,\ldots,L$, the $\ell$th transmitter is equipped with an \emph{encoder} $\mathcal{E}_\ell: \mathbb{Z}_{p}^{k_{\f,\ell}-k_{\c,\ell}} \rightarrow \mathbb{R}^n$ that maps its message into a \emph{channel input vector} $\mathbf{x}_\ell = \mathcal{E}_\ell(\mathbf{w}_{\ell})$ subject to the power constraint~\eqref{e:powerconstraint}. \defsym
\end{definition}

\begin{definition}[Decoder]\label{d:decoder}
Define $\kcmin \triangleq \min_{\ell} k_{\c,\ell}$, $\kfmax \triangleq \max_{\ell} k_{\f,\ell}$, and $k \triangleq \kfmax - \kcmin$. Also, define the coset 
\begin{equation}
\label{e:msgcoset}
\llb \wv_{\ell} \rrb \triangleq \left\{\wv \in \ZZ_{p}^{k}: \mathbf{w} = \begin{bmatrix} \rv \\ \wv_\ell \\ \zeros_{\kfmax-\kfl} \end{bmatrix} \text{~for some~} \rv \in \ZZ_p^{\kcl - \kcmin} \right\}  
\end{equation}
The receiver is equipped with a \emph{decoder} $\Dc:\mathbb{R}^{N_{\text{r}} \times n} \times \mathbb{R}^{N_{\text{r}} \times L} \times \mathbb{Z}^{L \times L} \rightarrow \mathbb{Z}_{p}^{L \times k}$ that takes as inputs the channel observation $\Ym$ from~\eqref{e:channeloutputmatrix}, the channel matrix $\Hm$, and the desired integer coefficient matrix $\Am$, and outputs an estimate $\mathbf{\hat{U}} = \Dc(\Ym,\Hm,\Am)$. Let $\mathbf{\hat{u}}\T_m$ denote the $m$th row of $\Um$. We say that decoding is \emph{successful} if $\mathbf{\hat{u}}_1 = \uv_1,\mathbf{\hat{u}}_2 = \uv_2,\ldots,\mathbf{\hat{u}}_L = \uv_L$ for some linear combinations of the form 
\begin{equation*}
\uv_m = \bigoplus_{\ell=1}^L q_{m, \ell} \mathbf{\tilde{w}}_\ell
\end{equation*} where the $q_{m,\ell} \in \ZZ_p$ are the entries of $\Qm = [\Am] \bmod{p}$ and $\mathbf{\tilde{w}}_\ell \in \llb \wv_{\ell} \rrb$. We say that the decoder makes an \emph{error} if it is not successful. We sometimes refer to $\uv_1,\uv_2,\ldots,\uv_L$ as linear combinations with {integer coefficient matrix} $\Am$.
\defsym\end{definition}

\begin{definition}[Computation Rate Region] \label{d:comprate}
	A \emph{computation rate region} is specified by a set-valued function $\Rc_\comp(\Hm,\Am)$ that maps each channel matrix $\Hm \in \mathbb{R}^{N_{r}\times L}$ and integer coefficient matrix $\Am \in \mathbb{Z}^{L \times L}$ to a subset of $\mathbb{R}_+^L$. The computation rate region described by a set-valued function ${\Rc_\comp}$ is \textit{achievable} if, for every rate tuple $(R_{1}, R_{2}, \ldots, R_{L})\in \mathbb{R}_{+}^{L}$, $\epsilon > 0$, and $n$ large enough, 
there exist  
\begin{itemize}
\item parameters $p, k_{\c,\ell}, k_{\f,\ell}$ satisfying ${\displaystyle \frac{k_{\f,\ell}-k_{\c,\ell}}{n} \log p }> R_{\ell} - \epsilon$ for $\ell=1,2,\ldots,L$,
\item encoders $\mathcal{E}_{1},\mathcal{E}_{2},\ldots,\mathcal{E}_{L}$,
\end{itemize}
such that, 
\begin{itemize}
\item for every channel matrix $\Hm \in \mathbb{R}^{N_{r}\times L}$ and
\item every integer matrix $\mathbf{A},$
satisfying $(R_{1}, R_{2}, \ldots, R_{L}) \in \Rc_\comp(\Hm,\mathbf{A}),$ 
\end{itemize}
there exists a decoder $\mathcal{D}$ with probability of decoding error at most $\epsilon$. \defsym\end{definition}

\begin{remark}
The usual approach to defining a rate region is to first define the notion of an achievable rate tuple, and then define the rate region as the set of all achievable rate tuples. Our definition does not have this structure because the encoders $\mathcal{E}_{1},\mathcal{E}_2,\ldots,\mathcal{E}_{L}$ are assumed to be ignorant of both the channel matrix $\mathbf{H}$ and the integer coefficient matrix $\mathbf{A}$. Thus, a rate tuple $(R_1,R_2,\ldots,R_L)$ selected by the encoders will not lead to successful decoding for all $(\Hm,\Am)$ pairs. Instead, we characterize the rate region via the set-valued function $\Rc_\comp(\Hm,\Am)$, which specifies the rate tuples that lead to successful decoding for each $(\Hm,\Am)$ pair.  \defsym
\end{remark}
\begin{remark}
In some cases, it may be possible to simplify the framework by setting all of the ``don't care'' entries to zero, i.e., setting $\rv = \zerov$ in~\eqref{e:msgcoset}. For example, if the encoders do not dither their lattice codewords prior to transmission in the proof of Theorem~\ref{t:comppara}, then it follows from Definition~\ref{d:linearlabeling} that this is possible. However, this will significantly complicate the proof, since the effective noise will not be independent from the desired linear combination. More generally, this problem statement is directly applicable for compute-and-forward over discrete memoryless networks~\cite{lfng15}. In that setting, the ``don't care'' entries play an important role for selecting codewords with the desired type, following the joint typicality encoding approach of Padakandla and Pradhan~\cite{pp13}. \defsym
\end{remark}

\section{Main Results} \label{s:mainresults}

In this section, we state our main coding theorems as well as provide intuitions and examples. Although the proofs of our achievability results rely on the existence of good nested lattice codes, we have deferred (nearly all) discussion of lattices to subsequent sections in order to make our main results more accessible. Our primary technical contribution is the generalization of the compute-and-forward framework to allow for unequal powers, multiple receive antennas, and recovering more than one linear combination (at a single receiver), all while maintaining a connection to $\ZZ_p^k$. We demonstrate the utility of our generalization of the compute-and-forward framework by applying it to the classical and compound Gaussian multiple-access channels.

Note that all of our results implicitly assume that the transmitters do not have access to channel state information. It is well-known that, with channel state information, the transmitters can steer the channel gains towards integer values, which can improve the end-to-end rates~\cite{nsgs09,nw12}. This can be captured within our framework by multiplying each channel gain by a scalar (chosen using channel state information). More generally, the achievable rates of linear beamforming strategies for multi-antenna transmitters can be captured within our framework by multiplying each transmitter's channel vector (on the right) by its beamforming vector. See~\cite{ncnc13ISIT} for an application to interference alignment and~\cite{hns14ISIT} for an application to uplink-downlink duality. 

\remove{The main contributions of this paper are Theorems \ref{t:comppara}, \ref{t:compsucc}, \ref{t:compprimitive}, \ref{t:compparamac}, \ref{t:compsuccmac}, \ref{t:compparamultiplerx} and \ref{t:compsuccmultiplerx}. Theorems \ref{t:comppara},\ref{t:compsucc} and \ref{t:compprimitive} describe achievable rate regions for the problem statement of Section \ref{sec:}. Theorems \ref{t:compparamac}, \ref{t:compsuccmac}, \ref{t:compparamultiplerx}, \ref{t:compsuccmultiplerx} apply the rate regions obtained in Theorems  \ref{t:comppara},\ref{t:compsucc} to various multiple access settings.} We now provide a high-level summary of our results:
\begin{itemize}
	\item \textbf{Parallel Computation:} Theorem~\ref{t:comppara} expands the computation rate region from~\cite{ng11IT} by permitting unequal power allocation across the transmitters. Each of the desired linear combinations is decoded independently of the others, which we refer to as ``parallel computation.''	
	\item \textbf{Successive Computation:} Theorem~\ref{t:compsucc} enlarges the computation rate region from~Theorem~\ref{t:comppara} by decoding the linear combinations one-by-one and employing successive cancellation. This ``successive computation'' technique can be viewed as a generalization of~\cite{nazer12,oen13} to the unequal power setting. Theorem~\ref{t:primitive} shows that it suffices to use ``primitive'' integer coefficient matrices, i.e., integer matrices with a unimodular completion.
	\item \textbf{Multiple-Access Sum Capacity within a Constant Gap:} Theorem~\ref{t:compparamac} shows that the sum of the $L$ highest parallel computation rates always lies within $L$ bits of the sum capacity of the underlying multiple-access channel. Furthermore, these computation rates can mapped to individual users, which leads to an operational interpretation as a multiple-access strategy. An implication of the theorem is that the parallel computation strategy, when combined with algebraic successive cancellation, is approximately optimal for multiple access.  From one perspective, Theorem \ref{t:compparamac} generalizes the compute-and-forward transform of~\cite{oen14} to the unequal power setting.	
	\item \textbf{Multiple-Access Sum Capacity:} Theorem~\ref{t:compsuccmac} shows that, for any unimodular integer coefficient matrix, the sum of the $L$ successive computation rates is exactly equal to the sum capacity of the underlying multiple-access channel. Under certain technical conditions, this can be employed as an optimal multiple-access strategy. Theorem \ref{t:compsuccmac} generalizes the successive integer-forcing scheme of~\cite{oen13} to the unequal power setting.
	\item \textbf{Multiple Receivers:} Theorems~\ref{t:compparamultiplerx} and~\ref{t:compsuccmultiplerx} give achievable rate regions for multiple receivers for parallel and successive computation, respectively.
	\remove{\item \textbf{Multiple Hops:} Theorem~\ref{t:multihop} shows that, for a multi-hop relay network consisting of $K$ transmitters, $K$ relays that each recover a single linear combination, and a single destination, no excess sum rate is required for the relay-to-destination links.}
	\end{itemize}

We now introduce some additional notation. Define $\Pm$ to be the diagonal matrix of the power constraints,
\begin{equation*}
\Pm  \triangleq \diag(P_{1}, P_{2}, \ldots, P_{L}) \ , 
\end{equation*} and let $\mathbf{F}\in \mathbb{R}^{L\times L}$ be any matrix that satisfies 
\begin{equation}
\label{e:F}
\Fm\T\Fm = \big(\Pm^{-1} + \Hm\T \Hm\big)^{-1} \ .
\end{equation} Note that $\Fm$ is not unique and can determined via several approaches, such as via its eigendecomposition or its Cholesky decomposition.

Recall that $\mathbf{A} \in \mathbb{Z}^{L \times L}$ is the desired integer coefficient matrix. Let $\av_m\T$ denote the $m$th row of $\Am$ and $a_{m,\ell}$ denote the $(m,\ell)$th entry. We will sometimes refer to $\uv_m$ as the \textit{linear combination with integer coefficient vector} $\av_m$. In certain scenarios (e.g., relaying, interference alignment), the receiver may wish to decode $M < L$ linear combinations. This can be explicitly represented in our framework by setting the last $M-L$ rows of $\Am$ to be zero but it will be convenient to develop more compact notation. Let $\Abm$ be an $M \times L$ integer matrix with $M < L$. We will implicitly take $\Rc_\comp(\Hm,\Abm)$ to mean $\Rc_\comp(\Hm,\Am)$ where
\begin{equation*}
\Am = \begin{bmatrix}
\Abm \\ \mathbf{O}_{(L - M) \times L} 
\end{bmatrix} \ . 
\end{equation*} 

We also recall the following basic result from linear algebra.

\begin{lemma}[Woodbury Matrix Identity] \label{l:woodbury}
For any (appropriately-sized) matrices $\Mm_1$, $\Mm_2$, $\Mm_3$, and $\Mm_4$, we have that
\begin{align*}
&\big(\Mm_1 + \Mm_2 \Mm_3 \Mm_4\big)^{-1} \\&= \Mm_1^{-1} - \Mm_1^{-1} \Mm_2 \big( \Mm_3^{-1} + \Mm_4 \Mm_1^{-1} \Mm_2\big)^{-1} \Mm_4 \Mm_1^{-1} 
\end{align*} \thmsym
\end{lemma} See, e.g.,~\cite[Theorem 18.2.8]{harville} for a proof. As an example, take $\Mm_1 = \Pm^{-1}$, $\Mm_2 =  \Hm\T$, $\Mm_3 = \Id$, and $\Mm_4 = \Hm$. It then follows from the Woodbury Identity that 
\begin{equation}
\big(  \Pm^{-1}  + \Hm\T \Hm \big)^{-1} = \Pm - \Pm \Hm\T \big(\Id + \Hm \Pm \Hm\T\big)^{-1} \Hm \Pm \ . \label{e:woodburyh}
\end{equation} We will make frequent use of this identity.
 
\subsection{Parallel Computation}
\label{s:mainpara}

We begin with a ``parallel computation'' strategy, in which the receiver decodes each of the desired linear combinations independently. To recover the $m$th linear combination, the receiver applies an equalization vector $\bv_m \in \RR^\ant$ to its observation to obtain the effective channel
\begin{align*}
\ytv_m &= \bv_m\T \Ym \\
&= \bv_m\T \Hm \Xm + \bv_m\T \Zm \\
&= \av_m\T \Xm ~ + ~\underbrace{\big( \bv_m \T \Hm - \av_m\T\big) \Xm + \bv_m\T \Zm}_{\text{effective noise}} \ , 
\end{align*} and then decodes to the closest lattice codeword. In Section~\ref{s:proofcomppara}, we will argue that, if the fine lattices associated with $\av_m\T \Xm$ can tolerate an effective noise variance of
\begin{align}
\sigma_{\para}^2(\Hm,\av_m,\bv_m) &\define \frac{1}{n} \ex\big\| \big( \bv_m \T \Hm - \av_m\T\big) \Xm + \bv_m\T \Zm \big\|^2  \nonumber \\
&= \| \bv_m \|^2 + \big\| \big(\bv_m\T \Hm - \av_m\T \big) \Pm^{1/2} \big\|^2 \ , \label{e:noisevarpara2} 
\end{align} then the linear combination with integer coefficient vector $\av_m$ can be successfully decoded. 

\begin{lemma}
\label{l:mmsepara}
The equalization vector $\bv_m \in \RR^\ant$ that minimizes the effective noise variance from~\eqref{e:noisevarpara2} is the MMSE projection vector
\begin{align*}
\bv_{\opt,m}\T = \av_m\T \Pm \Hm\T \big(\Id + \Hm \Pm \Hm\T \big)^{-1} \ . 
\end{align*} The minimal effective noise variance is
\begin{align}
\sigma^2_{\para}(\Hm,\av_m) &\define \sigma^2_{\para}(\Hm,\av_m,\bv_{\opt,m}) \label{e:minvarparadef} \\
&= \av_m\T \big(\Pm^{-1} + \Hm\T \Hm\big)^{-1} \av_m \nonumber \\ 
&= \| \Fm \av_m \|^2 \ , \label{e:minvarpara2}
\end{align} 
where $\mathbf{F}$ is any matrix that satisfies (\ref{e:F}). \thmsym
\end{lemma} See Appendix~\ref{app:proofmmsepara} for a proof.

The users that participate in the integer-linear combination $\av_m\T \Xm$ are simply those with non-zero integer coefficients, $a_{m,\ell} \neq 0$. These users should satisfy the rate constraints $R_\ell < 1/2\log^+\big(P_\ell / \sigma^2_{\para}(\Hm,\av_m)\big)$ in order for the receiver to directly decode the linear combination. Note that the receiver can also decode linear combinations \textit{indirectly}. Specifically, it can decode \textit{any} integer-linear combinations whose integer coefficient matrix $\Atm \in \ZZ^{L \times L}$ has the same rowspan as $\Am$, and then simply solve for $\Am \Xm$ from $\Atm \Xm$. Therefore, the achievable computation rate region involves a union over all integer matrices with the same rowspan, as captured by the following theorem.

\begin{theorem} \label{t:comppara}
For an AWGN network with $L$ transmitters, a receiver, and  power constraints $P_1, \ldots, P_L$, the following computation rate region is achievable,
\begin{align*}
&\mathcal{R}_{\comp}^{(\para)}(\Hm,\Am) = \bigcup_{\substack{\Atm \in \mathbb{Z}^{L \times L} \\ \rowspan(\Am)\subseteq \rowspan(\Atm) }} \hspace{-0.5in} \mathcal{R}_{\para}(\Hm,\Atm) 
\end{align*} where
\begin{align*}
&\mathcal{R}_{\para}(\Hm,\Atm) \define \Bigg\{ (R_1,\ldots,R_L) \in \mathbb{R}_+^L ~: \\
& \qquad  R_\ell \leq  \frac{1}{2}\log^+\bigg( \frac{P_\ell}{\sigma_{\para}^2(\Hm,\atv_m)} \bigg) ~\forall(m,\ell) \text{~s.t.~} \tilde{a}_{m,\ell} \neq 0 \Bigg\} 
\end{align*} and $\atv_m\T$ and $\tilde{a}_{m,\ell}$ are the $m$th row and $(m,\ell)$th entry of $\Atm$, respectively. \thmsym
\end{theorem} The achievability proof is presented in Section~\ref{s:proofcomppara}.

\begin{remark}
The computation rate region described in Theorem \ref{t:comppara}, when restricted to the special case of equal powers and a single antenna at the receiver, yields the rate region from \cite[Theorem 1]{ng11IT}.\footnote{Technically speaking, our expression of the achievable rate region slightly generalizes the region described in \cite{ng11IT} since we explicitly take a union over the set of integer matrices $\mathbf{\tilde{A}}$ that contain the rowspan of $\mathbf{A}$. This possibility is discussed in~\cite[Remark 7]{ng11IT} but not formally included in the statement of~\cite[Theorem 1]{ng11IT}.} \defsym
\end{remark}

The following lemma restricts the search space for integer vectors.
\begin{lemma} \label{l:normbound}
Let $\lambda_{\text{max}}\big(\Id + \Pm \Hm\T \Hm\big)$ denote the maximum eigenvalue of $\Id + \Pm \Hm\T \Hm$. Consider an integer matrix $\Atm \in \ZZ^{L \times L}$ and user index $\ell \in \{1,2,\ldots,L\}$. If, for some $m \in \{1,2,\ldots,L\}$, the $(m,\ell)$th entry of $\Atm$ satisfies
\begin{align*}
a_{m,\ell}^2 > \lambda_{\text{max}}\big(\Id + \Pm \Hm\T \Hm\big) \ ,
\end{align*} then $R_\ell = 0$ for any rate tuple $(R_1,\ldots,R_L) \in \mathcal{R}_{\para}(\Hm,\Atm)$. \thmsym
\end{lemma} The proof is deferred to Appendix~\ref{app:proofnormbound}.

\begin{example} \label{ex:comppara_sum} Consider a receiver that observes $Y = X_1 + \cdots + X_L + Z$ and wants to decode the linear combination with integer coefficient vector $\av_1 = \ones$ (i.e., the sum of the messages over $\ZZ_p$). (For $L = 2$, this corresponds to the multiple-access phase in a Gaussian two-way relay channel as studied by~\cite{wnps10,ncl10}.) Using~\eqref{e:woodburyh}, we get
\begin{align*}
\sigma_{\para}^2(\ones\T,\ones\T) &= \ones\T \Pm \ones  - \ones\T \Pm \ones \big(1 + \ones\T \Pm \ones\big)^{-1} \ones\T \Pm \ones \\
& =\frac{ \sum_{\ell=1}^L P_\ell } {1 + \sum_{\ell = 1}^L P_\ell} \ .
\end{align*} The resulting rate region is \begin{align*}
&\mathcal{R}_{\para}(\ones\T,\ones\T) \\ &= \Bigg\{ (R_1,\ldots,R_L) \in \mathbb{R}_+^L ~:~ R_\ell \leq  \frac{1}{2}\log^+\bigg( \frac{P_\ell}{\sum_{i=1}^L P_{i}} + P_{\ell} \bigg) \Bigg\} 
\end{align*} and is equal to that derived by Nam, Chung, and Lee~\cite{ncl11} for decoding the sum of lattice codewords over a multiple-access channel with unequal powers. 

It is well-known that this region can be expanded at low SNR by decoding the messages individually and then computing the sum. That is, the rowspace of $\tildebf{A} = \Id$ contains that of $\Am$ and yields the rate region
\begin{align*}
&\mathcal{R}_{\para}(\ones\T,\Id) \\
&= \Bigg\{ (R_1,\ldots,R_L) \in \mathbb{R}_+^L ~:~ R_\ell \leq  \frac{1}{2}\log\bigg( 1 + \frac{P_\ell}{\sum_{i\neq \ell} P_{i}}  \bigg) \Bigg\} \ .
\end{align*} The computation rate region can be further expanded by taking the union over all viable $\tildebf{A}$, e.g., by first decoding sums of subsets of the messages and then combining these. \defsym
\end{example}

\begin{example} \label{ex:comppara_mac}
Consider a receiver that wishes to recover all of the messages, $\Am = \Id$. Let $\deltav_m$ denote the $m$th column of $\Id$. Using~\eqref{e:woodburyh}, the effective noise variances are
\begin{align*}
\sigma_{\para}^2(\Hm,\deltav_m) &=\deltav_m\T  \big(\Pm- \Pm \Hm\T \big(\Id + \Hm \Pm \Hm\T\big)^{-1} \Hm \Pm \big)\deltav_m\\
&= P_m - P_m^2 \hv_m\T  \big(\Id + \Hm \Pm \Hm\T\big)^{-1} \hv_m \ . 
\end{align*}  The effective SNR of the $\ell$th user is
\begin{align*}
\frac{P_\ell}{\sigma_{\para}^2(\Hm,\deltav_\ell)} &= \frac{1}{1 - P_\ell \hv_\ell\T \big(\Id + \Hm \Pm \Hm\T\big)^{-1} \hv_\ell} \\
&= 1 + P_\ell \hv_\ell\T \Big( \Id + \sum\limits_{i\neq \ell} P_i \hv_i \hv_i\T \Big)^{-1} \hv_\ell
\end{align*} where the last step uses the Woodbury Matrix Identity (Lemma~\ref{l:woodbury}) with $\Mm_1 = 1$, $\Mm_2 = P_\ell^{1/2} \hv_\ell\T$, $\Mm_3 = \big( \Id + \sum_{i\neq \ell} P_i \hv_i \hv_i\T \big)^{-1}$, and $\Mm_4 = \Mm_2\T$. Finally, the rate region (for direct decoding) is 
\begin{align*}
&\mathcal{R}_{\para}(\Hm,\Id) =  \Bigg\{ (R_1,\ldots,R_L) \in \mathbb{R}_+^L ~:~ \\
&\qquad ~~~~R_\ell \leq  \frac{1}{2}\log\bigg(  1 + P_\ell \hv_\ell\T \Big( \Id + \sum_{i\neq \ell} P_i \hv_i \hv_i\T \Big)^{-1} \hv_\ell  \bigg) \Bigg\} \ , 
\end{align*} which matches the rates attainable via i.i.d.~Gaussian codebooks and treating interference as noise (see, e.g., \cite[Equation 8.69]{tseviswanath}). As shown in~\cite{zneg14} (for equal powers), the rate region for recovering all of the messages can be significantly expanded by \textit{integer-forcing decoding}, i.e.,  taking the union over all rank-$L$ matrices $\tildebf{A}$ in Theorem~\ref{t:comppara}. See Section~\ref{s:sumrate} for more details. \defsym \end{example}

As hinted in the example above, a receiver can recover all $L$ transmitted messages if the coefficient matrix $\Qm = [\Am] \bmod{p}$ has rank $L$ over $\ZZ_p$ and, under mild technical conditions, we can simply check if the integer coefficient matrix $\Am$ itself has rank $L$ over $\RR$. More generally, to recover the $m$th message $\wv_m$, the coefficient matrix must satisfy $\deltav_m\T \in \rowspan(\Qm)$ over $\ZZ_p$. It often more convenient to check whether $\delta_m\T \in \rowspan(\Am)$ over $\RR$ and the following lemma gives a sufficient condition on when this is allowable.

\begin{lemma}\label{l:solvablereal}
Consider the set of $L \times L$ integer matrices whose entries' magnitudes are upper bounded by a constant $a_{\text{max}}$. Let $\deltav_m\T$ denote the $m$th row of the $L \times L$ identity matrix. If $\deltav_m\T \in \rowspan(\Am)$, then any rate tuple $(R_1,\ldots,R_L)$ in the computation rate region for $\Am$ from Theorem~\ref{t:comppara} (or Theorem~\ref{t:compsucc} below) is also achievable for recovering message $\wv_m$. \thmsym
\end{lemma} Roughly speaking, for large enough\footnote{Recall that the field size $p$ can be chosen as large as desired according to Definition~\ref{d:comprate}.} $p$, it can be shown that $\deltav_m\T \in \rowspan(\Am)$ over the reals implies $\deltav_m\T \in \rowspan\big([\Am] \bmod{p}\big)$ over $\ZZ_p$ since the entries are integer-valued and bounded. The proof follows along the same lines as that of~\cite[Theorem 5]{oen14} and is omitted due to space considerations.

\begin{remark}
From Lemma~\ref{l:normbound}, it suffices to check integer matrices $\Am$ whose entries' magnitudes are upper bounded by $\lambda_{\text{max}}\big(\Id + \Pm \Hm\T \Hm\big)$ for Theorem~\ref{t:comppara}. Since we have imposed an upper bound on the maximum singular value of the channel matrix $\Hm$ in Section~\ref{s:channelmodel}, we automatically obtain an upper bound on the entries of all viable $\Am$. Furthermore, for scenarios where some probability of outage is permitted, it can be argued that it suffices for the probability density function of the largest singular value to have a vanishing tail. See~\cite[Remark 10]{ng11IT} for more details. \defsym
\end{remark}

In certain scenarios, it may be useful to recover the integer-linear combination of the codewords (rather than the linear combination of the messages). For instance, two relays in a network may wish to simultaneously transmit a linear function of the codewords to benefit from a coherent gain~\cite{nnad12}. Additionally, in a single-hop network, it is often more convenient to work directly with the codewords and ignore the finite field perspective. 

\begin{lemma} 
\label{l:realsum}
Under the nested lattice coding framework employed for Theorem~\ref{t:comppara} (and Theorem~\ref{t:compsucc} in the next subsection), if the linear combinations $\uv_1,\ldots,\uv_L$ with integer coefficient matrix $\Am$ can be successfully recovered, then the integer-linear combinations of the codewords $\av_1\T \Xm,\ldots,\av_L\T \Xm$ can be successfully recovered as well. \thmsym
\end{lemma} This follows directly from Lemma~\ref{l:realsumdirect} in Section~\ref{s:proofcompsucc}. As a quick example, consider the scenario from Example~\ref{ex:comppara_sum}. If the receiver can recover the modulo sum of the messages $\bigoplus_\ell \wtv_\ell$, then it can also recover the real sum of the codewords $\sum_\ell \xv_\ell$.

\begin{remark}\label{r:realsolvable} Notice that, for a single-hop network, Lemma~\ref{l:realsum} allows us to directly check recoverability conditions over $\RR$ instead of $\ZZ_p$. For instance, consider an integer matrix $\Am$ such that (i) $\deltav_m\T \in \rowspan(\Am)$ over $\RR$ and (ii) $\deltav_m\T \notin \rowspan\big([\Am]\bmod{p}\big)$ over $\ZZ_p$. The latter condition means that it is not possible to retrieve the $m$th message from the recovered linear combinations. However, using Lemma~\ref{l:realsum}, we can first recover the integer-linear combinations of the codewords $\Am \Xm$, solve for the $m$th codeword $\xv_m$ over $\RR$, and then infer the $m$th message $\wv_m$ from $\xv_m$. (This is not always possible in a multi-hop network since the destination may not have access to the channel observations that were used to recover the linear combinations.) \defsym
\end{remark}

\subsection{Successive Computation}
\label{s:mainsucc}

Consider a classical receiver that recovers individual codewords in a certain order. It is well-known that, once a codeword has been successfully decoded, it is beneficial to remove it from the channel observation so that subsequent decoding steps encounter less interference. Here, we explore an analogue of this successive interference cancellation (SIC) technique for compute-and-forward. 

Assume that the receiver has (correctly) decoded the linear combinations with integer coefficient vectors $\av_1, \ldots, \av_{m-1}$. These linear combinations can be used as side information at the receiver for two different forms of successive cancellation. The first reduces the effective noise variance and the second reduces the number of users that need to tolerate this effective noise.

\begin{remark}
Without loss of generality, we restrict ourselves to the decoding order $1,2,\ldots,L$. Any other decoding order can be mimicked by permuting the rows of the integer coefficient matrix $\Am$. \defsym
\end{remark}

We begin by showing how the side information can be employed to decrease the effective noise variance. From Lemma~\ref{l:realsum}, we know that the receiver has access to the integer-linear combinations of the codewords $\av_1\T \Xm, \ldots, \av_{m-1}\T \Xm$, which we will write concisely as $\Am_{m-1} \Xm$ where
\begin{align*}
\Am_{m-1} \define  \begin{bmatrix} \av_1\T \\ \vdots \\ \av_{m-1}\T \end{bmatrix} \ .
\end{align*} Operationally, the receiver forms the effective channel 
\begin{align*}
\ytv_m &= \bv_m\T \Ym + \cv_m\T \Am_{m-1} \Xm \\
&= \av_m\T \Xm~ + ~\underbrace{\big( \bv_m \T \Hm + \cv_m\T \Am_{m-1} - \av_m\T\big) \Xm + \bv_m\T \Zm}_{\text{effective noise}} \ , 
\end{align*} where $\bv_m \in \RR^{\ant}$ and $\cv_m \in \RR^{m-1}$ are equalization vectors. We denote the effective noise variance for successive computation by 
\begin{align}
&\sigma_{\succ}^2(\Hm,\av_m,\bv_m,\cv_m | \Am_{m-1}) \nonumber \\
&\define \frac{1}{n} \ex\big\| \big( \bv_m \T \Hm + \cv_m\T \Am_{m-1} - \av_m\T\big) \Xm + \bv_m\T \Zm \big\|^2   \nonumber \\
&= \| \bv_m \|^2 + \big\| \big(\bv_m\T \Hm + \cv_m\T \Am_{m-1} - \av_m\T \big) \Pm^{1/2} \big\|^2 \ . \label{e:noisevarsucc}
\end{align} Note that, so long as $\av_m\T$ is not orthogonal to $\rowspan(\Am_{m-1})$, we can select $\cv_m$ to obtain a strictly lower effective noise variance than possible via parallel computation~\eqref{e:noisevarpara2}.

\begin{lemma}
\label{l:mmsesucc} Assume that $\rank(\Am_{m-1}) = m-1$. (Otherwise, delete rows of $\Am_{m-1}$ until it has full rank and ignore the associated linear combinations in the decoding scheme.) The equalization vectors $\bv_m \in \RR^\ant$ and $\cv_m \in \RR^{m-1}$ that minimize the effective noise variance from~\eqref{e:noisevarsucc} are the MMSE projection vectors
\begin{align*}
\bv_{\opt,m}\T &= (\av_m\T - \cv_{\opt,m}\T \Am_{m-1} )\Pm \Hm\T \big(\Id + \Hm \Pm \Hm\T\big)^{-1} \\
\cv_{\opt,m}\T &= \av_m\T \Fm\T \Fm \Am_{m-1}\T \big( \Am_{m-1}  \Fm\T \Fm \Am_{m-1}\T \big)^{-1} \ ,
\end{align*} 
where $\mathbf{F}$ is a matrix that satisfies (\ref{e:F}). Let 
\begin{align}
\Nm_{m-1} &\define \Id - \Fm \Am_{m-1}\T \big( \Am_{m-1}  \Fm\T \Fm \Am_{m-1}\T \big)^{-1} \Am_{m-1} \Fm\T \label{e:nullspace}
\end{align} denote the projection matrix for the nullspace of $\Fm \Am_{m-1}\T$. The minimal effective noise variance is
\begin{align}
&\sigma^2_{\succ}(\Hm,\av_m | \Am_{m-1} ) \nonumber \\ &\define \sigma^2_{\succ}(\Hm,\av_m,\bv_{\opt,m},\cv_{\opt,m} | \Am_{m-1}) \label{e:minnoisevarsucc1}\\
&= \av_m\T \Fm\T \Nm_{m-1} \Fm \av_m \nonumber \\
&= \big\| \Nm_{m-1} \Fm  \av_m \big\|^2 \ . \nonumber
\end{align} \thmsym
\end{lemma} The proof can be found in Appendix~\ref{app:proofmmsesucc}.

The second form of successive cancellation utilizes the algebraic structure of the codebooks, and was originally proposed in~\cite{oen14}. Recall that, in the parallel computation scheme, every fine lattice that participates in $\av_m\T \Xm$ must be able to tolerate the associated effective noise. Clearly, if we knew some of the individual codewords, we could remove them from the channel observation and thus relax the noise tolerance constraints on their fine lattices. However, we only have access to certain linear combinations of the codewords, but it turns out that this is still enough side information to relax the noise tolerance constraints in a similar fashion. 

In Section~\ref{s:proofcompsucc}, we provide a detailed description of this \textit{algebraic successive cancellation} technique. At a high level, it can be viewed as performing Gaussian elimination over $\mathbb{Z}_p$ where row swaps are not permitted. The following definition will be used to specify valid user cancellation orders.

\begin{definition}[Admissible Mappings] \label{d:mappings}
Let $\Am$ denote an $L \times L$ integer matrix and let $\Ic \subset \{1,\ldots,L\} \times \{1,\ldots,L\}$ denote a set of index pairs. We say that $\Ic$ is an \textit{admissible mapping} for $\Am$ if there exists a real-valued, lower unitriangular\footnote{A \textit{unitriangular} matrix is a triangular matrix whose diagonal entries are equal to $1$.} matrix $\Lm \in \RR^{L \times L}$ such that the $(m,\ell)$th entry of $\Lm\Am$ is equal to zero for all $(m,\ell) \notin \Ic$. Let $\Mc(\Am)$ denote the set of admissible mappings for $\Am$. \defsym
\end{definition}

Within the context of our scheme, if $\Ic$ is an admissible mapping for $\Am$ and $(m,\ell) \in \Ic$, then user $\ell$ will not be cancelled out during the $m$th decoding step. Therefore, the $\ell$th transmitter must be able to tolerate the $m$th effective noise. The following theorem makes this notion precise.

\begin{theorem} \label{t:compsucc}
For an AWGN network with $L$ transmitters, a receiver, and power constraints $P_1, \ldots, P_L$, the following computation rate region is achievable,
\begin{align*}
\mathcal{R}_{\comp}^{(\succ)}(\Hm,\Am) &= \bigcup_{\substack{\Atm \in \mathbb{Z}^{L \times L} \\  \rowspan(\Am)\subseteq \rowspan(\Atm)}}  \hspace{-0.4in}\bigcup_{\Ic \in \Mc(\Atm)} \mathcal{R}_{\succ}(\Hm,\Atm, \Ic)  \nonumber
\end{align*} where
\begin{align}
&\mathcal{R}_{\succ}(\Hm,\Atm,\Ic) \define \Bigg\{ (R_1,\ldots,R_L) \in \mathbb{R}_+^L ~:  \label{e:compsuccregion} \\
 &\qquad ~~~~  R_\ell \leq  \frac{1}{2}\log^+\bigg( \frac{P_\ell}{\sigma_{\succ}^2(\Hm,\atv_m | \Atm_{m-1})} \bigg) ~\forall(m,\ell) \in \Ic  \Bigg \} \nonumber
\end{align} and $\atv_m\T$ and $\tilde{a}_{m,\ell}$ are the $m$th row and $(m,\ell)$th entry of $\Atm$, respectively. \thmsym
\end{theorem} The proof is presented in Section~\ref{s:proofcompsucc}.

\begin{corollary}
\label{cor:succisbetter} For every channel matrix $\Hm$ and integer coefficient matrix $\Am$, the parallel computation rate region is contained within the successive computation rate region, $\mathcal{R}^{(\para)}_{\comp}(\Hm,\Am) \subseteq \mathcal{R}^{(\succ)}_{\comp}(\Hm,\Am)$. Furthermore, there exist $\Hm$ and $\Am$ for which the subset relation is strict. \thmsym \end{corollary}

\begin{example}\label{ex:compsucc}
Consider three transmitters with equal power constraints, $P_1 = P_2 = P_3 = P$, and a receiver that observes $Y = 2 X_1 + X_2 + X_3 + Z$ and wants to decode the linear combinations with integer coefficient vectors $\av_1 = [1~\,1~\,1]\T$ and $\av_2 = [1 ~ -1 ~ -1]\T$. In our notation, this corresponds to channel matrix $\Hm = [2~1~1]$ and integer matrix $\Am = [\av_1 ~ \av_2 ~ \zerov]\T$. The effective noise variances for successively decoding these linear combinations are
\begin{align*}
\sigma_{\succ}^2(\Hm, \av_1) &= \frac{3P + 2P^2}{1 + 6P} \\
 \sigma_{\succ}^2(\Hm, \av_2 | \av_1) &= \frac{P(3 + 24P + 18P^2)}{8 + 38P + 36P^2} \ .
\end{align*} Following Definition~\ref{d:mappings}, the mappings 
\begin{align*}
\Ic_1 &= \big\{(1,1),\,(1,2),\,(1,3),\,(2,2),\,(2,3)\big\} \\
\Ic_2 &=  \big\{(1,1),\,(1,2),\,(1,3),\,(2,1)\big\} 
\end{align*} are admissible using the lower unitriangular matrices
\begin{align*}
\Lm_1 = \begin{bmatrix}
1&0&0\\
-1&1&0 \\
0&0&1
\end{bmatrix}
\qquad
\Lm_2 = \begin{bmatrix}
1&0&0\\
1&1&0 \\
0&0&1
\end{bmatrix} \ ,
\end{align*} respectively. This yields the rate regions
\begin{align*}
&\Rc_{\succ}(\Hm,\Am,\Ic_1) \\
&= \Bigg\{ (R_1,R_2,R_3) \in \RR_+^3 : \max_{\ell=1,2,3} R_\ell \leq \frac{1}{2} \log^+\bigg(\frac{1+6P}{3 + 2P}\bigg) \\
& \qquad \qquad \qquad \qquad \qquad R_1 \leq \frac{1}{2} \log^+\bigg(\frac{8 + 38P + 36P^2}{3 + 24P + 18P^2} \bigg) \Bigg \} \\
&\Rc_{\succ}(\Hm,\Am,\Ic_2) \\
&= \Bigg\{ (R_1,R_2,R_3) \in \RR_+^3 :  \max_{\ell=1,2,3} R_\ell \leq \frac{1}{2} \log^+\bigg(\frac{1+6P}{3 + 2P}\bigg),\\ 
&\qquad \qquad \qquad~~~\max_{\ell=2,3} R_\ell \leq \frac{1}{2} \log^+\bigg(\frac{8 + 38P + 36P^2}{3 + 24P + 18P^2} \bigg) \Bigg \} \ , 
\end{align*} which are part of the achievable computation rate region expressed in Theorem~\ref{t:compsucc}. Note that the linear combination with integer coefficient vector $\av_2$ cannot be directly decoded using Theorem~\ref{t:comppara} since $\sigma_{\para}^2(\Hm,\av_2) = 3P > P$. \defsym
\end{example}

\begin{example} \label{ex:compsucc_mac}
We return to the scenario of Example~\ref{ex:comppara_mac} wherein the receiver wants all of the messages, $\Am = \Id$. Clearly, the mapping $\Ic = \big\{(1,1),\ldots,(L,L)\big\}$ is admissible using the lower unitriangular matrix $\Lm = \Id$. Consider the $m$th decoding step. The receiver has successfully decoded the first $m-1$ messages corresponding to $\Am_{m-1} = [\deltav_1 ~\cdots~\deltav_{m-1}]\T$ where $\deltav_i$ is the $i$th column of $\Id$. By setting the $i$th entry of the equalization vector $\cv_m$ to be $\bv_m\T\hv_i$ (and the rest to $0$), the effective noise variance from~\eqref{e:noisevarsucc} will be reduced to $$ \| \bv_m \|^2 + \big\| \big( \bv_m\T\, \big[ \zerov~ \cdots ~\zerov ~\hv_m ~\cdots~\hv_L\big] - \deltav_m\T \big) \Pm^{1/2} \big\|^2 \ . $$ Following the same steps as in Example~\ref{ex:comppara_mac}, it can be shown that we can reach an effective SNR of
\begin{align*}
1 + P_\ell \hv_\ell\T \bigg( \Id + \sum\limits_{i = \ell+1}^L P_i \hv_i \hv_i\T \bigg)^{-1} \hv_\ell
\end{align*} for the $\ell$th user. Thus, the rate region (for successive decoding) is 
\begin{align*}
&\mathcal{R}_{\succ}(\Hm,\Id) =  \Bigg\{ (R_1,\ldots,R_L) \in \mathbb{R}_+^L ~: \\
& \qquad ~~R_\ell \leq  \frac{1}{2}\log\bigg(  1 + P_\ell \hv_\ell\T \bigg( \Id + \sum_{i=\ell+1}^L P_i \hv_i \hv_i\T \bigg)^{-1} \hv_\ell  \bigg) \Bigg\} \ . 
\end{align*} This is equal to the rate region attainable via i.i.d.~Gaussian codebooks and SIC decoding~\cite[Theorem 1]{vg97} under the lexicographic decoding order. The SIC rate region for any decoding order can be attained via Theorem~\ref{t:compsucc} by setting $\Atm$ to be the corresponding permutation matrix. 

As shown in~\cite{oen13} (for equal powers), the rate region can be enlarged via \textit{successive integer-forcing decoding}, i.e.,  taking the union over all full-rank matrices $\Atm$ in Theorem~\ref{t:compsucc}. Specifically, the successive integer-forcing rate region strictly contains the union of the SIC rate regions across all decoding orders. (Of course, when time-sharing across decoding orders is permitted then SIC can attain the entire capacity region.) \defsym
\end{example}

The statement of the computation rate region in Theorem~\ref{t:compsucc} takes a union over all integer matrices whose rowspan contains that of $\Am$. We now argue that it suffices to take this union over a certain subset of these matrices. As a motivating example, consider a receiver that wishes to recover a single linear combination with an integer coefficient vector $\av = [a_1 ~\ldots ~a_L]\T$ whose entries have a greatest common divisor larger than one, $\mathrm{gcd}(a_1,\ldots,a_L) = \alpha > 1$. If the receiver attempts to decode the corresponding integer-linear combination directly, it will encounter an effective noise variance of $\sigma^2_{\para}(\Hm,\av)= \| \Fm \av \|^2$ according to~\eqref{e:minvarpara2}. However, the receiver can instead decode the linear combination with integer coefficient vector $\atv = \alpha^{-1} \av$, which will yield a lower effective noise variance of $\sigma^2_{\para}(\Hm,\atv)= \| \Fm \atv \|^2 = \alpha^{-2} \| \Fm \av \|^2$. Afterwards, it can scale by $\alpha$ to recover its desired linear combination.

We now show to generalize this concept to decoding more than one linear combination. We will need the following definitions.

\begin{definition}[Unimodular Matrix]
A square integer matrix is \emph{unimodular} if its determinant equal to $+1$ or $-1$. \defsym
\end{definition} It can be shown that the inverse of a unimodular matrix is itself unimodular. 

\begin{definition}[Primitive Basis Matrix]
An integer matrix $\Am \in \mathbb{Z}^{L \times L}$ whose rank is equal to $M \leq L$ is said to be a \emph{primitive basis matrix}\footnote{Our choice of terminology is inspired by the fact that a matrix of this form can be viewed as the basis of a primitive sublattice (i.e., a sublattice that is formed by intersecting an $L$-dimensional lattice with a subspace of dimension $M < L$) and that any $L$-dimensional lattice basis is unimodular. See~\cite[Section 1.2]{bremner} for more details.} if it is of the form $$\Am = \begin{bmatrix}\Am_{M} \\ \mathbf{0}_{(L-M) \times L}\end{bmatrix}$$ where $\Am_{M}$ is a full-rank, $M \times L$ integer matrix, and there exists an integer matrix $\mathbf{B} \in \ZZ^{(L-M)\times L}$ such that the matrix
$$\begin{bmatrix}\Am_{M} \\ \mathbf{B} \end{bmatrix}$$
is unimodular. In other words, $\Am$ can be completed to a unimodular matrix. \defsym
\end{definition}The following theorem establishes that the computation rate region from Theorem~\ref{t:compsucc} is unchanged if we only take the union over primitive basis matrices.

\begin{theorem}
\label{t:primitive}
Let 
\begin{align*}
\mathcal{R}_{\comp}^{(\prim)}(\Hm,\Am) &= \bigcup_{\substack{\Atm \in \mathbb{Z}^{L \times L} \\ \Atm \text{~primitive basis} \\ \rowspan(\Am)\subseteq \rowspan(\Atm)}}  \hspace{-0.4in}\bigcup_{\Ic \in \Mc(\Atm)} \mathcal{R}_{\succ}(\Hm,\Atm, \Ic) 
\end{align*} where $\mathcal{R}_{\succ}(\Hm,\Atm, \Ic)$ is defined in~\eqref{e:compsuccregion}. This region is equal to the computation rate region from Theorem~\ref{t:compsucc}, 
\begin{align*}
\mathcal{R}_{\comp}^{(\prim)}(\Hm,\Am) = \mathcal{R}_{\comp}^{(\succ)}(\Hm,\Am) \ .
\end{align*} \thmsym
\end{theorem} The proof is deferred to Appendix~\ref{app:proofprimitive}.

\begin{corollary} \label{c:unimodular}
If the integer coefficient matrix $\Am \in \ZZ^{L\times L}$ has rank $L$, it suffices to take the union in Theorem~\ref{t:compsucc} over the set of all unimodular matrices, rather than the set of rank $L$ integer matrices. \thmsym
\end{corollary}

\subsection{Computation for Multiple-Access}
\label{s:sumrate}

At a first glance, it may seem that the lattice-based compute-and-forward framework developed above is a poor fit for multiple-access communication. Specifically, consider a receiver that observes the sum of the transmitted signals. From Example~\ref{ex:comppara_sum}, the receiver can decode the sum of the codewords at a high rate, but this alone is insufficient to discern the individual messages. On the other hand, if the transmitters employ (independent) random i.i.d.~codebooks, then all message tuples will be mapped to different sums with high probability. 

However, within the compute-and-forward framework, a receiver is not restricted to decoding a single linear combination. In fact, a natural multiple-access strategy is to decode $L$ linearly independent linear combinations and then solve for the underlying messages. This approach was first proposed in~\cite{oen14} for the parallel computation strategy with equal powers, and it was demonstrated that, when combined with algebraic successive cancellation, it always operates within a constant gap of the multiple-access sum capacity. Subsequent work~\cite{oen13} showed that, under certain technical conditions, the successive computation strategy with equal powers can operate at the exact multiple-access sum capacity. Below, we extend these results to the unequal power setting.

Recall that the multiple-access capacity region for our channel model is 
\begin{align}
&\Rc_{\text{MAC}}(\Hm) = \bigg\{ (R_1,\ldots,R_L) \in \RR_+^L :  \label{e:maccapacityregion} \\
&\qquad  \sum_{i \in \Sc} R_i \leq \frac{1}{2} \log \det \big( \Id + \Hm_{\Sc} \Pm_{\Sc} \Hm_{\Sc}\T \big) ~\forall \Sc \subseteq \{1,\ldots,L\} \bigg\}  \nonumber
\end{align} where $\Hm_{\Sc}$ refers to the submatrix consisting of the columns of $\Hm$ with indices in $\Sc$ and $\Pm_{\Sc}$ to the submatrix consisting of the entries of $\Pm$ whose row and column indices are in $\Sc$. The multiple-access sum capacity is simply $\frac{1}{2} \log \det\big(\Id + \Hm \Pm \Hm\T\big)$. Any rate tuple in the capacity region is achievable using i.i.d.~Gaussian codebooks at the transmitters combined with joint typicality decoding at the receiver~\cite[Section 15.3.1]{coverthomas}. 

In order to operate near the multiple-access sum capacity, we will need to uniquely map users to effective noise variances via the following definition.

\begin{definition} [Admissible Multiple-Access Mappings] \label{d:macmappings}
Let $\Ic \subset \{1,\ldots,L\} \times \{1,\ldots,L\}$ be an admissible mapping for integer matrix $\Am \in \ZZ^{L \times L}$ according to Definition~\ref{d:mappings} and let $\Lm \in \RR^{L \times L}$ be the associated lower unitriangular matrix. We say that $\Ic$ \textit{allows a multiple-access permutation} $\pi: \{1,2,\ldots,L\} \rightarrow \{1,2,\ldots,L\}$ if $\Lm \Am$ is upper triangular after column permutation by $\pi$. \defsym
\end{definition} 

Intuitively, if $\Ic$ allows a multiple-access permutation $\pi$, the scheme will remove the $(\pi^{-1}(m))$th codeword from its effective channel after the $m$th decoding step. This means that the $(\pi^{-1}(m))$th codeword only needs to tolerate the first $m$ effective noise variances. The details of this scheme are presented in Section~\ref{s:proofcompsucc}.

There is always at least one admissible multiple-access mapping for every full-rank matrix $\Am \in \ZZ^{L \times L}$. For instance, one can apply LU factorization with column permutation only to find an appropriate $\Lm$, admissible mapping $\Ic$, and permutation $\pi$. Also, recall from Remark~\ref{r:realsolvable} that if the receiver can recover $L$ linear combinations with a full-rank integer coefficient matrix $\Am \in \ZZ^{L \times L}$, then it can solve for the original messages.

\noindent\textbf{Parallel Computation for Multiple Access:} First, we note that, in the parallel computation strategy from Theorem~\ref{t:comppara}, each user must overcome the effective noise for all linear combinations in which it participates. A simple multiple-access strategy is to recover a rank-$L$ set of linear combinations that are then solved for the original messages. This corresponds to setting $\Am = \Id$ in Theorem~\ref{t:comppara}, leading to the following achievable rate region for multiple access:
\begin{align*}
&\mathcal{R}_{\comp}^{(\para)}(\Hm,\Id) = \Bigg\{ (R_1,\ldots,R_L) \in \mathbb{R}_+^L ~: \\
&\qquad ~~~~ R_\ell \leq  \max_{\substack{ \Atm \in \ZZ^{L \times L} \\ \rank(\Atm) = L}} \min_{m : a_{m,\ell} \neq 0} \frac{1}{2}\log^+\bigg( \frac{P_\ell}{\sigma_\para^2(\Hm,\atv_m)} \bigg)  \Bigg\} \ . 
\end{align*} This can be viewed as a generalization of the integer-forcing achievable rate from~\cite[Theorem 3]{zneg14} to unequal powers. As shown in~\cite{zneg14}, this strategy has a good ensemble average (e.g., for Rayleigh fading), but there are specific choices of $\Hm$ for which the sum rate lies arbitrarily far from the sum capacity. 

As argued in~\cite{oen14}, by augmenting parallel computation with algebraic successive cancellation, we can operate within a constant gap of the sum capacity. Here, we extend this result to the unequal power setting. This corresponds to applying Theorem~\ref{t:compsucc} and setting the equalization vectors for the side information to zero, $\cv_m = \zerov$, so that~\eqref{e:compsuccregion} is replaced with 
\begin{align*}
&\mathcal{R}_{\text{ASC}}(\Hm,\Atm,\Ic) = \Bigg\{ (R_1,\ldots,R_L) \in \mathbb{R}_+^L ~: \\
& \qquad ~~~~~R_\ell \leq  \frac{1}{2}\log^+\bigg( \frac{P_\ell}{\sigma_\para^2(\Hm,\atv_m)} \bigg) \text{ for all } (m,\ell) \in \Ic  \Bigg\} \ . 
\end{align*}

The next definition and lemma are taken from~\cite[Section VIII]{fsk13} and will be used to select appropriate integer vectors for parallel computation.

\begin{definition}[Dominant Solutions] \label{d:dominantsolutions} Let $\Fm$ be any matrix satisfying~\eqref{e:F}. A set of linearly independent integer vectors $\av^*_1,\ldots,\av^*_L \in \ZZ^L$ satisfying $\| \Fm \av^*_1\| \leq \cdots \leq \| \Fm \av^*_L\|$  is called a \textit{dominant solution} if, for any linearly independent integer vectors $\atv_1,\ldots,\atv_L \in \ZZ^L$ satisfying $\| \Fm \atv_1\| \leq \cdots \leq \| \Fm \atv_L\|$, we have that $\| \Fm \av^*_m \| \leq \| \Fm \atv_m \|$ for $m= 1,\ldots,L$. We will call an integer matrix $\Am^* \in \ZZ^{L \times L}$ a dominant solution if its rows $(\av^*_1)\T, \ldots, (\av^*_L)\T$ correspond to a dominant solution. \defsym
\end{definition}

\begin{lemma}[{\cite[Theorem 8]{fsk13}}]
For any $\Fm$ satisfying~\eqref{e:F}, there always exists a dominant solution $\av_1^*,\av_2^*, \ldots, \av_L^* \in \ZZ^L$ satisfying 
\begin{align*}
\av_1^* &= \argmin\big\{ \| \Fm \av \| : \av \in \ZZ^L \setminus \{\zerov\} \big\} \\
\av_2^* &= \argmin\big\{ \| \Fm \av \| : \av \in \ZZ^L,\ \av,\av_1^* \text{~linearly independent} \big\} \\
&~ \, \vdots\nonumber \\
\av_L^* &= \argmin\big\{ \| \Fm \av \| : \av \in \ZZ^L,\ \av,\av_1^*,\ldots,\av_{L-1}^* \\
&\qquad \qquad  \qquad \qquad \qquad \qquad \qquad \text{~linearly independent} \big\} \ .
\end{align*} \thmsym
\end{lemma} See~\cite{fsk13} for a proof as well as a greedy algorithm.

We can now show that the parallel computation strategy, when combined with algebraic successive cancellation, is approximately optimal for multiple access.

\begin{theorem}\label{t:compparamac}
For any dominant solution $\Am^* \in \ZZ^{L \times L}$ and admissible multiple-access mapping $\Ic$ with permutation $\pi$, the rate tuple 
\begin{align*}
R_\ell = \frac{1}{2}\log^+\bigg(\frac{P_\ell}{\sigma_{\para}^2\big(\Hm,\av^*_{\pi(\ell)}\big)} \bigg) \quad \ell = 1,\ldots,L
\end{align*} is achievable via the parallel computation strategy combined with algebraic successive cancellation, $(R_1,\ldots,R_L) \in \Rc_{\text{ASC}}(\Hm,\Am^*,\Ic)$. The sum of these rates is within a constant gap of the multiple-access sum capacity,
\begin{align*}
\sum_{\ell = 1}^L R_\ell \geq \frac{1}{2} \log \det \big( \Id + \Hm \Pm \Hm\T \big) - \frac{L}{2}\log(L) \ .
\end{align*} \thmsym
\end{theorem} The proof can be found in Appendix~\ref{app:proofcompparamac}.
 
\noindent\textbf{Successive Computation for Multiple Access:} It is well-known that the corner points of the Gaussian multiple-access capacity region can be attained using i.i.d.~Gaussian codebooks at the transmitters along with SIC decoding~\cite[Theorem 1]{vg97}. As we will argue below, successive computation enjoys a similar optimality property: there is always at least one integer matrix for which the successive computation rate region includes a rate tuple that attains the multiple-access sum capacity. For instance, as shown in Example~\ref{ex:compsucc_mac}, successive computation decoding can mimic SIC decoding and attain the corner points of the capacity region. Furthermore, the successive computation rate region often includes non-corner points that attain the sum capacity (i.e., rate tuples that lie on the interior of the dominant face of the multiple-access capacity region). These points are not directly accessible via SIC decoding (but can be attained by enhancing SIC with time-sharing~\cite[Section 4.4]{elgamalkim} or rate-splitting~\cite{ru96}).

Our optimality results stem from the following identity.
\begin{lemma} \label{l:unimodularsum}
For any unimodular matrix $\Am \in \ZZ^{L\times L}$ and permutation $\pi$, we have that
\begin{align} 
&\sum_{\ell =1}^L \frac{1}{2} \log\bigg(\frac{P_\ell}{\sigma_{\succ}^2\big(\Hm,\av_{\pi(\ell)} | \Am_{\pi(\ell)-1}\big)} \bigg) \nonumber \\
&= \frac{1}{2}\log\det\big(\Id + \Hm \Pm \Hm\T\big) \ . \label{e:unimodularsum} 
\end{align} \thmsym
\end{lemma} See Appendix~\ref{app:proofunimodularsum} for a proof.

At a first glance, it may appear that Lemma~\ref{l:unimodularsum} implies that \textit{every} unimodular matrix yields sum-rate optimal performance for successive computation. However, this lemma does not guarantee that the rates appearing in~\eqref{e:unimodularsum} are achievable. Specifically from~\eqref{e:compsuccregion}, for an admissible mapping $\Ic$ that allows multiple-access permutation $\pi$, decoding is possible if the $\ell$th user can tolerate effective noise of variance 
\begin{align*}
\max_{m: (m,\ell) \in \Ic} \sigma^2_{\succ}(\Hm,\av_m | \Am_{m-1}) \ .
\end{align*} In order to apply Lemma~\ref{l:unimodularsum} and show that Theorem~\ref{t:compsucc} attains the exact sum capacity, this maximum must be equal to $\sigma^2_{\succ}\big(\Hm,\av_{\pi(\ell)} | \Am_{\pi(\ell)-1}\big)$ for $\ell = 1,\ldots,L$. Moreover, the achievable rate expression in~\eqref{e:compsuccregion} are written in terms of the $\log^+$ function whereas the summands in~\eqref{e:unimodularsum} simply use the $\log$ function. Thus, each user's power should exceed its associated effective noise variance. Putting these two conditions together and applying Lemma~\ref{l:unimodularsum}, we arrive at the following optimality result for successive computation. 
  
\begin{theorem}\label{t:compsuccmac}
Let $\Am$ be an $L \times L$ unimodular matrix and let $\Ic$ be an admissible mapping that allows multiple-access permutation $\pi$. If, for $\ell = 1,\ldots,L$, we have that \begin{align}
\max_{m: (m,\ell) \in \Ic} \sigma^2_{\succ}(\Hm,\av_m | \Am_{m-1}) &= \sigma_{\succ}^2\big(\Hm,\av_{\pi(\ell)} \big| \Am_{\pi(\ell)-1}\big) \label{e:succnoisecondition} 
\end{align} and $P_\ell \geq \sigma_{\succ}^2\big(\Hm,\av_{\pi(\ell)} \big| \Am_{\pi(\ell)-1}\big)$, then the rate tuple
\begin{align*}
R_\ell = \frac{1}{2}\log\bigg( \frac{P_\ell}{\sigma_{\succ}^2\big(\Hm,\av_{\pi(\ell)} \big| \Am_{\pi(\ell)-1}\big)}\bigg) \quad \ell = 1,\ldots, L 
\end{align*} is achievable via the successive computation strategy, $(R_1,\ldots,R_L) \in \Rc_\succ(\Hm,\Am,\Ic)$. Moreover, the sum of these rates is equal to the multiple-access sum capacity,
\begin{align*}
\sum_{\ell=1}^L R_\ell =  \frac{1}{2} \log \det \big( \Id + \Hm \Pm \Hm\T \big) \ .
\end{align*} \thmsym
\end{theorem}

\begin{remark}
It is sometimes convenient to replace the condition in~\eqref{e:succnoisecondition} with the stricter condition that
\begin{align*}
\sigma_{\succ}^2(\Hm,\av_1) \leq \sigma_{\succ}^2(\Hm,\av_2 | \Am_1) \leq \cdots \leq \sigma_{\succ}^2(\Hm,\av_L | \Am_{L-1}) \ .
\end{align*} \defsym
\end{remark}

\begin{figure}[!ht]
\begin{center}

%
%
%
\definecolor{mycolor1}{rgb}{1.00000,0.00000,1.00000}%
\begin{tikzpicture}

\begin{axis}[%
width={0.4\textwidth},
height={0.4\textwidth},
scale only axis,
xmin=0,
xmax=2,
xlabel={\mbox{\large{$R_1$}}},
ymin=0,
ymax=2,
ylabel={\mbox{\large{$R_2$}}},
legend style={draw=black,fill=white,legend cell align=left}
]
\addplot [color=black,solid,line width=2.0pt]
  table[row sep=crcr]{%
0	1.66096404744368\\
0.382767373181489	1.66096404744368\\
1.5	0.54373142062517\\
1.5	0\\
};
\addlegendentry{Capacity Region};

\addplot [color=LineGreen,line width=3.0pt,mark size=6.0pt,only marks,mark=x,mark options={solid}]
  table[row sep=crcr]{%
1.36244638093282	0.590286122820911\\
0.993963583849713	0.958768919904014\\
};
\addlegendentry{Parallel Comp.};

\addplot [color=LineBlue,line width=2.5pt,mark size=6.0pt,only marks,mark=asterisk,mark options={solid}]
  table[row sep=crcr]{%
1.36244638093282	0.681285039692354\\
1.08496250072116	0.958768919904014\\
1.5	0.54373142062517\\
1.5	0.54373142062517\\
1.5	0.54373142062517\\
1.5	0.54373142062517\\
1.36244638093282	0.681285039692355\\
1.08496250072116	0.958768919904014\\
1.08496250072116	0.958768919904014\\
1.36244638093282	0.681285039692354\\
1.08496250072116	0.958768919904014\\
0.382767373181489	1.66096404744368\\
1.36244638093282	0.681285039692354\\
1.08496250072116	0.958768919904014\\
1.36244638093282	0.681285039692354\\
1.5	0.54373142062517\\
};
\addlegendentry{Successive Comp.};

\addplot [color=LineRed,line width=3.0pt,mark size=7.0pt,only marks,mark=o,mark options={solid}]
  table[row sep=crcr]{%
0.382767373181489	1.66096404744368\\
1.5	0.54373142062517\\
};
\addlegendentry{SIC};

\end{axis}
\end{tikzpicture}%
\end{center}
\caption{Plot of the Gaussian multiple-access capacity region as well as the dominant rate pairs for parallel computation, sum-capacity optimal rate pairs for successive computation, and SIC corner points for the channel matrix $\Hm = [1 ~\frac{3}{2}]$ and powers $P_1 = 7$ and $P_2 = 4$.}\label{f:twousermac}
\end{figure}
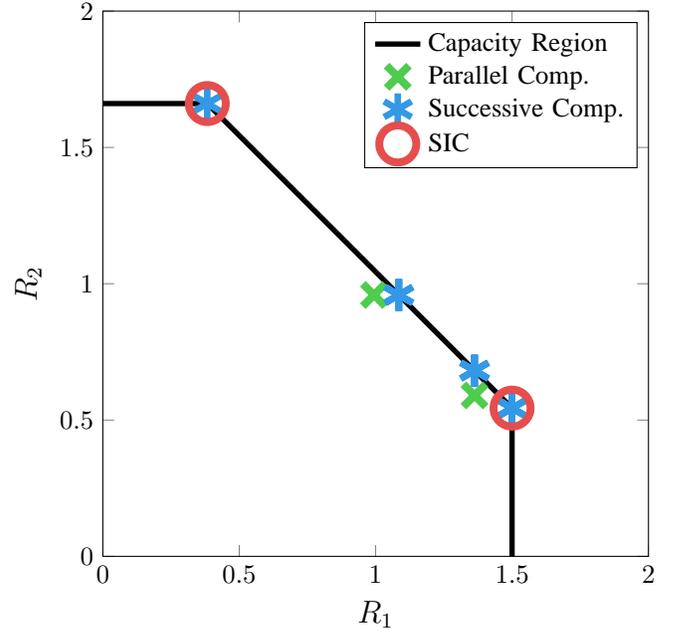

\begin{remark}
For any channel matrix $\Hm$ and power matrix $\Pm$, there always exists a unimodular matrix $\Am$ and multiple-access mapping $\Ic$ with permutation $\pi$ for which Theorem~\ref{t:compsuccmac} applies. For instance, as shown in Example~\ref{ex:compsucc_mac}, we can attain the performance for SIC decoding with order $\pi$ by settting $\Am$ to be the associated permutation matrix and $\Ic = \big\{ (1,\pi(1)),\ldots,(L,\pi(L))\}$. The next example demonstrates that the successive computation strategy can attain sum-capacity rate tuples that do not correspond to SIC corner points.  \defsym
\end{remark}

\noindent\textbf{Two-User Example:}
Consider a two-user multiple-access channel (i.e., $L = 2$) with channel matrix $\Hm = [\hv_1~\hv_2]$. 
For $\Hm = [1~\frac{3}{2}]$, $P_1 = 7$, and $P_2 = 4$, we have plotted, in Figure~\ref{f:twousermac}, the capacity region~\eqref{e:maccapacityregion} and marked the rate tuples achievable via SIC decoding, parallel computation for multiple access from Theorem~\ref{t:compparamac}, and successive computation for multiple access from Theorem~\ref{t:compsuccmac}. Specifically, SIC decoding yields the rates pairs $(0.3828,1.6610)$ and $(1.5000,0.5437)$. Successive computation can attain both of these rate pairs as well as $(1.0850,0.9588)$ and $(1.3624,0.6813)$. Finally, there are two dominant rate pairs for parallel computation: $(1.3624,0.5903)$ and $(0.9940,0.9588)$. Note that all successive computation rate pairs are sum-capacity optimal and that the parallel computation rate pairs are much closer to the sum-capacity than the (worst-case) bound of $2$ bits.\footnote{MATLAB code to generate this plot for any (two-user) choice of $\Hm$ and $\Pm$ is available on the first author's website.} \defsym

\begin{remark}
In independent and concurrent work, Zhu and Gastpar have developed a compute-and-forward approach to multiple-access~\cite{zg14}. Their main idea is to use the same fine lattice at each transmitter along with different coarse lattices. The second moments of the coarse lattices are set according to the desired rates, and each transmitter scales its lattice codeword to meet its power constraint. (Note that their approach does not establish a correspondence to a finite field.) Overall, they establish that a significant fraction of the multiple-access sum-capacity boundary is achievable via this approach. We note that the underlying compute-and-forward result for unequal powers~\cite[Theorem 1]{zg14} is a special case of Theorem~\ref{t:comppara}. \defsym
\end{remark}

\subsection{Multiple Receivers}\label{s:multiplerx}

So far, we have limited our discussion and results to single-receiver scenarios. Although this has allowed us to introduce our main ideas in a compact fashion, the compute-and-forward framework is the most useful in scenarios where there are \textit{multiple receivers} that observe interfering codewords. We now expand our problem statement to permit $K$ receivers, potentially with different demands. (See Figure~\ref{f:cfprobstatemultirx} for a block diagram.) As we will see, the parallel and successive computation rate regions can be expressed in terms as intersections of the corresponding rate regions for each receiver.

\begin{figure}[h]
\psset{unit=.77mm}
\begin{center}
\begin{pspicture}(7,2)(120,64)

\rput(14.5,55){$\wv_1$} \psline{->}(18,55)(23,55) \psframe(23,50)(33,60)
\rput(28,55){$\mathcal{E}_1$} \rput(38.5,58.75){$\xv_1\T$}
\psline[linecolor=black]{->}(33,55)(43,55)(48,55)(55,55)
\pscircle[fillstyle=solid,fillcolor=black](43,55){.7}
\psline[linecolor=black]{->}(43,55)(48,26)(55,26)

\rput(14.5,40){$\wv_2$} \psline{->}(18,40)(23,40) \psframe(23,35)(33,45)
\rput(28,40){$\mathcal{E}_2$} \rput(38.5,43.75){$\xv_2\T$}
\psline[linecolor=black]{->}(33,40)(43,40)(48,52)(55,52)
\pscircle[fillstyle=solid,fillcolor=black](43,40){.7}
\psline[linecolor=black]{->}(43,40)(48,23)(55,23)

\rput(15,15){$\wv_L$} \psline{->}(18.5,15)(23,15) \psframe(23,10)(33,20)
\rput(28,15){$\mathcal{E}_L$} \rput(38.5,18.75){$\xv_L\T$}
\psline[linecolor=black]{->}(33,15)(43,15)(48,44)(55,44)
\pscircle[fillstyle=solid,fillcolor=black](43,15){.7}
\psline[linecolor=black]{->}(43,15)(48,15)(55,15)

\rput(28,29){\large{$\vdots$}}
\rput(51,49.5){\small{$\vdots$}}
\rput(51,20.5){\small{$\vdots$}}
\rput(61.5,37){\large{$\vdots$}}
\rput(98,37){\large{$\vdots$}}

\rput(-8,9.5){
\psframe(63,32.5)(76,47.5) \rput(70,41){\large{$\mathbf{H}^{[1]}$}}
\psline{->}(76,40)(83,40)
\pscircle(85.5,40){2.5} \psline{-}(84.25,40)(86.75,40)
\psline{-}(85.5,38.75)(85.5,41.25) \psline{<-}(85.5,42.5)(85.5,47.5) \rput(87,50.5){$\Zm^{[1]}$}
\psline{->}(88,40)(100,40) \rput(94,43.5){$\Ym^{[1]}$}
\psframe(100,35)(112,45) \rput(106,40.5){$\mathcal{D}^{[1]}$}
\psline{->}(112,40)(117,40)
\rput(120,40){$\begin{array}{c}\uhv^{[1]}_1 \\ \vdots \\ \uhv^{[1]}_L\end{array}$}
}

\rput(-8,-19.5){
\psframe(63,32.5)(76,47.5) \rput(70,41){\large{$\mathbf{H}^{[K]}$}}
\psline{->}(76,40)(83,40)
\pscircle(85.5,40){2.5} \psline{-}(84.25,40)(86.75,40)
\psline{-}(85.5,38.75)(85.5,41.25) \psline{<-}(85.5,42.5)(85.5,47.5) \rput(88,50.5){$\Zm^{[K]}$}
\psline{->}(88,40)(100,40) \rput(94,43.5){$\Ym^{[K]}$}
\psframe(100,35)(112,45) \rput(106.5,40.5){$\mathcal{D}^{[K]}$}
\psline{->}(112,40)(117,40)
\rput(120,40){$\begin{array}{c}\uhv^{[K]}_1 \\ \vdots \\ \uhv^{[K]}_L\end{array}$}
}

\rput(65,4){$\uv_m^{[i]} = {\displaystyle \bigoplus_{\ell=1}^L}\  q_{m,\ell}^{[i]}  \wtv_{\ell} \  ,$}
\rput(98,4){$\wtv_\ell \in \llb \wv_\ell \rrb$}

\end{pspicture}
\end{center}
\caption{Block diagram for the compute-and-forward problem with multiple receivers. Each transmitter has a message $\wv_\ell$ whose elements are taken from $\ZZ_p$. This message is embedded into $\ZZ_p^k$ (by zero-padding), mapped into a codeword $\xv_\ell \in \RR^n$, and sent over the channel. Each receiver observes a noisy linear combination of these codewords, $\Ym^{[i]} = \Hm^{[i]} [\xv_1~\xv_2~\cdots~\xv_L]\T + \Zm^{[i]}$, and attempts to recover the linear combinations $\uv^{[i]}_1,\ldots,\uv^{[i]}_L$ of the coset representatives of the original messages.}
\label{f:cfprobstatemultirx}
\end{figure}

As in Section~\ref{s:channelmodel}, for $i = 1,\ldots,K$, the $i$th receiver is equipped with $N^{[i]}_{\text{r}}$ antennas and observes a noisy linear combination of the channel inputs,
\begin{align*}
\Ym^{[i]} = \sum_{\ell=1}^L \hv^{[i]}_\ell \xv_\ell\T + \Zm^{[i]} \label{e:chanmodelmulti}
\end{align*} where $\hv^{[i]}_\ell \in \RR^{N^{[i]}_{\text{r}}}$ is the channel vector between the $\ell$th transmitter and the $i$th receiver and $\Zm^{[i]}$ is elementwise i.i.d.~$\Nc(0,1)$. We group the channel vectors corresponding to the $i$th receiver into a channel matrix 
\begin{align*}
\Hm^{[i]} \triangleq \begin{bmatrix}\mathbf{h}^{[i]}_{1} & \mathbf{h}^{[i]}_{2} & \cdots & \mathbf{h}^{[i]}_{L}  \end{bmatrix} \ . 
\end{align*}

The problem statement is essentially the same as in Section~\ref{s:formalprobstate}. Following Definition~\ref{d:decoder}, we define a \textit{decoder} $\Dc^{[i]}: \mathbb{R}^{N^{[i]}_{\text{r}} \times n} \times \mathbb{R}^{N^{[i]}_{\text{r}} \times L} \times \mathbb{Z}^{L \times L} \rightarrow \mathbb{Z}_{p}^{L \times k}$ for the $i$th receiver that takes as inputs the channel observation $\Ym^{[i]}$, the channel matrix $\Hm^{[i]}$, and the desired integer coefficient matrix $\Am^{[i]}$, and outputs an estimate $\mathbf{\hat{U}}^{[i]} = \Dc^{[i]}(\Ym^{[i]},\Hm^{[i]},\Am^{[i]})$. Let $(\mathbf{\hat{u}}^{[i]}_m)\T$ denote the $m$th row of $\Um^{[i]}$. We say that decoding is \emph{successful} if $\mathbf{\hat{u}}^{[i]}_1 = \uv^{[i]}_1,\ldots,\mathbf{\hat{u}}^{[i]}_L = \uv^{[i]}_L$ for some linear combinations of the form 
\begin{equation*}
\uv^{[i]}_m = \bigoplus_{\ell=1}^L q^{[i]}_{m, \ell} \mathbf{\tilde{w}}_\ell
\end{equation*} where the $q^{[i]}_{m,\ell} \in \ZZ_p$ are the entries of $\Qm^{[i]} = [\Am^{[i]}] \bmod{p}$ and $\mathbf{\tilde{w}}_\ell \in \llb \wv_{\ell} \rrb$. An \textit{error} occurs if decoding is not successful for some $i \in \{1,\ldots,K\}$. 

\begin{remark} Note that we insist that the same coset representatives $\mathbf{\tilde{w}}_\ell$ are used across receivers. This is to ensure that the linear combinations from multiple receivers can later be put together to recover the original messages. \defsym
\end{remark}

We now adjust our definition of a computation rate region to accommodate multiple receivers. 

\begin{definition}[Multiple-Receiver Computation Rate Region] \label{d:compratemultiplerx}
A \emph{computation rate region for $K$ receivers} is specified by a set-valued function $\Rc_\comp\big(\Hm^{[1]},\ldots,\Hm^{[K]},\Am^{[1]},\ldots,\Am^{[K]}\big)$ that maps $K$-tuples of channel matrices $\big(\Hm^{[1]},\ldots,\Hm^{[K]}\big) \in \mathbb{R}^{N_{r}^{[1]} \times L} \times  \cdots \times \mathbb{R}^{N_{r}^{[K]} \times L}$ and $K$-tuples of integer coefficient matrices $\big(\Am^{[1]},\ldots,\Am^{[K]}\big) \in \mathbb{Z}^{L \times L} \times \cdots \times \mathbb{Z}^{L \times L}$ to a subset of $\mathbb{R}_+^L$. The computation rate region described by $\Rc_\comp$ is \textit{achievable} if, for every rate tuple $(R_{1}, R_{2}, \ldots, R_{L})\in \mathbb{R}_{+}^{L}$, $\epsilon > 0$, and $n$ large enough, 
there exist  
\begin{itemize}
\item parameters $p, k_{\c,\ell}, k_{\f,\ell}$ satisfying ${\displaystyle \frac{k_{\f,\ell}-k_{\c,\ell}}{n} \log p }> R_{\ell} - \epsilon$ for $\ell=1,2,\ldots,L$,
\item encoders $\mathcal{E}_{1},\ldots,\mathcal{E}_{L}$,
\end{itemize}
such that, 
\begin{itemize}
\item for all $K$-tuples of channel matrices $\big(\Hm^{[1]},\ldots,\Hm^{[K]}\big) \in \mathbb{R}^{N_{r}^{[1]} \times L} \times  \cdots \times \mathbb{R}^{N_{r}^{[K]} \times L}$ and
\item every  $K$-tuple of integer matrices $(\Am^{[1]},\ldots,\Am^{[K]}) $\\ satisfying $(R_{1}, \ldots, R_{L}) \in \Rc_\comp\big(\Hm^{[1]},\ldots,\Hm^{[K]},\Am^{[1]},\ldots,\Am^{[K]}\big)$ 
\end{itemize} there exist decoders $\mathcal{D}_{1},\ldots,\mathcal{D}_{K}$ with probability of decoding error at most $\epsilon$. \defsym \end{definition}

We can now state the achievable computation rate regions for parallel computation and successive computation. 

\begin{theorem}\label{t:compparamultiplerx}
For an AWGN network with $L$ transmitters satisfying power constraints $P_1, P_{2}, \ldots, P_L$, respectively, and $K$ receivers, the following computation rate region is achievable,
\begin{align*}
&\Rc_\comp^{(\para)}\big(\Hm^{[1]},\ldots,\Hm^{[K]},\Am^{[1]},\ldots,\Am^{[K]}\big) \\
&= \bigcap_{i=1}^K \Rc_{\comp}^{(\para)}\big(\Hm^{[i]},\Am^{[i]}\big)  
\end{align*} for the function $\Rc_{\comp}^{(\para)}\big(\Hm^{[i]},\Am^{[i]}\big)$ as defined in Theorem~\ref{t:comppara}. \thmsym
\end{theorem} The proof can be inferred from that of Theorem~\ref{t:comppara}: the encoders implement the same scheme as in the single-receiver case and each receiver implements the parallel computation decoder described in Section~\ref{s:proofcomppara}.

\begin{theorem}\label{t:compsuccmultiplerx}
For an AWGN network with $L$ transmitters satisfying power constraints $P_1, P_{2}, \ldots, P_L$, respectively, and $K$ receivers, the following computation rate region is achievable,
\begin{align*}
&\Rc_\comp^{(\succ)}\big(\Hm^{[1]},\ldots,\Hm^{[K]},\Am^{[1]},\ldots,\Am^{[K]}\big) \\
&= \bigcap_{i=1}^K \Rc_{\comp}^{(\succ)}\big(\Hm^{[i]},\Am^{[i]}\big)  
\end{align*} for $\Rc_{\comp}^{(\succ)}\big(\Hm^{[i]},\Am^{[i]}\big)$ as defined in Theorem~\ref{t:compsucc}. \thmsym
\end{theorem} Again, the proof can be inferred from that of Theorem~\ref{t:compsucc}: the encoders implement the same scheme as in the single-receiver case and each receiver implements the successive computation decoder described in Section~\ref{s:proofcompsucc}.

Unfortunately, the multiple-access sum-capacity optimality results do not transfer directly from the single-receiver setting. Nonetheless, in the multiple-receiver setting, computation for multiple-access outperforms i.i.d.~Gaussian encoding with SIC decoding even when time-sharing is allowed. We explore this phenomenon in the context of a compound multiple-access channel below.

\subsection{Case Study: Two-User Gaussian Compound Multiple-Access Channel}

We now take an in-depth look at the performance of SIC and compute-and-forward for a two-user Gaussian compound multiple-access channel. In our notation, this corresponds to $L = 2$ transmitters with messages $\wv_1$ and $\wv_2$, respectively, and $K = 2$ receivers that both want to recover $\wv_1$ and $\wv_2$. The capacity region is the intersection of the individual MAC capacity regions,
\begin{align*}
\Rc_{\text{CMAC}}\big(\Hm^{[1]},\Hm^{[2]}\big) = \Rc_{\text{MAC}}\big(\Hm^{[1]}\big) \cap \Rc_{\text{MAC}}\big(\Hm^{[2]}\big)  
\end{align*} and can be achieved via i.i.d.~Gaussian coding and joint typicality decoding. As discussed below in Remark~\ref{r:compoundmacinterference}, the compound MAC often appears in the context of $K$-user interference channels. In some scenarios, the transmitters may opt to induce interference alignment by using lattice codebooks instead of i.i.d.~Gaussian codebooks. This motivates the need for lattice-based decoding strategies.

In order for the $i$th receiver to successfully recover both messages with SIC decoding, the rates must fall within the SIC decoding region for the corresponding (two-user) multiple-access channel,
\begin{align*}
&\Rc_{\text{MAC}}^{(\text{SIC})}\big(\Hm^{[i]}\big) = \Rc_{\text{SIC},a}\big(\Hm^{[i]}\big)  \cup \Rc_{\text{SIC},b}\big(\Hm^{[i]}\big)  \\
&\Rc_{\text{SIC},a}\big(\Hm^{[i]}\big) = \Big\{ (R_1,R_2) \in \RR_+^2 : \\
&\qquad R_1 \leq \frac{1}{2} \log(1 + \big\| \hv_1^{[i]} \big\|^2 P_1),\nonumber \\
&\qquad R_2 \leq \frac{1}{2} \log\Big(1 + P_2 \big(\hv_2^{[i]}\big)\T \Big( \Id + P_1 \hv^{[i]}_1 \big(\hv^{[i]}_1\big)\T\Big)^{-1} \hv_2^{[i]} \Big) \Big\} \nonumber \\
&\Rc_{\text{SIC},b}\big(\Hm^{[i]}\big) =  \Big\{ (R_1,R_2) \in \RR_+^2 : \\
&\qquad  R_1 \leq \frac{1}{2} \log\Big(1 + P_1 \big(\hv_1^{[i]}\big)\T \Big( \Id + P_2 \hv^{[i]}_2 \big(\hv^{[i]}_2\big)\T\Big)^{-1} \hv_1^{[i]} \Big),\nonumber \\
&\qquad  R_2 \leq \frac{1}{2} \log(1 + \big\| \hv_2^{[i]} \big\|^2 P_2)\Big\} \ .\nonumber
\end{align*} Thus, for the compound multiple-access channel, SIC decoding combined with time-sharing can attain 
\begin{align*}
\Rc_{\text{CMAC}}^{(\text{SIC})}\big(\Hm^{[1]},\Hm^{[2]}\big) = \conv\Big(\Rc_{\text{MAC}}^{(\text{SIC})}\big(\Hm^{[1]}\big) \cap \Rc_{\text{MAC}}^{(\text{SIC})}\big(\Hm^{[2]}\big)\Big) 
\end{align*} where $\conv$ refers to the convex hull operation. Note that SIC decoding does not, in general, attain the sum-capacity even if aided by time-sharing. This is due to the fact that the corner points of the two multiple-access capacity regions do not coincide nor do the time-sharing ratios required to reach any other sum-capacity points.  

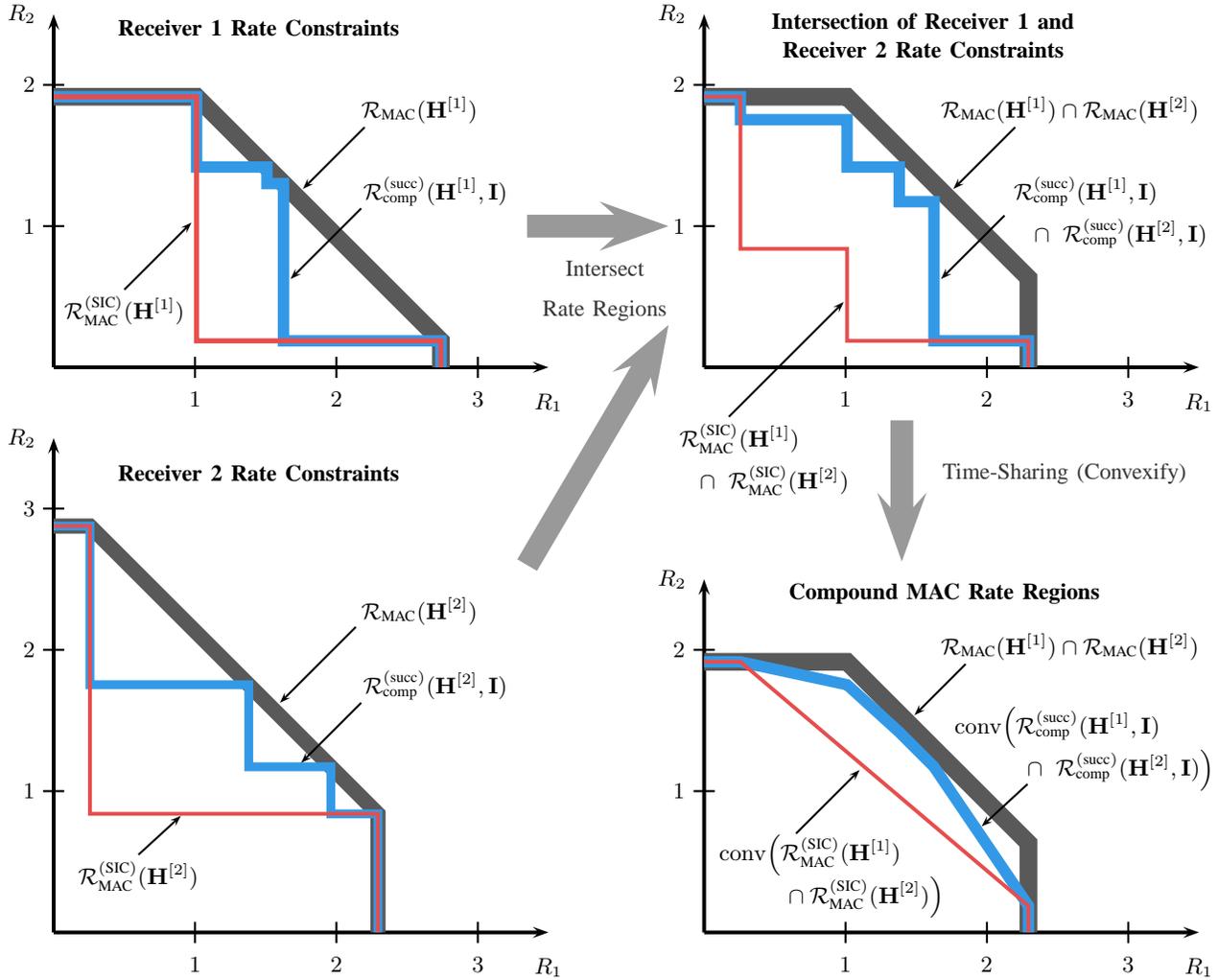
\begin{figure*}[!t]
\begin{center}
\psset{unit=.98mm}
\begin{pspicture}(0,-2)(168,130)
\small


\rput(3,80){
\psline[linewidth=1.5pt,linecolor=black]{->}(0,0)(70,0)  \rput(70,-5){$R_1$}
\psline[linewidth=1.5pt]{->}(0,0)(0,50)  \rput(-4.5,50){$R_2$}

\psline[linewidth=7pt,linecolor=black!65!white]{-}(0,38.3086)(20.2188,38.3086)(54.7768,3.7507)(54.7768,0)
\psline[linewidth=4.5pt,linecolor=LineBlue]{-}(0,38.3086)(20.2188,38.3086)(20.2188,28.3601)(30.1673,28.3601)(30.1673,26.0170)(32.5105,26.0170)(32.5105,3.7507)(54.7768,3.7507)(54.7768,0)
\psline[linewidth=2pt,linecolor=LineRed]{-}(0,38.3086)(20.2188,38.3086)(20.2188,3.7507)(54.7768,3.7507)(54.7768,0)

\psline(20,-1.5)(20,1.5) \rput(20,-4.5){\small{$1$}}
\psline(40,-1.5)(40,1.5) \rput(40,-4.5){\small{$2$}}
\psline(60,-1.5)(60,1.5) \rput(60,-4.5){\small{$3$}}
\psline(-1.5,20)(1.5,20) \rput(-3.5,20){\small{$1$}}
\psline(-1.5,40)(1.5,40) \rput(-3.5,40){\small{$2$}}

\rput(29,48){\textbf{Receiver 1 Rate Constraints}}
\psline{<-}(19.5,20.5)(10,11) \rput(10,8){$\mathcal{R}_{\text{MAC}}^{(\text{SIC})}(\Hm^{[1]})$}
\psline{<-}(35.25,25.25)(44,34) \rput(51.75,36.5){$\mathcal{R}_{\text{MAC}}(\Hm^{[1]})$}
\psline{<-}(33.5,12.5)(43.5,22.5) \rput(54,25){$\mathcal{R}_{\comp}^{(\succ)}(\Hm^{[1]},\Id)$}

}

\rput(3,0){
\psline[linewidth=1.5pt,linecolor=black]{->}(0,0)(70,0)  \rput(70,-5){$R_1$}
\psline[linewidth=1.5pt]{->}(0,0)(0,70)  \rput(-4.5,70){$R_2$}

\psline[linewidth=6pt,linecolor=black!65!white]{-}(0,57.5275)(5.1329,57.5275)(45.8736,16.7867)(45.8736,0)  
\psline[linewidth=3.5pt,linecolor=LineBlue]{-}(0,57.5275)(5.1329,57.5275)(5.1329,35.0617)(27.5987,35.0617)(27.5987,23.4483)(39.2121,23.4483)(39.2121,16.7867)(45.8736,16.7867)(45.8736,0)
\psline[linewidth=1.5pt,linecolor=LineRed]{-}(0,57.5275)(5.1329,57.5275)(5.1329,16.7867)(45.8736,16.7867)(45.8736,0)  

\psline(20,-1.5)(20,1.5) \rput(20,-4.5){\small{$1$}}
\psline(40,-1.5)(40,1.5) \rput(40,-4.5){\small{$2$}}
\psline(60,-1.5)(60,1.5) \rput(60,-4.5){\small{$3$}}
\psline(-1.5,20)(1.5,20) \rput(-3.5,20){\small{$1$}}
\psline(-1.5,40)(1.5,40) \rput(-3.5,40){\small{$2$}}
\psline(-1.5,60)(1.5,60) \rput(-3.5,60){\small{$3$}}

\rput(29,65){\textbf{Receiver 2 Rate Constraints}}
\psline{<-}(18,16.5)(12,11) \rput(12,8){$\mathcal{R}_{\text{MAC}}^{(\text{SIC})}(\Hm^{[2]})$}
\psline{<-}(32,32.5)(42.75,43.25) \rput(51.75,45.5){$\mathcal{R}_{\text{MAC}}(\Hm^{[2]})$}
\psline{<-}(35,24)(43.5,32.5) \rput(54,35){$\mathcal{R}_{\comp}^{(\succ)}(\Hm^{[2]},\Id)$}

}

\rput(95,80){
\psline[linewidth=1.5pt,linecolor=black]{->}(0,0)(70,0)  \rput(70,-5){$R_1$}
\psline[linewidth=1.5pt]{->}(0,0)(0,50)  \rput(-4.5,50){$R_2$}

\psline[linewidth=7pt,linecolor=black!65!white]{-}(0,38.3086)(20.2188,38.3086)(45.8736,12.6540)(45.8736,12.6540)(45.8736,0)
\psline[linewidth=4.5pt,linecolor=LineBlue]{-}(0,38.3086)(5.1329,38.3086)(5.1329,35.0617)(20.2188,35.0617)(20.2188,28.3601)(27.5987,28.3601)(27.5987,23.4483)(32.5105,23.4483)(32.5105,3.7507)(45.8736,3.7507)(45.8736,0)
\psline[linewidth=1.5pt,linecolor=LineRed]{-}(0,38.3086)(5.1329,38.3086)(5.1329,16.7867)(20.2188,16.7867)(20.2188,3.7507)(45.8736,3.7507)(45.8736,0)

\psline(20,-1.5)(20,1.5) \rput(20,-4.5){\small{$1$}}
\psline(40,-1.5)(40,1.5) \rput(40,-4.5){\small{$2$}}
\psline(60,-1.5)(60,1.5) \rput(60,-4.5){\small{$3$}}
\psline(-1.5,20)(1.5,20) \rput(-3.5,20){\small{$1$}}
\psline(-1.5,40)(1.5,40) \rput(-3.5,40){\small{$2$}}

\rput(31,49){\textbf{Intersection of Receiver 1 and}}
\rput(31,45){\textbf{Receiver 2 Rate Constraints}}
\psline{<-}(19.75,9.25)(3.5,-7) \rput(5,-10){$\mathcal{R}_{\text{MAC}}^{(\text{SIC})}(\Hm^{[1]})$}
\rput(9,-16){$~\cap~\mathcal{R}_{\text{MAC}}^{(\text{SIC})}(\Hm^{[2]})$}
\psline{<-}(35.25,25.25)(44,34) \rput(51.75,36.5){$\mathcal{R}_{\text{MAC}}(\Hm^{[1]}) \cap \mathcal{R}_{\text{MAC}}(\Hm^{[2]})$}
\psline{<-}(33.5,12.5)(42.5,21.5) \rput(54,25){$\mathcal{R}_{\comp}^{(\succ)}(\Hm^{[1]},\Id)$}
\rput(58,19){$~\cap~\mathcal{R}_{\comp}^{(\succ)}(\Hm^{[2]},\Id)$}
}

\rput(95,0){
\psline[linewidth=1.5pt,linecolor=black]{->}(0,0)(70,0)  \rput(70,-5){$R_1$}
\psline[linewidth=1.5pt]{->}(0,0)(0,50)  \rput(-4.5,50){$R_2$}
\psline[linewidth=7pt,linecolor=black!65!white]{-}(0,38.3086)(20.2188,38.3086)(45.8736,12.6540)(45.8736,12.6540)(45.8736,0)
\psline[linewidth=4.5pt,linecolor=LineBlue]{-}(0,38.3086)(5.1329,38.3086)(20.2188,35.0617)(27.5987,28.3601)(32.5105,23.4483)(45.8736,3.7507)(45.8736,0)

\psline[linewidth=1.5pt,linecolor=LineRed]{-}(0,38.3086)(5.1329,38.3086)(45.8736,3.7507)(45.8736,0)

\psline(20,-1.5)(20,1.5) \rput(20,-4.5){\small{$1$}}
\psline(40,-1.5)(40,1.5) \rput(40,-4.5){\small{$2$}}
\psline(60,-1.5)(60,1.5) \rput(60,-4.5){\small{$3$}}
\psline(-1.5,20)(1.5,20) \rput(-3.5,20){\small{$1$}}
\psline(-1.5,40)(1.5,40) \rput(-3.5,40){\small{$2$}}

\rput(34,48){\textbf{Compound MAC Rate Regions}}
\psline{<-}(22,23)(13,14) \rput(15,11){$\conv\Big(\mathcal{R}_{\text{MAC}}^{(\text{SIC})}(\Hm^{[1]})$}
\rput(22,5){$~\cap\mathcal{R}_{\text{MAC}}^{(\text{SIC})}(\Hm^{[2]})\Big)$}
\psline{<-}(29.75,30.75)(37,38) \rput(51.75,40.5){$\mathcal{R}_{\text{MAC}}(\Hm^{[1]}) \cap \mathcal{R}_{\text{MAC}}(\Hm^{[2]})$}
\psline{<-}(39.25,15)(45.75,21) \rput(50,29){$\conv\Big(\mathcal{R}_{\comp}^{(\succ)}(\Hm^{[1]},\Id)$}
\rput(58,23){$~\cap~\mathcal{R}_{\comp}^{(\succ)}(\Hm^{[2]},\Id)\Big)$}
}

\psline[linewidth=9pt,linecolor=LineGray]{->}(70,100)(90,100)
\psline[linewidth=9pt,linecolor=LineGray]{->}(70,52)(90,86)
\rput(81,94){\textcolor{black!80!white}{Intersect}}
\rput(81,88){\textcolor{black!80!white}{Rate Regions}}

\psline[linewidth=9pt,linecolor=LineGray]{->}(123,72.5)(123,52.5)
\rput(146,65){\textcolor{black!80!white}{Time-Sharing (Convexify)}}

\end{pspicture}
\end{center}
\caption{Step-by-step illustration for determining the capacity region, successive computation rate region, and SIC rate region for a compound multiple-access channel with two transmitters with powers $P_1 = 4$ and $P_2 = 3$, respectively, and two receivers with channel matrices $\Hm^{[1]} = [3.3~~2.1]$ and $\Hm^{[2]} = [2.4 ~~ 4.2]$, respectively. The top left depicts the rate constraints that must be satisfied for the first receiver to successfully decode both messages and the bottom left depicts the corresponding rate constraints for the second receiver. The top right shows the intersection of these rate constraints and the bottom right shows the achievable rate regions for the compound multiple-access channel that comes from convexification of the intersected regions, which corresponds operationally to time-sharing. Note that no convexification is needed for the capacity region $\mathcal{R}_{\text{MAC}}(\Hm^{[1]}) \cap \mathcal{R}_{\text{MAC}}(\Hm^{[2]})$ since it is the intersection of two convex rate regions.}

\label{f:compoundmacregion}
\end{figure*}

Successive computation does not reach the sum capacity for similar reasons. Namely, the sum-capacity rate pairs that can be directly attained with successive computation (see Theorem~\ref{t:compsuccmac}) will differ across receivers as will the required time-sharing ratios for other sum-capacity points. Using the achievable computation rate region from Theorem~\ref{t:compsuccmultiplerx} combined with time-sharing and setting $\Am^{[1]} = \Am^{[2]} = \Id$, we get that 
\begin{align*}
&\Rc_{\text{CMAC}}^{\succ}\big(\Hm^{[1]},\Hm^{[2]}\big) \\
&= \conv\Big(\Rc_{\comp}^{(\succ)}\big(\Hm^{[1]},\Id \big) \ \cap \  
\Rc_{\comp}^{(\succ)}\big(\Hm^{[2]},\Id \big)  \Big)
\end{align*} is achievable for the compound multiple-access channel. The flexibility to optimize over full-rank integer matrices $\Atm$ yields a larger rate region than SIC, as shown below.

\begin{example} In Figure~\ref{f:compoundmacregion}, we have illustrated how these rate regions are calculated for a compound multiple-access channel with $\Hm^{[1]} = [3.3~2.1]$, $\Hm^{[2]} = [2.4~4.2]$, $P_1 = 4$, and $P_2 = 3$. The corner points of the multiple-access capacity region with respect to receiver $1$ are $(1.0109,1.9154)$ and $(2.7388, 0.1875)$. Successive computation achieves these corner points as well as the rate pairs $(1.5084,1.4180)$ and $(1.6255,1.3008)$. With respect to receiver $2$, the corner points are $(0.2566,2.8764)$ and $(2.2937,0.8393)$ and successive computation achieves these corner points as well as the rate pairs $(1.3799,1.7531)$ and $(1.9606,1.1724)$. As expected, after intersecting the individual multiple-access regions and time-sharing, neither SIC nor successive computation reach the sum capacity.
\end{example}

\begin{remark} In recent work, Wang~\etal\cite{wsk14} have demonstrated that, for this scenario, a variation of SIC that encodes messages across multiple blocks and employs sliding-window decoding can match the performance of joint typicality decoding.  \defsym
\end{remark}

\begin{remark}\label{r:compoundmacinterference} The compound multiple-access channel is an important building block for understanding the capacity region of multi-user interference networks. For instance, in the strong interference regime, the capacity region of a two-user interference channel corresponds exactly to that of a two-user compound multiple-access channel~\cite{hk81,sato81}. In this two-user setting, successive computation is inferior to i.i.d.~random coding with joint typicality decoding. However, in an interference channel with three or more users, there is the possibility of interference alignment~\cite{mmk08,cj08}. For instance, in the $K$-user symmetric Gaussian interference channel, the sum capacity in the strong interference regime can be approximated by that of a two-user symmetric Gaussian compound multiple-access channel~\cite{oen14}. To induce alignment at the receivers, the scheme from~\cite{oen14} employs the same lattice codebook at each transmitter and has each receiver decode its desired message indirectly, by first recovering two independent linear combinations via compute-and-forward. In recent work~\cite{ncnc13ISIT}, we have shown that lattice interference alignment is possible in any setting where the beamforming vectors are aligned ``stream-by-stream'' so long as the codewords are allowed to have unequal powers. 
\end{remark}

\section{Nested Lattice Codes} \label{s:lattices}
In this section, we describe the nested lattice codes that will be the building blocks of our encoding and decoding schemes. We begin with some basic lattice definitions in Section \ref{s:latticedefinitions}, present nested lattice constructions and properties in Section \ref{s:latticeconstructions}, and discuss mappings to and from $\ZZ_p^k$ in Section~\ref{s:linearlabeling}.

\subsection{Lattice Definitions} \label{s:latticedefinitions}

We now review some properties of lattices that will be useful for our code constructions and refer interested readers to the textbook of Zamir~\cite{zamir} for a comprehensive treatment of lattices for coding. A \textit{lattice} $\L$ is a discrete additive subgroup of $\RR^n$ that is closed under addition and reflection, i.e., for any $\lambdav_1, \lambdav_2 \in \L$, we have that $-\lambdav_1, -\lambdav_2 \in \L$ and $\lambdav_1 + \lambdav_2 \in \L$. Note that this implies that the zero vector $\mathbf{0}$ is always an element of the lattice. 

Let 
\begin{align*}
Q_\L(\xv) \define \argmin_{\lambdav \in \L} \| \xv - \lambdav \|
\end{align*} denote the nearest neighbor quantizer for $\L$. Using this, we define the (fundamental) \textit{Voronoi region} $\Vc$ of $\L$ to be the set of all points in $\RR^n$ which are quantized to the zero vector (breaking ties in a systematic fashion). We also define the \textit{modulo operation}, which outputs the error from quantizing $\xv$ onto $\L,$ as
\begin{align*}
[ \xv ] \modl  \define \xv - Q_\L(\xv) \ .
\end{align*} The modulo operation satisfies a \textit{distributive law},
\begin{align*} 
\big[ a [\xv] \modl + b[\yv] \modl \big] \modl = [ a \xv + b \yv ] \modl  \ ,
\end{align*} for any integers $a,b \in \ZZ$. The \textit{second moment} of a lattice, denoted as $\sigma^{2}(\L)$, is the second moment per dimension of the norm of a random vector that is drawn uniformly over the fundamental Voronoi region $\Vc$, that is, 
\begin{align*}
\sigma^2(\L) \define \frac{1}{n} \int_{\Vc} \| \xv \|^2 \frac{1}{\Vol(\Vc)} d\xv \ ,
\end{align*}
where $\mbox{Vol}(\Vc)$ denotes the volume of $\Vc$.

The following lemma will be useful in characterizing the distributions of dithered lattice codewords. 
\begin{lemma}[Crypto Lemma] \label{l:crypto}
Let $\L$ be a lattice, $\xv$ a random vector with an arbitrary distribution over $\RR^n$, and $\dv$ a random vector that is independent of $\xv$ and uniform over $\Vc$. It follows that the random vector $[ \xv + \dv ] \bmod{\L}$ is independent of $\xv$ and uniform over $\Vc$. \thmsym
\end{lemma} See~\cite[Ch. 4.1]{zamir} for a detailed discussion and proof.

We say that lattice $\L_{\c}$ is \textit{nested} in lattice $\L_{\f}$ if $\L_{\c} \subset \L_{\f}$. The lattice $\L_{\f}$ is often referred to as the {\it fine} lattice and $\L_{\c}$ as the {\it coarse} lattice. The coarse lattice induces a \textit{partition} of the fine lattice into \textit{cosets} of the form $\lambdav + \L_{\c}$ for $\lambdav \in \L_{\f}$. The set of all such cosets is written as $\L_{\f} / \L_{\c} \triangleq \big\{\lambdav + \L_{\c} : \lambdav \in \L_{\f}\big\}$. (The notation $\L_{\f} / \L_{\c}$ refers to the fact that this is a quotient group.) It will often be useful to represent each coset by a single element, i.e., a \textit{coset representative}. We will represent each coset by its minimum norm element to obtain a set of coset representatives,
\begin{align*}
[ \L_{\f} / \L_{\c} ] \triangleq \L_{\f} \bmod{\L_{\c}} \ . 
\end{align*} 

A \textit{nested lattice codebook} $\Lc$ is generated using a nested lattice pair $\L_{\c} \subset \L_{\f}$. The codebook $\mathcal{L}$ comprises all elements of the fine lattice that fall within the Voronoi region $\Vc_{\c}$ of the coarse lattice, $\Lc = \L_{\f} \cap \Vc_{\c}$. The rate of the codebook is 
\begin{align*}
\frac{1}{n} \log \big| \Lc \big| = \frac{1}{n} \log \bigg(\frac{\Vol(\Vc_{\c})}{\Vol({\cal V}_{\f})} \bigg)
\end{align*} where $\Vc_{\f}$ is the Voronoi region of $\L_{\f}$. It can be shown that the set $\L_\f \cap \mathcal{V}_\c$ is equal to the set of minimum-norm coset representatives $[\L_\f/\L_\c]$. This implies that the codebook $\mathcal{L}$ can be interpreted algebraically (i.e., as the set of coset representatives $[\Lambda_{\f}/\Lambda_{\c}]$) and geometrically (i.e., as the set of all elements of the fine lattice $\Lambda_{\f}$ that are in the Voronoi region of the coarse lattice $\Lambda_{\f}$). Loosely speaking, the geometric properties of our lattice constructions will be useful in ensuring that the codewords are spaced sufficiently far apart in $\mathbb{R}^{n}$ and that the power constraints are satisfied (Theorem \ref{t:mainlattice}). Similarly, the algebraic interpretation will be useful for constructing a linear mapping between our nested lattice codebooks and $\ZZ_p^k$, the vector space from which the messages are drawn (Theorem~\ref{t:linearlabeling}).

Finally, note that any nested lattice pair $\L_\c \subset \L_\f$ satisfies the following quantization property:
\begin{align}
\big[ Q_{\L_{\f}}(\xv) \big] \modl_\c = \Big[ Q_{\L_{\f}}\big([\xv]\modl_\c \big) \Big] \modl_\c \ . \label{e:nestedquantization}
\end{align}

\subsection{Nested Lattice Constructions} \label{s:latticeconstructions}
We will employ the nested lattice construction of Ordentlich and Erez~\cite{oe16} as part of our achievability scheme. We will design nested lattice codebooks using the same parameters in our problem statement, $n,p,k_{\c,\ell},k_{\f,\ell},$ for $\ell = 1,\ldots,L$. Consider the prime $p$ and the corresponding finite field\footnote{The field $\mathbb{Z}_{p}$ is considered here rather than a generic prime-sized finite field since we will use the natural mapping from $\mathbb{Z}_{p}$ to $\mathbb{Z}$ to lift our codes from the finite field to reals.} $\mathbb{Z}_p$. The key idea is to take a series of nested linear codes of length $n$ over $\mathbb{Z}_{p}$ and then lift the codes from $\mathbb{Z}_{p}^{n}$ to $\mathbb{R}^{n}$ using Construction A to obtain a series of nested lattices. 

We now specify parameters\footnote{These parameter choices are made to simplify the existence proofs. For instance, the prime $p$ is chosen to grow with $n$ so that the channel input distributions will look nearly Gaussian. In practice, one could take $p$ to be relatively small and accept the rate loss associated with $p$-ary inputs to a Gaussian channel.} for which sequences of good nested lattices exist. Let $P_{\text{max}} \triangleq \max_\ell P_\ell$ and $V_n$ be the volume of an $n$-dimensional ball of radius $1$. Following the construction in~\cite{oe16}, for a given blocklength $n$, powers $P_\ell$, and noise tolerances $\sigma_{\eff,\ell}^2$, we will set $p$ to be the largest prime between $\frac{1}{2}n^{3/2}$ and $n^{3/2}$, which is guaranteed to exist for $n > 1$ by Bertrand's Postulate\footnote{For any $m >3$,  Bertrand's Postulate, as proven by Chebyshev~\cite{chebyshev52}, states that there exists a prime between $m$ and $2m$.}, and
\begin{align}
\gamma &= 2 \sqrt{nP_{\text{max}}2^{\alpha}} \nonumber \\
k_{\c,\ell} &= \frac{n}{2\log{p}} \Bigg( \log\bigg(\frac{P_{\text{max}}}{P_\ell}\bigg) +  \log\bigg(\frac{4}{V_n^{2/n}}\bigg) + \alpha \Bigg) \label{eq:coarse_parameter} \\
k_{\f,\ell} &= \frac{n}{2\log{p}} \Bigg( \log\bigg(\frac{P_{\text{max}}}{\sigma_{\eff,\ell}^2}\bigg) +  \log\bigg(\frac{4}{V_n^{2/n}}\bigg) + \alpha \Bigg)  \label{eq:fine_parameter}
\end{align} where $\alpha > 0$ will be chosen as part of Theorem~\ref{t:mainlattice}.

Recall that $\kfmax \triangleq \max_{\ell} \kfl$ and consider a $\kfmax \times n$ matrix $\mathbf{G}$ over $\mathbb{Z}_{p}.$ Let $\mathbf{G}_{\c,\ell}$ denote the matrix consisting of the first $k_{\c,\ell}$ rows of $\mathbf{G}$ and let $\mathbf{G}_{\f,\ell}$ denote the matrix consisting of the first $k_{\f,\ell}$ rows of $\mathbf{G}$ for $\ell=1,\ldots,L$. Define $\mathcal{C}_{\c,\ell}$ and $\mathcal{C}_{\f,\ell}$ to be the vector spaces generated by taking the columns of $\mathbf{G}_{\c,\ell}\T$ and $\mathbf{G}_{\f,\ell}\T$ as a basis, respectively: 
\begin{align*}
\mathcal{C}_{\c,\ell} &= \Big\{\mathbf{G}_{\c,\ell}\T \wv:  \mathbf{w}\in \mathbb{Z}_{p}^{k_{\c,\ell}} \Big\} \\
\mathcal{C}_{\f,\ell} &= \Big\{\mathbf{G}_{\f,\ell}\T \wv: \mathbf{w}\in \mathbb{Z}_{p}^{k_{\f,\ell}} \Big\} \ . 
\end{align*} Note that $\mathcal{C}_{\c,\ell},\mathcal{C}_{\f,\ell},\ell = 1,\ldots,L$ can also be viewed as an ensemble of nested linear codes.

Define the mapping $\phi: \ZZ_p \rightarrow \RR$ as
\begin{align*}
\phi(w) \triangleq \gamma p^{-1} w  
\end{align*} along with the inverse map 
\begin{align*}
\bar{\phi}(\kappa) \triangleq [ \gamma^{-1} p \kappa] \bmod{p} \ , 
\end{align*} which is only defined over the domain $\gamma p^{-1} \ZZ$. When applied to vectors, these mappings operate elementwise.

 Following (a scaled version\footnote{Construction A was originally proposed in~\cite{conwaysloane} as the vectors of the integer lattice $\ZZ^n$ whose modulo-$p$ reduction are elements of the linear codebook $\CC \subset \ZZ_p^n$, i.e., $\big\{\lambdav \in \ZZ^n : [ \lambdav]\bmod{p} \in \Cc\big\}$.} of) Construction A, we create the lattices 
\begin{align*}
\Lambda_{\c, \ell} &= \Big\{\lambdav \in \gamma p^{-1} \ZZ^n : \bar{\phi}(\lambdav) \in \mathcal{C}_{\c,\ell} \Big\} \\
\Lambda_{\f, \ell} &= \Big\{\lambdav \in \gamma p^{-1} \ZZ^n : \bar{\phi}(\lambdav)  \in \mathcal{C}_{\f,\ell} \Big\}
\end{align*} Note that, by construction, $\lambdav \in \L_{\c,\ell}$ (or $\L_{\f,\ell}$) if and only if $\bar{\phi}(\lambdav) \in \Cc_{\c,\ell}$ (or $\Cc_{\f,\ell}$). We will refer to $\bar{\phi}(\lambdav)$ as the \textit{corresponding linear codeword} of $\lambdav$.

We denote the Voronoi regions of $\Lambda_{\c,\ell}$ and $\Lambda_{\f,\ell}$ by $\mathcal{V}_{\c,\ell}$ and $\mathcal{V}_{\f,\ell}$, respectively. All $2L$ lattices in this ensemble are nested with respect to the permutation that places the parameters $k_{\c,\ell}$ and $k_{\f,\ell}$ in increasing order (i.e., they form a nested lattice chain). In particular, since $k_{\c,\ell} < k_{\f,\ell}$, the nested lattice codebook $\mathcal{L}_{\ell}$ can be constructed using $\Lambda_{\c,\ell}$ as the coarse lattice and $\Lambda_{\f,\ell}$ as the fine lattice,
\begin{align*}
\mathcal{L}_{\ell} \triangleq \Lambda_{\f, \ell} \cap \mathcal{V}_{\c,\ell} \ . 
\end{align*}

The theorem below restates key existence results from~\cite{oe16} in a form that is convenient for our achievability proofs. At a high level, the theorem guarantees that there exists a generator matrix $\mathbf{G}$, such that, for $\ell = 1,\ldots,L$, the submatrices $\mathbf{G}_{\c,\ell}, \mathbf{G}_{\f,\ell}$ are full rank and that each resulting nested lattice codebook $\Lc_\ell$ satisfies its power constraint $P_\ell$, tolerates effective noise with variance up to $\sigma_{\eff,\ell}^2 < P_\ell$, and has rate close to $\frac{1}{2} \log(P_\ell / \sigma_{\eff,\ell}^2)$.

\begin{theorem}[{\cite[Theorem 2]{oe16}}]\label{t:mainlattice}
Consider any choices of powers $P_\ell > 0$ and effective noise tolerances $0 < \sigma_{\eff,\ell}^2 < P_\ell$ for $\ell=1,\ldots,L$. For any $\epsilon > 0$ and $n$ large enough, there is a constant $\alpha > 0$ such that for the choices of $k_{\c,\ell},k_{\f,\ell}$ in \eqref{eq:coarse_parameter}-\eqref{eq:fine_parameter}, there exists a matrix $\mathbf{G} \in \ZZ_p^{k_{\f} \times n}$, such that, for $\ell = 1,\ldots,L$,
\begin{itemize}
\setlength{\itemindent}{0em}
\item[(a)] the submatrices $\mathbf{G}_{\c,\ell}, \mathbf{G}_{\f,\ell}$ are full rank.
\item[(b)] the coarse lattices $\L_{\c,\ell}$ have second moments close to the power constraint $$P_\ell - \epsilon < \sigma^2(\Lambda_{\c,\ell}) \leq P_\ell \ . $$
\item[(c)] the lattices tolerate the prescribed level of effective noise. Specifically, consider any linear mixture of Gaussian and Voronoi-shaped noise of the form $\mathbf{z}_{\eff} = \beta_0 \mathbf{z}_0 + \sum_{\ell = 1}^L \beta_\ell \mathbf{z}_\ell$ where $\beta_0,\beta_1,\ldots,\beta_L \in \mathbb{R}$, $\mathbf{z}_0 \sim \mathcal{N}(\mathbf{0},\mathbf{I})$, $\mathbf{z}_\ell \sim \mathrm{Unif}(\mathcal{V}_{\c,\ell})$, and the noise components $\zv_0,\zv_1,\ldots,\zv_L$ are independent of each other and $\lambdav$. Then, for any fine lattice point $\lambdav \in \Lambda_{\f,m}$, 
$$ \mathbb{P}\Big(  Q_{\Lambda_{\f,m}}(\lambdav + \mathbf{z}_{\eff} ) \neq \lambdav  \Big) < \epsilon$$ if $\beta_0^2 + \sum_{\ell = 1}^L \beta_\ell^2 P_\ell \leq \sigma_{\eff,m}^2$. Similarly, for any coarse lattice point $\lambdav \in \Lambda_{\c,m}$, 
$$ \mathbb{P}\Big(  Q_{\Lambda_{\c,m}}(\lambdav + \mathbf{z}_{\eff} ) \neq \lambdav  \Big) < \epsilon$$ if $\beta_0^2 + \sum_{\ell = 1}^L \beta_\ell^2 P_\ell \leq P_m^2$.
\item[(d)] the nested lattice codebooks $\Lc_\ell =  \Lambda_{\f,\ell} \cap \mathcal{V}_{\c,\ell}$ have appropriate rates $$  \frac{1}{n} \log \big| \Lc_\ell \big| = \frac{k_{\f,\ell} - k_{\c,\ell}}{n} \log{p}  > \frac{1}{2} \log\bigg( \frac{P_\ell}{\sigma^2_{\eff,\ell}}\bigg) - \epsilon \ .  $$
\end{itemize} \thmsym
\end{theorem} Note that (a) is established in the proof of~\cite[Theorem 1]{oe16}, which is then used to establish~\cite[Theorem 2]{oe16} as a corollary.

\begin{remark}
In Theorem~\ref{t:mainlattice}, we have only stated lattice properties that are essential for our achievability proofs. In many cases, it can be shown that nested lattices satisfying stronger versions of these properties exist, which in turn could be used to relax some of the assumptions in our problem statement. For instance, one can select lattices that are tuned for non-Gaussian channel noise. See~\cite{elz05,oe16,zamir} for more details. \defsym
\end{remark}

\subsection{Linear Labeling} \label{s:linearlabeling}

We now show how this ensemble of nested lattice codebooks can be connected to computing linear combinations of messages over $\ZZ_p$. Roughly speaking, we would like to map messages to nested lattice codewords so that our desired linear combinations can be directly recovered from appropriate integer-linear combinations of the lattice codewords. The original compute-and-forward framework~\cite{ng11IT} directly employed the Construction A mapping from a linear code to a lattice code. Subsequent work by Feng \etal\cite{fsk13} developed a richer set of algebraic connections as well as guidelines for selecting codes and constellations that are amenable to compute-and-forward. In the process,~\cite{fsk13} proposed the concept of a \textit{linear labeling} as an elegant way to map between an algebraic message space and a nested lattice code. We adopt this approach for our expanded framework.

\begin{figure*}[!t]
	\includegraphics[width=\textwidth]{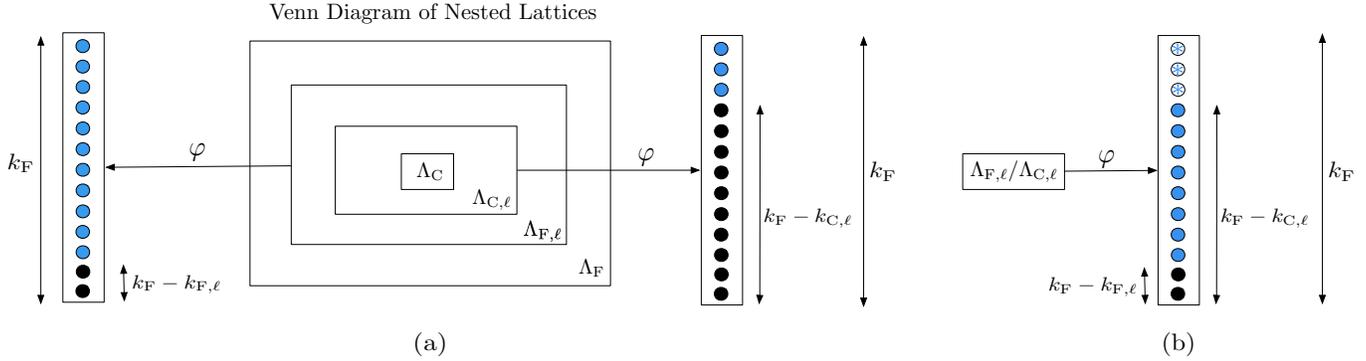}
	\caption{(a) Depiction of the linear labeling $\varphi$ of Theorem \ref{t:linearlabeling} via signal levels for a user $\ell$ with $k_{\textrm{F}}=13$, $k_{\f,\ell}=11$, and $k_{\c,\ell}=3$. We use black circles to represent zeros and lightly shaded (blue) circles to represent occupied symbols. On the left-hand side, we see that the linear labeling of a lattice point from $\Lambda_{\f,\ell}$ must have zeros in its last $k_{\f} - k_{\f,\ell}$ entries. Similarly, on the right-hand side, we see that the linear labeling of a lattice point from $\Lambda_{\c,\ell}$ must have zeros in its last $k_{\f} - k_{\c,\ell}$ entries.  (b) Depiction of the linear labeling  $\varphi$ of Theorem \ref{t:linearlabeling} of an element of $\Lambda_{\f,\ell}/ \Lambda_{\c,\ell}$. The $*$ elements represent ``don't care'' entries; for a given choice of signal levels represented by the blue circles, different realizations of the ``don't care'' levels correspond to different representatives of the same coset.
	}
	
\label{fig:lin_labeling}
\end{figure*}

Let $\ell_{\text{min}} = \argmin_\ell k_{\c,\ell}$ and $\ell_{\text{max}} = \argmax_\ell k_{\f,\ell}$. It will be convenient to define $\L_{\c} \triangleq \L_{\c,\ell_{\text{min}}}$ and $\L_{\f} \triangleq \L_{\f,\ell_{\text{max}}}$ as the coarsest and finest lattices in the ensemble, respectively. It follows that $\L_{\c} \subseteq \L_{\c,\ell} \subset \L_{\f,\ell} \subseteq \L_{\f}$ for $\ell = 1,\ldots,L$. Recall that $\kcmin \triangleq \min_{\ell} k_{\c,\ell}$, $\kfmax \triangleq \max_{\ell} k_{\f,\ell}$, and $k \triangleq \kfmax - \kcmin$.

\begin{definition}\label{d:linearlabeling}
A mapping $\varphi: \L_{\f} \rightarrow \ZZ_p^{k}$ is called a \textit{linear labeling} if it satisfies the following two properties:
\begin{itemize}
\item[(a)] A lattice point $\lambdav$ belongs to $\L_{\f,\ell}$ if and only if the last $\kfmax - k_{\f,\ell}$ components of its label $\varphi(\lambdav)$ are equal to $0$. Similarly, a lattice point $\lambdav$ belongs to $\L_{\c,\ell}$ if and only if the last $\kfmax - k_{\c,\ell}$ components of its label $\varphi(\lambdav)$ are equal to $0$.

\item[(b)] For all $a_\ell \in \ZZ$ and $\lambdav_\ell \in \L_{\f}$, we have that
\begin{align*}
\varphi\bigg( \sum_{\ell = 1}^L a_\ell \lambdav_\ell \bigg) = \bigoplus_{\ell = 1}^L q_\ell \,\varphi(\lambdav_\ell)
\end{align*} where $q_\ell = [ a_\ell ] \bmod{p}$. 
\end{itemize} \defsym
\end{definition}

Our proposed linear labeling stems directly from the fact that, for any $\lambdav \in \L_{\f}$, the corresponding linear codeword can be expressed as $\bar{\phi}(\lambdav) = \Gm\T \vv$ for some vector $\vv \in \ZZ_p^{\kfmax}$. Assuming that $\Gm$ is full rank, then this vector is unique.

\begin{theorem}\label{t:linearlabeling}
Assume that $\Gm$ is full rank. Let $\varphi: \L_{\f} \rightarrow \ZZ_p^k$ be the function that maps each $\lambdav \in \L_{\f}$ to the vector $\varphi(\lambdav)$ that consists of the last $k$ components of the unique vector $\vv$ satisfying $\bar{\phi}(\lambdav) = \Gm\T \vv$. Then, $\varphi$ is a linear labeling. \thmsym
\end{theorem} See Appendix~\ref{app:prooflinearlabeling} for a proof. We have depicted Theorem \ref{t:linearlabeling} through signal levels in Figure \ref{fig:lin_labeling}.

It will also be useful to have an explicit inverse $\bar{\varphi}: \ZZ_p^k \rightarrow \L_{\f}$ for the linear labeling $\varphi$ from Theorem~\ref{t:linearlabeling}. Specifically, define 
\begin{align*}
\bar{\varphi}(\wv) \triangleq \phi\bigg(\Gm\T\begin{bmatrix} \zerov_{\kcmin} \\ \wv \end{bmatrix} \bigg) \ . 
\end{align*} It follows that $\varphi\big(\bar{\varphi}(\wv)\big) = \wv$.

We now have all the ingredients we need to construct coding schemes for compute-and-forward. The next two sections develop coding schemes for parallel and successive computation, respectively.

\section{Parallel Computation Achievability: Proof of Theorem \ref{t:comppara}} 
\label{s:proofcomppara}

We now provide a detailed description of the encoding and decoding strategies used to achieve the parallel computation rate region from Theorem~\ref{t:comppara}. Notice that, for a given integer coefficient matrix $\Am$,  we determine the computation rate region $\Rc_{\comp}^{(\para)}(\Hm,\Am)$ by taking the union over all integer matrices $\Atm$ whose rowspan contains that of $\Am$. This has a clear operational meaning within our scheme: the receiver recovers linear combinations with integer coefficient matrix $\Atm$ and then solves these for the desired linear combinations with integer coefficient matrix $\Am$. As a first step, we will show how to directly recover linear combinations with integer coefficient matrix $\Am$ (for notational simplicity,) and derive conditions under which the probability of error can be driven to zero in Lemma~\ref{l:compparadirect}. This will serve as a building block in the proof of Theorem~\ref{t:comppara}. 

We begin with a high-level description of the encoding steps used to map the finite field messages onto dithered lattice codewords as well as the decoding steps used to estimate integer-linear combinations of lattice codewords, which are then mapped back to linear combinations of the messages. Take any choice of rates $R_1,\ldots, R_L$ and parameter $\epsilon > 0$, and apply Theorem~\ref{t:mainlattice} to select good nested lattices. Afterwards, apply Theorem~\ref{t:linearlabeling} to obtain a linear labeling $\varphi$ and its inverse $\bar\varphi$. The encoding and decoding process will make use of random\footnote{The use of random dither vectors should not be viewed as common randomness, but rather as part of the random coding proof. In Appendix~\ref{app:fixeddithers}, we will show that it suffices to use fixed dither vectors.} dither vectors that are generated independently and uniformly over the Voronoi regions of the coarse lattices, $\dv_\ell \sim \mathrm{Unif}(\Vc_{\c,\ell})$. 

\begin{figure}[h!]
\psset{unit=.77mm}
\begin{center}
\begin{pspicture}(11,30)(116,55)

\rput(12.5,40){$\wv_\ell$} \psline{->}(16,40)(23,40) \psframe(23,35)(41,45)
\rput(32,42.25){\small{Zero}}\rput(32,37.75){\small{Padding}} 
\psline[linecolor=black]{->}(41,40)(47,40)
\psframe(47,35)(57,45) \rput(52,40){$\bar{\varphi}$}
\psline{->}(57,40)(69,40)
\rput(63,43){$\lambdav_\ell$}
\pscircle(71.5,40){2.5} \psline{-}(72.75,40)(70.25,40)
\psline{-}(71.5,38.75)(71.5,41.25) \psline{<-}(71.5,42.5)(71.5,47.5) \rput(71.5,50){$\dv_\ell$}
\psline{->}(74,40)(82,40)
\psframe(82,35)(105,45)
\rput(93.5,40){$\modl_{\c,\ell}$}
\psline{->}(105,40)(112,40)
\rput(115.5,40){$\xv_\ell$}

\psframe[linewidth=2pt,linestyle=dashed,linecolor=LineBlue](19,25)(108,55)
\rput(64,30){$\ell$th Encoder}

\end{pspicture}
\end{center}
\caption{Block diagram for the $\ell$th encoder for both parallel computation and successive computation.}
\label{f:compparaenc}
\end{figure}
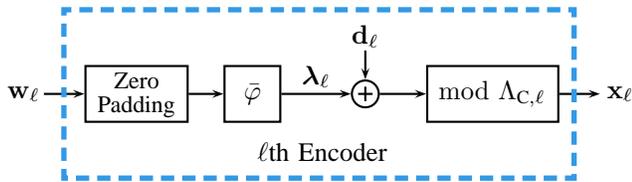

The $\ell$th encoder begins by adding $k_{\c,\ell} - \kcmin$ leading zeros and $\kfmax - k_{\f,\ell}$ trailing zeros to its message $\wv_\ell \in \ZZ_p^{k_{\f,\ell} - k_{\c,\ell}}$. The resulting length-$k$ vector is mapped onto a lattice point $\lambdav_\ell \in \Lc_\ell$ using the inverse $\bar{\varphi}$ of the linear labeling followed by taking modulo $\L_{c,\ell}$. Afterwards, the encoder dithers $\lambdav_\ell$ using $\dv_\ell$ to obtain its channel input $\xv_\ell$. These encoding operations are illustrated in Figure~\ref{f:compparaenc} and formally written out in~\eqref{e:compparaencoding}. Note that encoding does not depend on either the channel matrix $\Hm$ or the choice of integer coefficient vector $\av_m$.

\noindent\textbf{Encoding:}
\begin{subequations}\label{e:compparaencoding}
\begin{align}
\lambdav_\ell &= \left[ \bar{\varphi}\left( \begin{bmatrix} \zerov_{k_{\c,\ell} - \kcmin} \\ \wv_\ell \\ \zerov_{\kfmax - k_{\f,\ell}} \end{bmatrix} \right)\right] \modl_{\c,\ell} \label{e:compparaencoding1} \\
 \xv_\ell &= [ \lambdav_\ell + \dv_\ell ] \modl_{\c,\ell} \label{e:compparaencoding2}
\end{align}
\end{subequations}

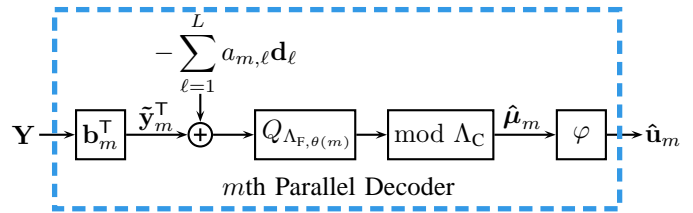
\begin{figure}[h!]
\psset{unit=.68mm}
\begin{center}
\begin{pspicture}(36,30)(164,65)

\rput(37,40){$\Ym$}
\psline[linecolor=black]{->}(40,40)(47,40)
\psframe(47,35)(57,45) \rput(52,40){$\bv_m\T$}
\psline{->}(57,40)(69,40)
\rput(63,43.5){$\ytv_m\T$}
\pscircle(71.5,40){2.5} \psline{-}(72.75,40)(70.25,40)
\psline{-}(71.5,38.75)(71.5,41.25) \psline{<-}(71.5,42.5)(71.5,48) \rput(76.5,56){$\displaystyle -\sum_{\ell=1}^L a_{m,\ell} \dv_\ell$}
\psline{->}(74,40)(82,40)
\psframe(82,35)(102,45)
\rput(92.5,40){$Q_{\L_{\f,\theta(m)}}$}
\psline{->}(102,40)(108,40)
\psframe(108,35)(129,45)
\rput(118.5,40){$\modl_{\c}$}
\psline{->}(129,40)(141,40)
\rput(134.5,43.5){$\muhv_m$}
\psframe(141,35)(151,45)
\rput(146,40){$\varphi$}
\psline{->}(151,40)(158,40)
\rput(162,40){$\uhv_m$}

\psframe[linewidth=2pt,linestyle=dashed,linecolor=LineBlue](42.5,25)(154,65)
\rput(98.5,30){$m$th Parallel Decoder}

\end{pspicture}
\end{center}
\caption{Block diagram for the $m$th decoder for parallel computation.}
\label{f:compparadec}
\end{figure}

To recover the linear combination $\uv_m$ with integer coefficient vector $\av_m$, the receiver first applies the equalization vector $\bv_m \in \RR^{\ant}$ to its channel observation $\Ym$, to obtain the effective channel output $\ytv_m$. The receiver then attempts to recover the integer-linear combination 
\begin{equation} \label{e:integerlinear}
\muv_m = \bigg[ \sum_{\ell = 1}^L a_{m,\ell} \lambdatv_\ell \bigg] \modl_{\c}
\end{equation} where 
\begin{equation} \label{e:lambdatilde}
\lambdatv_\ell \define \lambdav_\ell - Q_{\Lambda_{\c,\ell}}(\lambdav_\ell + \dv_\ell)  
\end{equation} is a lattice point in the same coset $\L_{\f,\ell} / \L_{\c,\ell}$ as the $\ell$th user's lattice codeword $\lambdav_\ell$. An estimate $\muhv_m$ of $\muv_m$ is obtained by first subtracting an integer-linear combination of the dithers from $\ytv_m$, then quantizing onto the fine lattice $\L_{\f,\theta(m)}$ where 
\begin{equation*}
\theta(m) \define  \argmax \big\{ k_{\f,\ell} : \ell \in \{1,\ldots,L\} \text{~s.t.~} [a_{m,\ell}] \bmod{p} \neq 0 \big\}
\end{equation*} denotes the index of the finest lattice amongst those that participate in the integer-linear combination $\muv_m$, and finally taking $\modl_{\c}$. To make an estimate $\uhv_m$ of the desired linear combination $\uv_m$, the receiver simply applies the linear labeling $\varphi$ to $\muhv_m$. These decoding operations are illustrated in Figure~\ref{f:compparadec} and formally written out in~\eqref{e:compparadecoding}.

\noindent\textbf{Decoding:}
\begin{subequations}\label{e:compparadecoding}
\begin{align}
\ytv_m\T &= \bv_m\T \Ym \label{e:compparadecoding1} \\
\muhv_m &= \bigg[ Q_{\L_{\f,\theta(m)}}\bigg(\ytv_m - \sum_{\ell=1}^L a_{m,\ell} \, \dv_\ell \bigg) \bigg] \modl_{\c} \label{e:compparadecoding2} \\
\uhv_m &= \varphi\big(\muhv_m\big) \label{e:compparadecoding3}
\end{align} 
\end{subequations}

The following lemma describes rates that are attainable for the coding strategy above. 

\begin{lemma}\label{l:compparadirect}
Consider any choice of rates $R_1,\ldots,R_L$ and parameter $\zeta  > 0$ and assume that the transmitters and receiver employ the encoding and decoding operations from~\eqref{e:compparaencoding}-\eqref{e:compparadecoding}. For $n$ and prime $p$ large enough, there exists a generator matrix $\Gm \in \ZZ_p^{\kfmax \times n}$ and corresponding nested lattice codes $\Lc_1,\ldots,\Lc_L$ with rates at least $R_1- \zeta,\ldots,R_L - \zeta$, respectively, such that, for any choice of channel matrix $\Hm \in \RR^{\ant \times L}$ and  integer matrix $\Am \in \ZZ^{L\times L}$, the receiver can recover the linear combinations $\uv_1,\ldots,\uv_L$ with probability of error at most $\zeta$ so long as
\begin{align}
R_\ell < \frac{1}{2}\log^+\left( \frac{P_\ell} { \sigma_{\para}^2(\Hm,\av_m,\bv_m) } \right) \label{e:compparadirectcondition}
\end{align} for all $(m,\ell)$ such that $[a_{m,\ell}]\bmod{p} \neq 0$ for some choice of equalization vectors $\bv_1,\ldots,\bv_L \in \RR^{\ant}$. \thmsym
\end{lemma}

\begin{IEEEproof} First, we will select an ensemble of good nested lattice codebooks. Set $\sigma_{\eff,\ell}^2 = P_\ell 2^{-2 R_\ell}$ for $\ell = 1,\ldots,L$. Invoke Theorem~\ref{t:mainlattice} with these $\sigma_{\eff,1}^2,\ldots,\sigma_{\eff,L}^2$ and parameter $\epsilon = \zeta/L$ to obtain a generator matrix $\Gm \in \ZZ_p^{\kfmax \times n}$ and associated nested lattice codebooks $\Lc_1,\ldots,\Lc_L$ with rates at least $R_1-\epsilon, \ldots,R_L - \epsilon$, respectively. Use Theorem~\ref{t:linearlabeling} to obtain a linear labeling $\varphi: \L_{\f} \rightarrow \ZZ_p^k$ and its inverse $\bar{\varphi}: \ZZ_p^k \rightarrow \L_{\f}$.

For $\ell = 1,\ldots,L$, generate an independent random dither vector according to $\dv_\ell \sim \mathrm{Unif}(\Vc_{\c,\ell})$ and make it available to the $\ell$th transmitter and the receiver. Each transmitter employs~\eqref{e:compparaencoding} to generate its lattice codeword $\lambdav_\ell$ and channel input $\xv_\ell$. From the Crypto Lemma, we know that $\xv_\ell$ is independent of $\lambdav_\ell$ and uniform over $\Vc_{\c,\ell}$. Therefore, it follows from Theorem~\ref{t:mainlattice}(b) that each transmitter satisfies its power constraint~\eqref{e:powerconstraint}.

For $m = 1,\ldots, L$, the receiver selects an equalization vector $\bv_m \in \RR^{\ant}$ and generates an effective channel output $\ytv_m$ via~\eqref{e:compparadecoding1}. After subtracting an integer combination of the dither vectors, we obtain an integer combination of lattice codewords plus effective noise:
\begin{align*}
&\ytv_m - \sum_{\ell = 1}^L a_{m,\ell}\, \dv_\ell \\
&= \sum_{\ell = 1}^L a_{m,\ell} (\xv_\ell - \dv_\ell) + \zv_{\para,m} \\
&= \sum_{\ell = 1}^L a_{m,\ell} \big( \lambdav_\ell + \dv_\ell - Q_{\L_{\c,\ell}}(\lambdav_\ell + \dv_\ell) - \dv_\ell \big) + \zv_{\para,m} \\
&= \sum_{\ell=1}^L a_{m,\ell} \lambdatv_\ell + \zv_{\para,m} 
\end{align*}where
\begin{align*}
\zv_{\para,m}\T &\triangleq (\bv_m\T \Hm - \av_m\T) \Xm + \bv_m\T \Zm \ .
\end{align*}

The linear label of $\lambdatv_\ell$ satisfies
\begin{align*}
\varphi(\lambdatv_\ell) = \varphi(\lambdav_\ell) \ominus \varphi\big(Q_{\L_{\c,\ell}}(\lambdav_\ell + \dv_\ell) \big) 
\end{align*} where $\ominus$ denotes subtraction over $\ZZ_p$. By Definition~\ref{d:linearlabeling}(a), since $Q_{\L_{\c,\ell}}(\lambdav_\ell + \dv_\ell) \in \L_{\c,\ell}$, the last $\kfmax -\kcl$ components of the label $\varphi\big(Q_{\L_{\c,\ell}}(\lambdav_\ell + \dv_\ell)\big)$ are zero. Therefore, $\varphi(\lambdatv_\ell)$ agrees with $\varphi(\lambdav_\ell)$ on its last $\kfmax - \kcl$ components, i.e., it may not agree on the first $k - (\kfmax - \kcl) = \kfmax - \kcmin - \kfmax + \kcl = \kcl - \kcmin$ components. This implies that $\varphi(\lambdatv_\ell) \in \llb \wv_\ell \rrb$ as defined in~\eqref{e:msgcoset}. By Definition~\ref{d:linearlabeling}(b), $\varphi\big(\sum_\ell a_{m,\ell} \, \lambdatv_\ell \big) = \bigoplus_\ell q_{m,\ell} \,\varphi(\lambdatv_\ell)$ where $q_{m,\ell} = [a_{m,\ell}]\bmod{p}$ so the linear label can be viewed as a linear combination $\uv_m = \varphi\big(\sum_\ell a_{m,\ell} \, \lambdatv_\ell \big)$ with integer coefficient vector $\av_m$. This also implies, via Definition~\ref{d:linearlabeling}(a), that $\sum_\ell a_{m,\ell} \lambdatv_\ell \in \L_{\f,\theta(m)}$, which will be useful in the next step.

Applying~\eqref{e:compparadecoding2}, the receiver makes an estimate $\muhv_m$ of the integer-linear combination $\muv_m$. By Theorem~\ref{t:mainlattice}(c), we have that $\pr(\muhv_m \neq \muv_m) < \epsilon$ if 
\begin{align} \label{e:paranoisethreshold}
\sigma_{\text{para}}^2(\Hm,\av_m,\bv_m) < \sigma_{\text{eff},\theta(m)}^2
\end{align} where $\sigma_{\text{para}}^2(\Hm,\av_m,\bv_m)$ is defined in~\eqref{e:noisevarpara2}. Afterwards, the receiver applies the linear labeling as in~\eqref{e:compparadecoding3} to obtain its estimate $\uhv_m$. If $\muhv_m = \muv_m$, then $\uhv_m = \uv_m$ since $\varphi(\muv_m) = \varphi\big(\sum_\ell a_{m,\ell} \lambdatv_\ell\big) \ominus \varphi\big(Q_{\L_\c}\big(\sum_\ell a_{m,\ell} \lambdatv_\ell\big)\big)$ and the second term is equal to $\zerov_k$ since it is a linear labeling of an element of $\L_{\c}$ (i.e., the last $\kfmax - \kcmin = k$ components of its label are zero). 

Note that the condition~\eqref{e:paranoisethreshold} is equivalent to insisting that $R_\ell < \frac{1}{2} \log^+\big(P_\ell / \sigma_{\text{para}}^2(\Hm,\av_m,\bv_m)\big)$ for all $m$ and $\ell$ such that $[a_{m,\ell}]\bmod{p} \neq 0$. Applying the union bound, we get that $\pr\big(\cup_m \{ \uhv_m \neq \uv_m \} \big) < L\epsilon = \zeta$. Finally, it is argued in Appendix~\ref{app:fixeddithers} that it suffices to use fixed dither vectors.
\end{IEEEproof}

We are now ready to prove Theorem~\ref{t:comppara}.

\noindent\textit{Proof of Theorem~\ref{t:comppara}:} Choose rates $R_1,\ldots,R_L$ and select nested lattice codebooks via Lemma~\ref{l:compparadirect}. Each transmitter employs the encoding strategy from Lemma~\ref{l:compparadirect} (which does not depend on the channel matrix $\Hm$ nor the integer matrix $\Am$). For a given channel matrix $\Hm$, say that the receiver wishes to decode linear combinations with integer coefficient matrix $\Am$ where $(R_1,\ldots,R_L) \in \mathcal{R}_{\comp}^{(\para)}(\Hm,\Am)$. This implies that there exists some integer matrix $\Atm$ satisfying $\rowspan(\Atm) \subseteq \rowspan(\Am)$ and $(R_1,\ldots,R_L) \in \mathcal{R}_{\para}(\Hm,\Atm)$. The receiver applies the decoding strategy from Lemma~\ref{l:compparadirect} with integer matrix $\Atm$ and optimal equalization vectors $\bv_{\opt,m}$ chosen via Lemma~\ref{l:mmsepara}. Recall from~\eqref{e:minvarparadef} that $\sigma^2_{\para}(\Hm,\atv_m) \define \sigma^2_{\para}(\Hm,\atv_m,\bv_{\opt,m})$ so~\eqref{e:compparadirectcondition} matches the rate constraints from $\mathcal{R}_{\para}(\Hm,\Atm)$. $\blacksquare$

\section{Successive Computation Achievability: Proof of Theorem~\ref{t:compsucc}} \label{s:proofcompsucc}

In this section, we show how to improve the decoding process using successive cancellation, culminating in a proof of Theorem~\ref{t:compsucc}. The successive computation rate region $\Rc_{\comp}^{(\para)}(\Hm,\Am)$ involves a union over all integer matrices $\Atm$ whose rowspan contains that of $\Am$ as well as a union over all admissible mappings $\Mc(\Atm)$ for each $\Atm$. As before, the union over integer matrices means that the receiver can first recover linear combinations with integer coefficient matrix $\Atm$, and then solve these for the desired linear combinations with integer coefficient matrix $\Am$. Each admissible mapping $\Ic \in \Mc(\Atm)$ corresponds to a specific successive cancellation order for the codewords, as we will explain in detail below. We will begin by showing how to directly recover linear combinations with coefficient matrix $\Am$ and admissible mapping $\Ic$ (for notational simplicity), and, in Lemma~\ref{l:compsuccdirect}, state conditions under which the probability of error can be made to vanish in the blocklength. Afterwards, we will use this lemma to prove Theorem~\ref{t:compsucc}. Note that the decoding order for linear combinations is fixed to be lexicographic for notational convenience; other orders can be reached by exchanging rows of $\Am$ (which is taken care of by the union over integer matrices that include the rowspan of $\Am$).

We start with a high-level overview of our encoding and decoding process. As in Section~\ref{s:proofcomppara}, we fix rates $R_1,\ldots,R_L$ and parameter $\epsilon > 0$ and then invoke Theorems~\ref{t:mainlattice} and~\ref{t:linearlabeling} to select good nested lattices as well as a linear labeling $\varphi$ and its inverse $\bar{\varphi}$. Next, generate random dither vectors independently and uniformly over the Voronoi regions of the coarse lattices, $\dv_\ell \sim \mathrm{Unif}(\Vc_{\c,\ell})$. 

The encoding process is identical to that in the parallel computation case. We summarize the operations in~\eqref{e:compsuccencoding} and refer the reader to Section~\ref{s:proofcomppara} for a detailed discussion and Figure~\ref{f:compparaenc} for a block diagram.
\newline~\newline
\noindent\textbf{Encoding:}
\begin{subequations}\label{e:compsuccencoding}
\begin{align}
\lambdav_\ell &= \left[ \bar{\varphi}\left( \begin{bmatrix} \zerov_{k_{\c,\ell} - \kcmin} \\ \wv_\ell \\ \zerov_{\kfmax - k_{\f,\ell}} \end{bmatrix} \right)\right] \modl_{\c,\ell} \label{e:compsuccencoding1} \\
 \xv_\ell &= [ \lambdav_\ell + \dv_\ell ] \modl_{\c,\ell} \label{e:compsuccencoding2}
\end{align}
\end{subequations}

The first decoding step~\eqref{e:compsuccdecodefirst1}-\eqref{e:compsuccdecodefirst3} is quite similar to that of parallel computation (since there is not yet any side information to exploit). The decoder first recovers an integer-linear combination of lattice codewords $\muv_1$ as defined in~\eqref{e:integerlinear} and then applies the linear labeling. Let
\begin{equation*}
\vartheta(m) \define  \argmax \big\{ k_{\f,\ell} : \ell \in \{1,\ldots,L\} \text{~s.t.~} (m,\ell) \in \Ic \big\}
\end{equation*} denote the index of finest lattice that will participate in the $m$th integer-linear combination  under the admissible mapping $\Ic$. Note that $\vartheta(1) = \theta(1)$ since no successive cancellation takes place in the first round. 
\newline~\newline
\noindent\textbf{Decoding, m$\mathbf{\ =1}$:}
\begin{subequations}\label{e:compsuccdecodefirst}
\begin{align}
\ytv_1\T &= \bv_1\T \Ym \label{e:compsuccdecodefirst1} \\
\muhv_1 &= \bigg[ Q_{\L_{\f,\vartheta(1)}}\bigg(\ytv_1 - \sum_{\ell = 1}^L a_{1,\ell}\, \dv_\ell \bigg) \bigg] \modl_{\c} \label{e:compsuccdecodefirst2} \\
\uhv_1 &= \varphi\big(\muhv_1\big)  \label{e:compsuccdecodefirst3} \\
\chihv_1 &= \bigg[ \muhv_1 + \sum_{\ell=1}^L a_{1,\ell}\, \dv_\ell \bigg] \modl_\c \label{e:compsuccdecodefirst4}\\
\shv_1 &= Q_{\L_{\c}}\big(\ytv_1 - \chihv_1\big) + \chihv_1 \label{e:compsuccdecodefirst5}
\end{align}
\end{subequations}
After making its estimate $\uhv_1$, the decoder attempts to reconstruct the integer-linear combination of channel inputs $\av_1\T \Xm$ from $\muhv_1$ and $\ytv_1$ in~\eqref{e:compsuccdecodefirst4}-\eqref{e:compsuccdecodefirst5}. The following lemma characterizes when this process succeeds.

\begin{lemma}\label{l:realsumdirect}
Assume a receiver has access to an observation of the form $\ytv\T = \av\T \Xm + \zv_{\text{eff}}\T$ where $\av \in \ZZ^L$ and $\zv_{\text{eff}} \in \Vc_{\c}$, dithers $\dv_1,\ldots,\dv_L$, and the integer-linear combination $\muv = \big[ \sum_\ell a_\ell \lambdatv_\ell] \bmod{\L_{\c}}$ where $a_\ell$ is the $\ell$th entry of $\av$ and $\lambdatv_\ell = \lambdav_\ell - Q_{\L_{\c,\ell}}(\lambdav_\ell + \dv_\ell)$. Then, by calculating
\begin{align*}
\chiv &= \bigg[ \muv + \sum_{\ell=1}^L a_\ell\, \dv_\ell \bigg] \modl_\c \\
\sv &= Q_{\L_{\c}}\big(\ytv - \chiv\big) + \chiv \ , 
\end{align*} the receiver can obtain the integer-linear combination of the channel inputs, $\sv\T = \av\T \Xm$.
\end{lemma}
\begin{IEEEproof}
By the distributive law,
\begin{align}
\chiv &= \bigg[\sum_{\ell=1}^L a_\ell \big(\lambdatv_\ell +  \dv_\ell\big) \bigg] \modl_\c \nonumber \\
&= \bigg[\sum_{\ell=1}^L a_\ell \Big(\lambdav_\ell +  \dv_\ell - Q_{\L_{\c,\ell}}(\lambdav_\ell + \dv_\ell)\Big) \bigg] \modl_\c \nonumber \\
&= \bigg[\sum_{\ell=1}^L a_\ell \xv_\ell \bigg] \modl_{\c}\ . \label{e:chimod}
\end{align} Therefore, 
\begin{align*}
\sv\T &= Q_{\L_{\c}}\big(\ytv\T - \chiv\T\big) + \chiv\T \\
&= Q_{\L_{\c}}\Big(\av\T \Xm + \zv_{\eff}\T - \big[ \av\T \Xm \big] \modl_{\c} \Big) + \chiv\T\\
&= Q_{\L_{\c}}\Big( Q_{\L_{\c}}\big(\av\T \Xm \big) + \zv_{\eff}\T \Big) + \chiv\T \\
&\overset{(i)}{=} Q_{\L_{\c}}\big(\av\T \Xm \big) +  \big[ \av\T \Xm \big] \modl_{\c}  \\
&= \av\T \Xm 
\end{align*} where (i) uses the fact that $\zv_{\text{eff}} \in \Vc_{\c}$ as well as~\eqref{e:chimod}.
\end{IEEEproof}

Thus, if $\muhv_1 = \muv_1$, Lemma~\ref{l:realsumdirect} will allow us to argue that $\shv_1\T = \av_1\T \Xm$, which can be used for~\textit{successive computation} as proposed by~\cite{nazer12}, i.e., creating better effective channels for subsequent linear combinations. In general, at the $m$th decoding step, we will have access to $\Am_{m-1} \Xm$ where $\Am_{m-1}$ is the submatrix consisting of the first $m-1$ rows of $\Am$, assuming all previous decoding steps are correct.  

The second ingredient in our decoding process is \textit{algebraic successive cancellation} as proposed by~\cite{oen14}. The main idea is that, at decoding step $m$, it is possible to use linear combinations from steps $1$ through $m-1$, to cancel out some of the codewords participating in the integer-linear combination $\muv_m$ without changing the effective noise variance. This in turn reduces the noise tolerance constraints placed on the fine lattices associated with the codewords and increases the overall rate region. Before we proceed, we need the following lemma that connects the definition of an admissible mapping to the existence of a matrix over $\ZZ_p$ that can be used for algebraic successive cancellation.

\begin{lemma}\label{l:algebraicsic} Let $\Ic$ be an admissible mapping for $\Am \in \ZZ^{L \times L}$. For prime $p$ large enough, there exists a lower unitriangular matrix $\Lbm \in \ZZ_p^{L \times L}$ such that, the $(m,\ell)$th entry of $\Abm = [\Lbm \Am] \bmod{p}$ is equal to zero (i.e., $\bar{a}_{m,\ell} = 0$)  for all $(m,\ell) \neq \Ic$. Furthermore, $\Lbm$ has a lower triangular inverse $\Lbm^{(\text{inv})} \in \ZZ_p^{L \times L}$ satisfying $[\Am] \bmod{p}  = [\Lbm^{(\text{inv})} \Abm] \bmod{p}$.
\end{lemma} 
\begin{IEEEproof}
By Definition~\ref{d:mappings}, since $\Ic$ is an admissible mapping, there exists a real-valued, lower unitriangular matrix $\Lm \in \RR^{L \times L}$ such that the $(m,\ell)$th entry of $\Lm \Am$ is equal to zero for all $(m,\ell) \neq \Ic$. It follows from~\cite[Appendix A]{oen14} that, for $p$ large enough, there exists a lower unitriangular matrix $\Lbm \in \ZZ_p^{L \times L}$ satisfying the same criterion for $\Abm = [\Lbm \Am] \bmod{p}$. Finally, since $\Lbm$ is lower unitriangular, it has a lower unitriangular inverse over $\ZZ_p$.
\end{IEEEproof}

An immediate consequence of Lemma~\ref{l:algebraicsic} is that we can use preceding linear combinations to eliminate lattice codewords according to the admissible mapping,
\begin{align} 
\nuv_m &= \bigg[ \muv_m + \sum_{i=1}^{m-1} \bar{l}_{m,i} \muv_i \bigg] \modl_{\c} \nonumber \\
&= \bigg[ \sum_{\ell = 1}^L \bar{a}_{m,\ell} \, \lambdatv_\ell \bigg] \modl_{\c} \label{e:reducedequation}
\end{align} where $\bar{l}_{m,i}$ is the $(m,i)$th entry of $\Lbm$ chosen via Lemma~\ref{l:algebraicsic} and $\bar{a}_{m,\ell}$ is the $(m,\ell)$ entry of $\Abm = [\Lbm \Am] \bmod{p}$. Similarly, using the inverse $\Lbm^{(\text{inv})}$ of $\Lbm$, we can return to the original integer-linear combinations, 
\begin{align} \label{e:rebuiltequation}
\bigg[\nuv_m +  \sum_{i=1}^{m-1} \bar{l}_{m,i}^{(\text{inv})} \nuv_i \bigg] \modl_{\c} = \muv_m
\end{align} where $\bar{l}_{m,i}^{(\text{inv})}$ is the $(m,i)$th entry of $\Lbm^{(\text{inv})}$.

\begin{figure*}[!t]
\psset{unit=.77mm}
\begin{center}
\begin{pspicture}(15,5)(244,65)

\rput(3,0){
\rput(16,40){$\Ym$}
\psline[linecolor=black]{->}(19,40)(27,40)
\psframe(27,35)(37,45) \rput(32,40){$\bv_m\T$}
\psline{->}(37,40)(47,40)
\pscircle(49.5,40){2.5} \psline{-}(48.25,40)(50.75,40)
\psline{-}(49.5,38.75)(49.5,41.25) \psline{<-}(49.5,42.5)(49.5,52) \rput(50.25,56){$\mathbf{\hat{S}}_{m-1}$}
}

\psline{->}(55,40)(69,40)
\pscircle[fillstyle=solid,fillcolor=black](62,40){1}
\rput(62,44.5){$\ytv_m\T$}
\pscircle(71.5,40){2.5} \psline{-}(72.75,40)(70.25,40)
\psline{-}(71.5,38.75)(71.5,41.25) \psline{<-}(71.5,42.5)(71.5,48) \rput(76.5,56){$\displaystyle -\sum_{\ell=1}^L a_{m,\ell} \dv_\ell$}

\rput(30,0){
\psline{->}(44,40)(69,40)
\pscircle(71.5,40){2.5} \psline{-}(72.75,40)(70.25,40)
\psline{-}(71.5,38.75)(71.5,41.25) \psline{<-}(71.5,42.5)(71.5,48) \rput(78.5,56){$\displaystyle  \sum_{i=1}^{m-1} \bar{l}_{m,i} \muhv_i $}
}

\rput(30,0){
\psline{->}(74,40)(82,40)
\psframe(82,35)(101,45)
\rput(92,40){$Q_{\L_{\f,\theta(m)}}$}
\psline{->}(101,40)(109,40)
\psframe(109,35)(129,45)
\rput(119,40){$\modl_{\c}$}
\psline{->}(129,40)(141,40)
\rput(134.5,43.5){$\nuhv_m$}
\rput(72,0){
\pscircle(71.5,40){2.5} \psline{-}(72.75,40)(70.25,40)
\psline{-}(71.5,38.75)(71.5,41.25) \psline{<-}(71.5,42.5)(71.5,48) \rput(79,56){$\displaystyle  \sum_{i=1}^{m-1} \bar{l}_{m,i}^{(\text{inv})} \muhv_i $}
\psline{->}(74,40)(82,40)
}
\rput(13,0){
\psframe(141,35)(161,45)
\rput(151,40){$\modl_{\c}$}
\psline{->}(161,40)(175,40)
\pscircle[fillstyle=solid,fillcolor=black](168,40){1}
\rput(167.5,44.5){$\muhv_m$}
}
\rput(47,0){
\psframe(141,35)(151,45)
\rput(146,40){$\varphi$}
\psline{->}(151,40)(158,40)
\rput(162,40){$\uhv_m$}
}
}

\psline{->}(62,40)(62,27.5)
\pscircle(62,25){2.5} \psline{-}(60.75,25)(63.25,25)
\psline{->}(59.5,25)(52,25)
\psframe(52,20)(41,30)
\rput(46.5,25){$Q_{\L_{\c}}$}
\psline{->}(41,25)(34.5,25)
\pscircle(32,25){2.5} \psline{-}(30.75,25)(33.25,25)
\psline{-}(32,23.75)(32,26.25)
\psline{->}(29.5,25)(22,25)
\rput(19,25){$\shv_m$}

\psline{->}(211,40)(211,25)(176,25)
\psline{->}(173.5,17)(173.5,22.5)
\rput(176,12){$\displaystyle  \sum_{\ell=1}^{L} a_{m,\ell} \dv_\ell $}
\pscircle(173.5,25){2.5} \psline{-}(172.25,25)(174.75,25)
\psline{-}(173.5,23.75)(173.5,26.25)
\psline{->}(171,25)(159,25)
\psframe(159,20)(139,30) \rput(149,25){$\modl_{\c}$}
\psline{->}(139,25)(64.5,25)
\rput(133,29.5){$\chihv_m$}
\pscircle[fillstyle=solid,fillcolor=black](133,25){1}
\psline{->}(133,25)(133,15)(32,15)(32,22.5)

\psframe[linewidth=2pt,linestyle=dashed,linecolor=LineBlue](26,2)(232,65)
\rput(92,6){$m$th Successive Decoder}

\end{pspicture}
\end{center}
\caption{Block diagram for the $m$th decoder for successive computation.}
\label{f:compsuccdec}
\end{figure*}
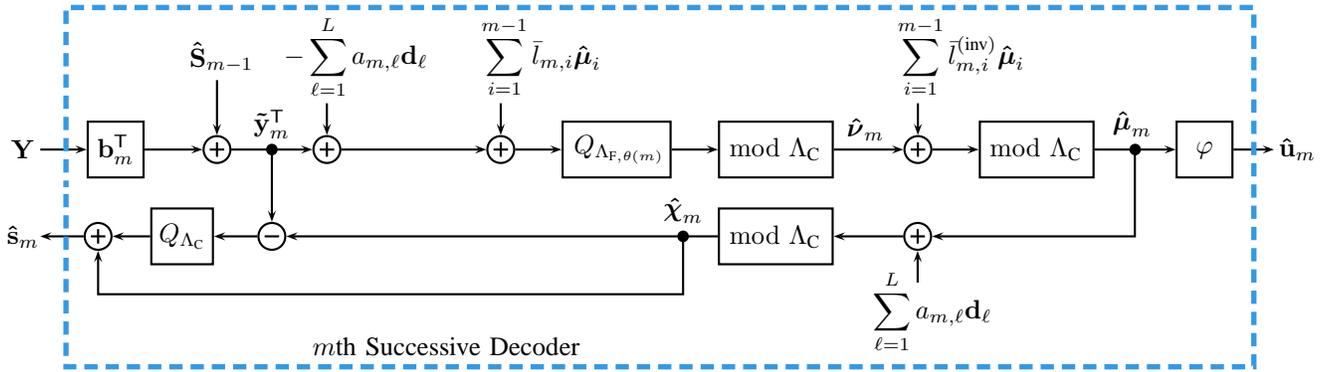

Overall, for $m \geq 2$, the $m$th successive decoding step begins by assembling the estimates of the integer-linear combinations of channel inputs from the previous $m-1$ decoding steps into a matrix $\Shm_{m-1}$. Assuming prior steps are correct, we have that $\Shm_{m-1} = \Am_{m-1} \Xm$. It then applies equalization vectors $\bv_m \in \RR^{\ant}$ and $\cv_m \in \RR^{m-1}$ to its observation $\Ym$ and side information $\Shm_{m-1}$, respectively, and adds the results together to form its effective channel observation $\ytv_m$. Next, it applies algebraic successive cancellation, removes the dithers, and quantizes onto the appropriate fine lattice. After quantizing, it reverses the algebraic successive cancellation to obtain an estimate $\muhv_m$ of the integer-linear combination $\muv_m$ from~\eqref{e:integerlinear}. Finally, the decoder uses the linear labeling to make its estimate of the desired linear combination and follows the steps from Lemma~\ref{l:realsumdirect} to estimate the corresponding integer-linear combination of the channel inputs. These operations are depicted in Figure~\ref{f:compsuccdec} and formally expressed in~\eqref{e:compsuccdecoding}.
\newline~\newline
\noindent\textbf{Decoding, m$\mathbf{\ \geq 2}$:}
\begin{subequations} \label{e:compsuccdecoding}
\begin{align}
&\Shm_{m-1} \define [\shv_1 ~\cdots~\shv_{m-1}]\T \\
&\ytv_m = \bv_m\T \Ym + \cv_m\T \Shm_{m-1}  \label{e:compsuccdecoding1} \\
&\nuhv_m = \bigg[ Q_{\L_{\f,\vartheta(m)}}\bigg(\ytv_m + \sum_{i=1}^{m-1} \bar{l}_{m,i} \muhv_i - \sum_{\ell = 1}^L a_{m,\ell}\, \dv_\ell \bigg) \bigg] \modl_{\c} \label{e:compsuccdecoding2} \\
&~~~\muhv_m = \bigg[ \nuhv_m +  \sum_{i=1}^{m-1} \bar{l}^{(\text{inv})}_{m,i} \nuhv_i \bigg] \modl_{\c} \label{e:compsuccdecoding3} \\
&~~~\uhv_m = \varphi\big(\muhv_m\big) \label{e:compsuccdecoding4} \\
&~~~\chihv_m = \bigg[ \muhv_m + \sum_{\ell=1}^L a_{m,\ell}\, \dv_\ell \bigg] \modl_\c \label{e:compsuccdecoding5}\\
&~~~\shv_m = Q_{\L_{\c}}\big(\ytv_m - \chihv_m \big) + \chihv_m \label{e:compsuccdecoding5} 
\end{align} 
\end{subequations}

The following lemma captures the rates achievable for directly recovering the linear combinations with coefficient matrix $\Am$ via successive computation.

\begin{lemma}\label{l:compsuccdirect}
Consider any choice of rates $R_1,\ldots,R_L$ and parameter $\zeta  > 0$ and assume that the transmitters and receiver employ the encoding and decoding operations from~\eqref{e:compsuccencoding},~\eqref{e:compsuccdecodefirst}, and~\eqref{e:compsuccdecoding}. For $n$ and prime $p$ large enough, there exists a generator matrix $\Gm \in \ZZ_p^{\kfmax \times n}$ and corresponding nested lattice codes $\Lc_1,\ldots,\Lc_L$ with rates at least $R_1- \zeta,\ldots,R_L - \zeta$, respectively, such that, for any choice of channel matrix $\Hm \in \RR^{\ant \times L}$, integer matrix $\Am \in \ZZ^{L\times L}$, and admissible mapping $\Ic$, the receiver can recover the linear combinations $\uv_1,\ldots,\uv_L$ with probability of error at most $\zeta$ so long as
\begin{align}
R_\ell < \frac{1}{2}\log^+\left( \frac{P_\ell} { \sigma_{\succ}^2(\Hm,\av_m,\bv_m,\cv_m | \Am_{m-1}) } \right) ~~~ \forall (m,\ell) \in \Ic \label{e:compsuccdirectcondition}
\end{align} for some choice of equalization vectors $\bv_m \in \RR^{\ant}$ and $\cv_m \in \RR^{m-1}$. \thmsym
\end{lemma}
\begin{IEEEproof} The codebook generation process is nearly identical to that in the proof of Lemma~\ref{l:compparadirect}, except that we set  $\epsilon = \zeta / (2L)$ (and keep the choice $\sigma_{\eff,\ell}^2 = P_\ell 2^{-2 R_\ell}$). The encoding process is also identical, and so the power constraint is met as in the proof of Lemma~\ref{l:compparadirect}. Finally, the decoding step~\eqref{e:compsuccdecodefirst} to recover $\uv_1$ is identical as well and, as argued in the proof of Lemma~\ref{l:compparadirect}, $\uhv_1 = \uv_1$ with probability of error at most $\epsilon$. We condition on the event that $\uhv_1 = \uv_1$ for the remainder of the proof.

To show that $\shv_1 = \av_1\T \Xm$, we need to argue that the effective noise $(\bv_m\T \Hm - \av_m\T) \Xm + \bv_m\T \Zm$ is contained in the Voronoi region of the coarsest lattice $\Vc_{\c}$ so that we can invoke Lemma~\ref{l:realsumdirect}. If at least one rate is non-zero, then we have that $\sigma_{\text{para}}^2(\Hm,\av_m,\bv_m) < P_{\text{max}}$. From Theorem~\ref{t:mainlattice}(c), it follows that the effective noise leaves $\Vc_{\c}$ with probability at most $\epsilon$. Therefore, the decoder can recover $\av_1\T \Xm$ with probability at most $2 \epsilon$.

We now proceed to argue by induction. Assume that decoding has been successful for steps $1$ through $m-1$ with total probability of error at most $2(m-1) \epsilon$. We would like to argue that step $m$ is successful with probability of error at most $2m\epsilon$. Define the successive effective noise as 
\begin{align*}
\zv_{\succ,m}\T &\triangleq (\bv_m\T \Hm - \av_m\T) \Xm + \cv_m\T \Shm_{m-1} + \bv_m\T \Zm \\ 
&= (\bv_m\T \Hm + \cv_m\T \Am_{m-1} - \av_m\T) \Xm + \bv_m\T \Zm
\end{align*} where the last step uses the correct decoding assumption, $\Shm_{m-1} = [ \sv_1 ~\cdots~\sv_{m-1}]\T = \Am_{m-1} \Xm$.

The following equations show that the argument inside the quantizer in~\eqref{e:compsuccdecoding3} can be written as $\nuv_m$ plus noise:
\begin{align*}
&\bigg[\ytv_m + \sum_{i=1}^{m-1} \bar{l}_{m,i} \muhv_i - \sum_{\ell = 1}^L a_{m,\ell}\, \dv_\ell \bigg] \modl_{\c} \\
&\overset{(i)}{=} \bigg[\ytv_m + \sum_{i=1}^{m-1} \bar{l}_{m,i} \muv_i - \sum_{\ell = 1}^L a_{m,\ell}\, \dv_\ell \bigg] \modl_{\c} \\
&= \bigg[\sum_{\ell = 1}^L a_{m,\ell} (\xv_\ell - \dv_\ell)  + \sum_{i=1}^{m-1} \bar{l}_{m,i} \muv_i  + \zv_{\succ,m} \bigg] \modl_{\c} \\
&\overset{(ii)}{=} \bigg[\muv_m  + \sum_{i=1}^{m-1} \bar{l}_{m,i} \muv_i + \zv_{\succ,m} \bigg] \modl_{\c} \\
&\overset{(iii)}{=} \big[\nuv_m  + \zv_{\succ,m} \big] \modl_{\c} 
\end{align*}where $(i)$ uses the assumption that prior decoding steps are correct, $(ii)$ uses the distributive law, and $(iii)$ follows from~\eqref{e:reducedequation}. Combining this with the nested quantization property from~\eqref{e:nestedquantization}, we find that
\begin{align*}
\nuhv_m = \Big[ Q_{\L_{\f,\vartheta(m)}}(\nuv_m + \zv_{\succ,m}) \Big] \modl_{\c} \ . 
\end{align*} Following a similar labeling argument as in the proof of Lemma~\ref{l:compparadirect}, it can be shown that $\nuv_m \in \L_{\f,\vartheta(m)}$. Thus, by Theorem~\ref{t:mainlattice}(c), we have that $\pr(\nuhv_m \neq \nuv_m) < \epsilon$ if 
\begin{align} \label{e:succnoisethreshold}
\sigma_{\succ}^2(\Hm,\av_m,\bv_m,\cv_m | \Am_{m-1}) < \sigma_{\text{eff},\vartheta(m)}^2
\end{align} where $\sigma_{\succ}^2(\Hm,\av_m,\bv_m,\cv_m | \Am_{m-1})$ is defined in~\eqref{e:noisevarsucc}.
Notice that, if $\nuhv_m = \nuv_m$, then $\muhv_m = \muv_m$ by~\eqref{e:rebuiltequation}, $\uhv_m = \uv_m$ by the linear labeling argument in the proof of Lemma~\ref{l:compparadirect}. If at least one rate is non-zero, then $\sigma_{\succ}^2(\Hm,\av_m,\bv_m,\cv_m | \Am_{m-1}) < P_{\text{max}}$ and we can apply Theorem~\ref{t:mainlattice} to establish that the effective noise $\zv_{\succ,m}$ leaves $\Vc_{\c}$ with probability at most $\epsilon$. Assuming this is the case, we can invoke Lemma~\ref{l:realsumdirect} to show that $\shv_m\T = \av_m\T \Xm$. To complete the induction step, we apply the union bound to show that $\pr\big(\cup_{i=1}^m \{ \uhv_m \neq \uv_m \} \big) < 2m \epsilon$. After $L$ decoding steps, we have probability of error at most $2L\epsilon = \zeta$ as desired. Note that the condition in~\eqref{e:succnoisethreshold} is equivalent to requiring that $R_\ell < \frac{1}{2} \log^+\big(P_\ell / \sigma_{\succ}^2(\Hm,\av_m,\bv_m,\cv_m | \Am_{m-1})\big)$ for all $(m,\ell) \in \Ic$. Finally, it is shown in Appendix~\ref{app:fixeddithers} that it suffices to use fixed dither vectors.
\end{IEEEproof}

We are now ready to prove Theorem~\ref{t:compsucc}.

\noindent\textit{Proof of Theorem~\ref{t:compsucc}:} 
Choose rates $R_1,\ldots,R_L$ as well as nested lattice codebooks via Lemma~\ref{l:compsuccdirect}. The transmitters use the encoding strategy from Lemma~\ref{l:compsuccdirect} (which does not depend on the channel matrix $\Hm$ nor the integer matrix $\Am$). For a given channel matrix $\Hm$, assume the receiver wants linear combinations with integer coefficient matrix $\Am$. This implies that there exists an integer matrix $\Atm$ satisfying $\rowspan(\Am) \subseteq \rowspan(\Atm)$ and admissible mapping $\Ic \in \Mc(\Atm)$ such that $(R_1,\ldots,R_L) \in \Rc_\succ(\Hm,\Atm,\Ic)$. Therefore, the receiver can use the decoding strategy from Lemma~\ref{l:compsuccdirect} with integer matrix $\Atm$, admissible mapping $\Ic$, and optimal equalization vectors $\bv_{\opt,m}$ and $\cv_{\opt,m}$ chosen via Lemma~\ref{l:mmsesucc}. Recall from~\eqref{e:minnoisevarsucc1} that~$\sigma^2_{\succ}(\Hm,\av_m | \Am_{m-1} ) \define \sigma^2_{\succ}(\Hm,\av_m,\bv_{\opt,m},\cv_{\opt,m} | \Am_{m-1})$ so~\eqref{e:compsuccdirectcondition} matches the rate constraints from $\mathcal{R}_{\succ}(\Hm,\Atm,\Ic)$. \hfill$\blacksquare$

\section{Conclusions} \label{s:conclusions}

Although compute-and-forward was originally proposed as a relaying strategy~\cite{ng11IT}, recent efforts have demonstrated that it can be useful in the context of interference alignment~\cite{oen14,ncnc13ISIT,fn13ISIT} and low-complexity MIMO transceivers~\cite{zneg14,hc12,hc13,shv13,oen13,oe15}. The aim of this paper was to create a unified framework that captures techniques (e.g., unequal power allocation, successive decoding) that are useful in these settings. Follow-up efforts have employed this expanded framework to develop a notion of uplink-downlink duality for integer-forcing~\cite{hns14ISIT} as well as investigate compute-and-forward for discrete memoryless networks~\cite{lfng15}. 

As mentioned earlier, the results of this paper are directly applicable to complex-valued channels by working with a real-valued decomposition of the channel, which corresponds to approximating the complex channel gains with Gaussian integers. However, recent efforts have demonstrated the advantages of working directly over the complex field, e.g., by building lattices from Eisenstein integers~\cite{thbn15}. An interesting direction for future work is to generalize the results of Huang \etal \cite{thbn15} to create an expanded compute-and-forward framework for Eisenstein integers. More generally, it is of interest to generalize the algebraic framework of Feng \etal\cite{fsk13} to permit unequal powers.

Another issue for future study is the development of optimization techniques for use within the expanded compute-and-forward framework. It is well-established (see, for instance, \cite{fsk13,zneg14}) that the LLL basis reduction algorithm~\cite{lll82} and its variants are an attractive low-complexity solution for the problem of identifying good integer coefficient vectors within the original compute-and-forward framework. These techniques are a natural fit for the parallel computation strategy in Theorem~\ref{t:comppara}. For the successive computation strategy in Theorem~\ref{t:compsucc}, it has been shown~\cite[Theorem 3]{oen13} that the problem of minimizing the effective noise variances can be linked to Korkin-Zolotarev basis reduction~\cite{miccianciogoldwasser}. This can in turn be approximated by multiple applications of the LLL algorithm. More broadly, applications such as integer-forcing interference alignment~\cite{ncnc13ISIT} require simultaneous optimization of beamforming, equalization, and integer coefficient vectors. New heuristics and approximation algorithms are needed to characterize the performance of this strategy versus existing alignment strategies. See~\cite{ehn15} for a recent approach that builds on uplink-downlink duality.

\section*{Acknowledgments}

The authors gratefully acknowledge valuable discussions with U. Erez, C. Feng, M. Gastpar, U. Niesen, O. Ordentlich, and R. Zamir that helped shape this work. In particular, the authors would like to thank C. Feng for his help in adapting the linear labeling results of~\cite{fsk13} to form Section~\ref{s:linearlabeling}.

\appendices

\section{Proof of Lemma~\ref{l:mmsepara}}
\label{app:proofmmsepara}

Rewriting~\eqref{e:noisevarpara2}, we have that
\begin{align}
\sigma_{\para}^2(\Hm,\av_m,\bv_m)  &= \bv_m\T \bv_m +  \big( \bv_m \T \Hm - \av_m\T\big) \Pm  \big(  \Hm\T \bv_m - \av_m\big) \nonumber \\
&= \bv_m\T \big( \Id + \Hm \Pm \Hm\T \big) \bv_m - \av_m\T \Pm \Hm\T \bv_m \nonumber \\
&~~~ - \bv_m\T \Hm \Pm \av_m + \av_m\T \Pm \av_m \ , \label{e:noisevarpara3}
\end{align} which has a positive definite Hessian with respect to $\bv_m$. Therefore, the minimizing value $\bv_{\opt,m}$ is found by setting the derivative equal to zero:
\begin{align*}
&2 \bv_{\opt,m}\T \big(\Id + \Hm \Pm \Hm\T\big) - 2 \av_m\T \Pm \Hm\T = \zeros  \\
&\bv_{\opt,m}\T = \av_m\T \Pm \Hm\T \big( \Id + \Hm \Pm \Hm\T \big)^{-1}  \ .
\end{align*} Plugging this back into~\eqref{e:noisevarpara3}, we get that
\begin{align*}
&\min_{\bv_m \in \RR^\ant} \sigma^2_{\para}(\Hm, \av_m, \bv_m) \\
&= \av_m\T \big(\Pm - \Pm \Hm\T \big( \Id + \Hm \Pm \Hm\T \big) \Hm \Pm \big) \av_m 
\end{align*} from which the result follows by applying~\eqref{e:woodburyh} and~\eqref{e:F}. \hfill$\blacksquare$

\section{Proof of Lemma~\ref{l:normbound}}
\label{app:proofnormbound}

Assume, for the sake of the contradiction, that $\tilde{a}_{m,\ell}^2 > \lambda_{\text{max}}\big(\Id + \Pm \Hm\T \Hm\big)$ for some $m$ and that there is a rate tuple in $\Rc_{\para}(\Hm,\Atm)$ for which $R_\ell > 0$. Starting from the rate expression in Theorem~\ref{t:comppara}, $R_\ell > 0$ implies that
\begin{align*}
P_\ell &> \sigma_{\para}^2(\Hm,\atv_m) \\
&= \atv_m\T \big(\Pm^{-1} + \Hm\T \Hm \big)^{-1} \atv_m \\
&= \atv_m\T \Pm \big( \Id + \Pm \Hm\T \Hm \big)^{-1} \atv_m \\
&\overset{\text{(i)}}{\geq} \atv_m\T \Pm \atv_m \lambda_{\text{min}}\Big(\big(\Id + \Pm \Hm\T \Hm\big)^{-1}\Big) \\
&= \frac{ \atv_m\T \Pm \atv_m }{ \lambda_{\text{max}}\big( \Id + \Pm \Hm\T \Hm \big)} \\
&\geq \frac{ P_\ell \tilde{a}_{m,\ell}^2}{ \lambda_{\text{max}}\big( \Id + \Pm \Hm\T \Hm \big)} \\
&\implies  \lambda_{\text{max}}\big( \Id + \Pm \Hm\T \Hm \big) \geq \tilde{a}_{m,\ell}^2
\end{align*} where in $(i)$ $\lambda_{\text{min}}\Big(\big(\Id + \Pm \Hm\T \Hm\big)^{-1}\Big)$ refers to the minimum eigenvalue of $\big(\Id + \Pm \Hm\T \Hm\big)^{-1}$ and the inequality is due to the Min-Max Theorem~\cite[Theorem 4.2.2]{hornjohnson} for symmetric matrices. Thus, a contradiction has been reached, which establishes the desired bound on $\tilde{a}_{m,\ell}^2$.\hfill$\blacksquare$

\section{Proof of Lemma~\ref{l:mmsesucc}}
\label{app:proofmmsesucc}

From~\eqref{e:noisevarsucc}, it follows that 
\begin{align}
&\sigma_{\succ}^2(\Hm,\av_m,\bv_m,\cv_m | \Am_{m-1}) \nonumber \\
&= \bv_m\T \big(\Id + \Hm \Pm \Hm\T \big) \bv_m ~-~ 2 \bv_m\T \Hm \Pm \big(\av_m - \Am_{m-1}\T \cv_m \big) \nonumber \\
&~~~ + ~  \big(\av_m\T - \cv_m\T \Am_{m-1}\big) \Pm \big(\av_m - \Am_{m-1}\T \cv_m\big)  \label{e:noisevarsucc2} \ .
\end{align} This expression has a positive definite Hessian with respect to the vector $[\bv_m\T~ \cv_m\T]$. Therefore, the optimizing vector can be found by setting the derivative equal to zero. We start by solving for $\bv_{\opt,m}$ in terms of $\cv_{\opt,m}$.

Taking the derivative of~\eqref{e:noisevarsucc2} with respect to $\bv_m$ and setting it equal to zero, we obtain
\begin{align*}
2 \big(\Id + \Hm \Pm \Hm\T \big) \bv_m - 2 \Hm\Pm \big( \av_m - \Am_{m-1}\T \cv_m \big) = \zerov \ .
\end{align*} It follows that 
\begin{align*}
\bv_{\opt,m}\T = \big(\av_m\T - \cv_m\T \Am_{m-1} \big) \Pm \Hm\T \big(\Id + \Hm \Pm \Hm\T\big)^{-1} \ .  
\end{align*} 

Plugging back into~\eqref{e:noisevarsucc2} and canceling terms, we get
\begin{align}
&\sigma_{\succ}^2(\Hm,\av_m,\bv_{\opt,m},\cv_m | \Am_{m-1})  \nonumber \\
&=\scalemath{0.85}{\big(\av_m\T  - \cv_m\T \Am_{m-1} \big) \big( \Pm - \Pm \Hm\T \big( \Id + \Hm \Pm \Hm\T \big)^{-1} \Hm \Pm \big) \big(\av_m - \Am_{m-1}\T \cv_m \big)} \nonumber \\
&= \big(\av_m\T - \cv_m\T \Am_{m-1} \big) \Fm\T \Fm  \big(\av_m - \Am_{m-1}\T \cv_m \big) \ . \label{e:noisevarsucc3}
\end{align} where the last step uses~\eqref{e:woodburyh} and~\eqref{e:F}. Next, we take the derivative with respect to $\cv_m$ and set it equal to zero,
\begin{align*}
2 \Am_{m-1} \Fm\T \Fm \Am_{m-1}\T \cv_m - 2 \Am_{m-1} \Fm\T \Fm \av_m = \zerov  
\end{align*} from which it follows that
\begin{align*}
\cv_{\opt,m}\T = \av_m\T \Fm\T \Fm \Am_{m-1}\T \big( \Am_{m-1} \Fm\T \Fm \Am_{m-1} \big)^{-1} \ . 
\end{align*}

Finally, we plug back into~\eqref{e:noisevarsucc3} and cancel terms to obtain 
\begin{align*}
&\sigma_{\succ}^2(\Hm,\av_m,\bv_{\opt,m},\cv_{\opt,m} | \Am_{m-1}) \\
&= \av_m\T \big( \Id - \Fm \Am_{m-1}\T \big( \Am_{m-1} \Fm\T \Fm \Am_{m-1}\T \big)^{-1} \Am_{m-1} \Fm\T \big) \Fm \av_m \ . 
\end{align*} Note that $\Fm \Am_{m-1}\T \big( \Am_{m-1} \Fm\T \Fm \Am_{m-1}\T \big)^{-1} \Am_{m-1} \Fm\T$ is the projection matrix for the subspace spanned by $\Fm \Am_{m-1}\T$ and $\Nm_{m-1} =  \Id - \Fm \Am_{m-1}\T \big( \Am_{m-1} \Fm\T \Fm \Am_{m-1}\T \big)^{-1} \Am_{m-1} \Fm\T \big) \Fm$ is the projection matrix for the corresponding nullspace. Since projection matrices are idempotent (i.e, $\Nm_{m-1}^2 = \Nm_{m-1}$) and $\Nm_{m-1}$ is symmetric, it follows that
\begin{align*}
\sigma_{\succ}^2(\Hm,\av_m,\bv_{\opt,m},\cv_{\opt,m} | \Am_{m-1}) &= \av_m\T \Fm\T \Nm_{m-1} \Fm \av_m \\
&= \av_m\T \Fm\T \Nm_{m-1}\T \Nm_{m-1} \Fm \av_m \\
&= \big\| \Nm_{m-1} \Fm \av_m \big\|^2 \ . \end{align*}\hfill$\blacksquare$

\section{Proof of Theorem~\ref{t:primitive}}
\label{app:proofprimitive}

The inclusion $\Rc_{\comp}^{(\prim)}(\Hm,\Am) \subseteq \Rc_{\comp}^{(\succ)}(\Hm, \Am)$ follows directly from the fact that the union in the former computation rate region is taken over a subset of the matrices $\Atm$ used in the union in the latter. We now turn to argue that $\Rc_{\comp}^{(\prim)}(\Hm,\Am) \supseteq \Rc_{\comp}^{(\succ)}(\Hm, \Am)$. In particular, we will show that for any integer matrix $\Atm$, there exists a primitive basis matrix $\Atm_\prim$ with the same rowspan such that $\Rc_{\succ}(\Hm, \Atm,\Ic) \subseteq \Rc_{\succ}(\Hm, \Atm_\prim,\Ic)$ for any admissible mapping $\Ic$ with respect to $\Atm$. Without loss of generality, we assume that $\Atm$ is of the form
\begin{align}
\Atm = \begin{bmatrix}
\Atm_M \\
\zeros_{(L-M) \times L}
\end{bmatrix} \label{e:nonprimitive}
\end{align} where the submatrix $\Atm_M$ has $M \leq L$ rows and is full rank. 

\begin{lemma}\label{l:primitivetransform}
For any integer matrix $\Atm$ of the form~\eqref{e:nonprimitive}, there exists a rank $M$ primitive basis matrix  $\Atm_\prim = \begin{bmatrix} \Atm_{\prim,M} \\ \zeros_{(L-M)\times L} \end{bmatrix}$ such that $\Atm_M = \Tm \Atm_{\prim,M}$ where $\Tm$ is an $M \times M$ lower triangular integer matrix with strictly positive diagonal entries. \thmsym
\end{lemma} The proof follows directly from~\cite[Corollary 1.24]{bremner}.

Let $\Atm_\prim$ be a primitive basis matrix chosen using the lemma above. By assumption, there exists a lower unitriangular matrix $\Lm$ that shows that $\Ic$ is admissible for $\Atm$. Specifically, for each index pair $(i,j) \notin \Ic$, we have that $\lv_i\T \atv_j = 0$ where $\lv_i\T$ is the $i$th row of $\Lm$. We would like to show that $\Ic$ is admissible for $\Atm_\prim$ as well. Define 
$$\tildebf{L} = \Lm \begin{bmatrix} \Tm & \zeros_{M \times (L-M)} \\ \zeros_{(L-M) \times M} & \Id_L \end{bmatrix}$$ and note that $\tildebf{L}$ is lower triangular and $\tildebf{l}_i\T \atv_{\prim,j} = 0$ where $\tildebf{l}_i\T$ is the $i$th row of $\tildebf{L}$ and $\atv_{\prim,j}$ is the $j$th column of $\Atm_\prim$. Finally, we renormalize the rows to obtain a unitriangular matrix $\Lm_{\prim}$ whose $i$th row is equal to $\lv_{\prim,i}\T = (\tilde{l}_{i,i})^{-1} \tildebf{l}_{i}\T$. We still have that  $\lv_{\prim,i}\T \atv_{\prim,j} = 0$ so $\Ic$ is admissible for $\Atm_{\prim}$. 

To complete the proof, we need to argue that the effective noise variances can only decrease by using $\Atm_\prim$ instead of $\Atm$. Let $t_{i,j}$ denote the $(i,j)$th entry of $\Tm$ and $\Nm_{\prim,m-1}$ the nullspace projection matrix~\eqref{e:nullspace} for $\Atm_{\prim}$. Starting from~\eqref{e:minnoisevarsucc1}, we have that
\begin{align*}
&\sigma_{\succ}^2(\Hm,\atv_m | \Atm_{m-1}) \\
&= \min_{\bv_m} \|\bv_m\|^2 + \big\| \big(\bv_m\T \Hm - \cv_{\opt,m}\T \Atm_{m-1} - \atv_m\T\big) \Pm^{1/2}\big\|^2 \\ 
&\overset{(i)}{=} \scalemath{0.95}{\min_{\bv_m} \|\bv_m\|^2 + \big\| \big(\bv_m\T \Hm - \tildebf{c}\T \Atm_{\prim,m-1} - t_{m,m} \atv_{\prim,m}\T\big) \Pm^{1/2}\big\|^2} \\ 
&\geq \scalemath{0.91}{\min_{\bv_m,\cv_m} \|\bv_m\|^2 + \big\| \big(\bv_m\T \Hm - \cv_{m}\T \Atm_{\prim,m-1} - t_{m,m} \atv_{\prim,m}\T\big) \Pm^{1/2}\big\|^2} \\ 
&= \sigma_{\succ}^2(\Hm,t_{m,m} \atv_{\prim,m} | \Atm_{m-1}) \\
&= \big\| \Nm_{\prim,m-1} \Fm t_{m,m} \atv_{\prim,m} \big\|^2 \\
&\overset{(ii)}{\geq} \big\| \Nm_{\prim,m-1} \Fm \atv_{\prim,m} \big\|^2 \\
&= \sigma_{\succ}^2(\Hm,\atv_{\prim,m} | \Atm_{\prim,m-1}) 
\end{align*} where $(i)$ relies on the fact that $\cv_{\opt,m}\T \Atm_{m-1} - \atv_m\T =  \tildebf{c}\T \Atm_{\prim,m-1} - t_{m,m} \atv_{\prim,m}\T$ for some choice of $\tildebf{c}$ which is shown below and $(ii)$ uses the fact that $t_{m,m} \geq 1$ from Lemma~\ref{l:primitivetransform}. 

Let the $j$th entry of $\tildebf{c}$ be $\tilde{c}_j = \sum_{i=j}^{m-1} t_{i,j} c_{\opt,m,i} - t_{m,j}$ where $c_{\opt,m,i}$ is the $i$th entry of $\cv_{\opt,m}$. It follows that 
\begin{align*}
&\tildebf{c}\T \Atm_{\prim,m-1} - t_{m,m} \atv_{\prim,m}\T \\
&= \sum_{j=1}^{m-1} \tilde{c}_j \atv_{\prim,j}\T - t_{m,m} \atv_{\prim,m}\T \\
&= \sum_{j=1}^{m-1} \bigg( \sum_{i=j}^{m-1}  t_{i,j} c_{\opt,m,i} - t_{m,j}\bigg) \atv_{\prim,j}\T   - t_{m,m} \atv_{\prim,m}\T \\
&= \sum_{j=1}^{m-1} \bigg( \sum_{i=j}^{m-1}  t_{i,j} c_{\opt,m,i} \bigg) \atv_{\prim,j}\T  -   \sum_{j=1}^m t_{m,j} \atv_{\prim,j}\T \\
&= \sum_{i=1}^{m-1}  c_{\opt,m,i} \sum_{j=1}^{i} t_{i,j}  \atv_{\prim,j}\T  -   \sum_{j=1}^m t_{m,j} \atv_{\prim,j}\T \\
&\overset{(i)}{=} \sum_{i=1}^{m-1}  c_{\opt,m,i}   \atv_{i}\T  -  \atv_m\T \\
&= \cv_{\opt,m}\T \Atm_{m-1} - \atv_m\T 
\end{align*} where $(i)$ uses the fact that $\Atm_M = \Tm \Atm_{\prim,M}$ and that $\Tm$ is lower triangular. \hfill$\blacksquare$

\section{Proof of Theorem~\ref{t:compparamac}} \label{app:proofcompparamac}

Our proof follows along the same lines as the proof for the equal power setting~\cite[Theorem 3]{oen14}. The following definition and theorem specialize basic results from the geometry of numbers~\cite{cassels} to the Euclidean case.

\begin{definition}[Successive Minima]\label{d:successiveminima}
Let $\Lambda$ be a full-rank lattice in $\RR^L$ . For $m = 1,\ldots,L$, the $m$th \textit{successive minimum} $\lambda_m$ of $\Lambda$ corresponds to the radius of the smallest Euclidean ball centered at the origin that captures $m$ linearly independent lattice points,
\begin{align*}
\lambda_m \define \inf \Big\{ r > 0 : \dim \mathrm{span} \big( \Bc(\zerov,r) \cap \Lambda \big) = m \Big\} \ .
\end{align*}
\end{definition}

The following lemma is a special case of Minkowski's Second Theorem~\cite[p.156]{cassels}.
\begin{lemma}[{{\cite[Theorem 1.5]{miccianciogoldwasser}}}]
Let $\mathbf{F} \in \RR^{L \times L}$ be a full-rank matrix and let $\Lambda = \mathbf{F} \ZZ^L$ be the resulting full-rank lattice. The product of the successive minima is upper bounded as follows:
\begin{align*}
\prod_{m = 1}^L \lambda_m^2 \leq L^L \big| \det(\mathbf{F}) \big|^2
\end{align*}
\end{lemma} 

Now, let $\Lambda_{\text{channel}} = \Fm \ZZ^L$ be the full-rank lattice generated by $\Fm$ from~\eqref{e:F}.  Notice that the lengths of a dominant solution $\av_1^*,\ldots,\av_L^*$ correspond exactly to the successive minima of this lattice, $\lambda_m = \| \Fm \av_m^*\|$. Furthermore, recall that the determinant of a lattice is defined as the absolute value of the determinant of any basis for the lattice~\cite[Definition 2.1.2]{zamir} so that $\det(\Lambda_{\text{channel}}) = |\det(\Fm)|$.

We now lower bound the sum of the rates from the theorem statement.

\begin{align*}
&\sum_{\ell = 1}^L \frac{1}{2} \log^+ \bigg( \frac{P_\ell}{\sigma_{\para}^2\big(\Hm,\av_{\pi(\ell)}^*\big)}\bigg) \\
&\geq \sum_{\ell = 1}^L \frac{1}{2} \log \bigg( \frac{P_\ell}{\sigma_{\para}^2\big(\Hm,\av_{\pi(\ell)}^*\big)}\bigg) \\
&= \frac{1}{2}\log\bigg(\prod_{\ell=1}^L P_\ell\bigg) - \frac{1}{2} \log \bigg( \prod_{\ell=1}^L \sigma_{\para}^2\big(\Hm,\av_{\pi(\ell)}^*\big) \bigg) \\
&= \frac{1}{2}\log\det(\Pm) - \frac{1}{2} \log \bigg( \prod_{m=1}^L \| \Fm \av_m^* \|^2  \bigg) \\
&\overset{(i)}{\geq} \frac{1}{2}\log\det(\Pm) - \frac{1}{2} \log \big(L^{L} | \det(\Fm)|^2 \big) \\
&= \frac{1}{2}\log\det(\Pm) - \frac{1}{2} \log \det(\Fm\T\Fm) - \frac{L}{2} \log(L) \\
&= \frac{1}{2}\log\det(\Pm) - \frac{1}{2} \log \det\Big(\big(\Pm^{-1} + \Hm\T \Hm \big)^{-1} \Big) - \frac{L}{2} \log(L) \\
&= \frac{1}{2}\log\det\big(\Id + \Pm \Hm\T \Hm\big)  - \frac{L}{2} \log(L) \\
&\overset{(ii)}{=} \frac{1}{2}\log\det\big(\Id +  \Hm \Pm \Hm\T \big) - \frac{L}{2} \log(L)
\end{align*} where $(i)$ is due to Minkowski's Second Theorem from above and $(ii)$ is due to Sylvester's Determinant Identity~\cite{sylvester1851}: for any square matrices $\Mm_1$ and $\Mm_2$, 
\begin{align} \label{e:sylvester}
\det\big(\Id + \Mm_1 \Mm_2\big) = \det\big(\Id + \Mm_2 \Mm_1 \big) \ .
\end{align}
\hfill $\blacksquare$

\section{Proof of Lemma~\ref{l:unimodularsum}} \label{app:proofunimodularsum}

First, note that the vectors $\Fm\av_1,\ldots,\Fm \av_L$ form a basis of $\RR^L$ since $\Fm$ and $\Am$ have rank $L$. Denote the Gram-Schmidt orthogonalization of this basis by $\gv_1^*,\ldots,\gv_L^*$ where 
\begin{align*}
\gv_m^* = \Nm_{m-1} \Fm \av_m 
\end{align*} and $\Nm_{m-1}$ refers to the projection matrix for the nullspace of $\Fm \Am_{m-1}\T$ defined in~\eqref{e:nullspace}. Define $\Gm^* = [ \gv_1^*~\cdots~\gv_L^*]\T$. It can be shown~\cite[Theorem 3.4]{bremner} that $\det(\Gm^*) = \det(\Fm \Am\T)$. 

Since the Gram-Schmidt vectors are orthogonal, we have that
\begin{align*}
\det\big( \Gm^* (\Gm^*)\T \big) = \prod_{m=1}^L \| \gv_m^* \|^2 = \prod_{m=1}^L \sigma_{\succ}^2(\Hm,\av_m | \Am_{m-1} ) \ .
\end{align*} It follows that
\begin{align*}
&\sum_{\ell = 1}^L \frac{1}{2} \log \bigg( \frac{P_\ell}{\sigma_{\succ}^2\big(\Hm,\av_{\pi(\ell)} \big| \Am_{\pi(\ell) -1 } \big)}\bigg) \\
&=\frac{1}{2} \log\bigg( \prod_{\ell=1}^L P_\ell \bigg) - \frac{1}{2} \log\bigg( \prod_{m=1}^L  \sigma_{\succ}^2(\Hm,\av_m | \Am_{m-1} ) \bigg) \\
&= \frac{1}{2} \log\det(\Pm) - \frac{1}{2}\log\det\big( \Gm^* (\Gm^*)\T \big) \\
&= \frac{1}{2} \log\det(\Pm) - \frac{1}{2}\log\det(\Am \Fm\T \Fm \Am\T) \\
&= \frac{1}{2} \log\det(\Pm) - \frac{1}{2}\log\det\Big(\big( \Pm^{-1} + \Hm\T \Hm \big)^{-1}\Big) - \log\det(\Am) \\
&\overset{(i)}{=} \frac{1}{2} \log\det\big(\Id + \Pm \Hm\T \Hm\big)  \\
&\overset{(ii)}{=} \frac{1}{2} \log\det\big(\Id + \Hm \Pm \Hm\T\big) 
\end{align*} where $(i)$ uses the fact that $\Am$ is unimodular so $\det(\Am) = 1$ and $(ii)$ uses Sylvester's Determinant Identity from~\eqref{e:sylvester}. \hfill $\blacksquare$

\section{Proof of Theorem~\ref{t:linearlabeling}}
\label{app:prooflinearlabeling}

The following lemma will be useful for the proof.

\begin{lemma}\label{l:corresponding}
For $\ell = 1,\ldots,L$, let $\lambdav_\ell \in \L_{\f}$ be lattice codewords with corresponding linear codewords $\bar{\phi}(\lambdav_\ell)$. Then, for any integer combination $\muv = \sum_{\ell =1}^L a_\ell \lambdav_\ell$ where $a_\ell \in \ZZ$, the corresponding linear codeword $\bar{\phi}(\muv)$ satisfies
\begin{align*}
\bar{\phi}(\muv) = \bigoplus_{\ell=1}^L q_\ell \, \bar{\phi}(\lambdav_\ell)
\end{align*} where $q_\ell = [a_\ell]\bmod{p}$. \thmsym
\end{lemma}
\begin{IEEEproof}
\begin{align*}
\bar{\phi}(\muv) &= \bar{\phi}\bigg(\sum_{\ell=1}^L a_\ell \lambdav_\ell \bigg) \\
&= \bigg[ \gamma^{-1} p \sum_{\ell=1}^L a_\ell \lambdav_\ell \bigg] \bmod{p} \\
&= [\gamma^{-1} p a_1 \lambdav_{1} ]\bmod{p} \, \oplus  \, \cdots \, \oplus \, [\gamma^{-1} p a_L \lambdav_{L} ]\bmod{p} \\
&= [a_1]\bmod{p} \cdot [\gamma^{-1} p  \lambdav_{1} ]\bmod{p} \, \oplus \, \\
&~~~~ \cdots~ \oplus \, [a_L]\bmod{p} \cdot [\gamma^{-1} p \lambdav_{L} ]\bmod{p} \\
&= q_1 \, \bar{\phi}(\lambdav_1) \, \oplus \, \cdots \, \oplus \, q_L \, \bar{\phi}(\lambdav_L)
\end{align*}
\end{IEEEproof}

We now establish that the proposed labeling $\varphi$ satisfies Definition~\ref{d:linearlabeling}(a). Recall that $\lambdav \in \L_{\f,\ell}$ if and only if the corresponding linear codeword $\bar{\phi}(\lambdav) \in \Cc_{\f,\ell}$. Let $\vv \in \ZZ_p^{\kfmax}$ be the unique vector satisfying $\bar{\phi}(\lambdav) = \Gm\T \vv$. If $\bar{\phi}(\lambdav) \in \Cc_{\f,\ell}$, then the last $\kfmax - k_{\f,\ell}$ components of $\vv$ must be equal to $0$, meaning that the last $\kfmax - k_{\f,\ell}$ components of the label $\varphi(\lambdav)$ are equal to $0$. Similarly, $\lambdav \in \L_{\c,\ell}$ if and only if the last $\kfmax - k_{\c,\ell}$ components of the label $\varphi(\lambdav)$ are equal to $0$.

Now we turn to establish that $\varphi$ satisfies Definition~\ref{d:linearlabeling}(b). Let $\muv = \sum_{\ell = 1}^L a_\ell \lambdav_\ell$. For each $\lambdav_\ell \in \L_{\f}$, let $\vv_\ell \in \ZZ_p^{\kfmax}$ denote the unique vector that satisfies $\bar{\phi}(\lambdav_{\ell}) = \Gm\T \vv_\ell$. From Lemma~\ref{l:corresponding}, we have that
\begin{align*}
\bar{\phi}(\muv) &= \bigoplus_{\ell=1}^L q_\ell \, \bar{\phi}(\lambdav_\ell) \\
&= \bigoplus_{\ell=1}^L q_\ell \Gm\T \vv_\ell \\
&= \Gm\T \bigoplus_{\ell=1}^L q_\ell \vv_\ell
\end{align*} where $q_\ell = [a_\ell]\bmod{p}$. Let $\tv \in \ZZ_p^{\kfmax}$ be the unique vector satisfying $\bar{\phi}(\muv) = \Gm\T \tv$. Since $\Gm$ is full rank, it follows that 
\begin{align*}
\tv = \bigoplus_{\ell=1}^L q_\ell \vv_\ell \ . 
\end{align*} Finally, since the labels $\varphi(\muv)$ and $\varphi(\lambdav_\ell)$ consist of the last $k$ elements of $\tv$ and $\vv_\ell$, respectively, we have that
\begin{align*}
\varphi(\muv) = \bigoplus_{\ell=1}^L q_\ell \, \varphi(\lambdav_\ell)
\end{align*} as desired. \hfill $\blacksquare$

\section{Fixed Dithers} \label{app:fixeddithers}

Assume that we have established Lemma~\ref{l:compparadirect} or~\ref{l:compsuccdirect} using random dither vectors. Specifically, each dither vector $\dv_\ell$ is independently generated according to a uniform distribution over $\Vc_\ell$. We would like to establish that there exist fixed dither vectors that can achieve the same rate region. Select any power constraints, channel matrix, and integer matrix as well as any rate tuple in the achievable region. Let $\xv_\ell(\wv_\ell, \dv_\ell)$ denote the channel input sequence for message vector $\wv_\ell \in \ZZ_p^{\kfl - \kcl}$ and dither vector $\dv_\ell \in \Vc_{\c,\ell}$. For a given realization of $\dv_{\ell}$, the average power is
$$P_{\text{avg},\ell}(\dv_\ell) = \frac{1}{p^{\kfl - \kcl}} \sum_{\wv_\ell } \big\| \xv_\ell(\wv_\ell, \dv_\ell) \big\|^2  $$ where the average is taken with respect to a uniform distribution over possible message vectors. Let $0 < \gamma < 1$ be a parameter to be specified later. Using the fact that $\ex[P_{\text{avg},\ell}(\dv_\ell)] \leq P_\ell$, it follows from Markov's inequality that
$$\pr\bigg( P_{\text{avg},\ell}(\dv_\ell) < \frac{P_{\ell}}{1 - \gamma^{1/L}} \bigg) > \gamma^{1/L} \ . $$ Using the independence of the dithers, we thus have that
\begin{align}
\pr\bigg( \bigcap_{\ell = 1}^L \bigg\{ P_{\text{avg},\ell}(\dv_\ell) < \frac{P_{\ell}}{1 - \gamma^{1/L}} \bigg\} \bigg) > \gamma \ . \label{e:powerprobability}
\end{align}

For a given realization of $\dv_1,\ldots,\dv_L$, the average probability of error is 
\begin{align*}
&p_{\text{error}}(\dv_1,\ldots,\dv_L) \\
&= \frac{1}{p^{\sum_\ell \kfl - \kcl}} \sum_{\wv_1,\ldots,\wv_L} \mathbbm{1}( \uhv_\ell \neq \uv_\ell~\text{for some~}\ell) 
\end{align*} where the average is taken with respect to a uniform distribution over possible message vectors. Using the fact that $\ex[p_{\text{error}}(\dv_1,\ldots,\dv_L)] \leq \epsilon$, we have from Markov's inequality that
$$\pr\bigg(p_{\text{error}}(\dv_1,\ldots,\dv_L) < \frac{2\epsilon}{\gamma} \bigg) > 1 - \frac{\gamma}{2} \ . $$ Combining this with~\eqref{e:powerprobability}, we know that with probability $\gamma/2$, each power is at most $\frac{1}{1 - \gamma^{1/L}}$ times larger than its original target $P_\ell$ and the average error probability is at most $\frac{2 \epsilon}{\gamma}$. Therefore, there exist fixed dither vectors that satisfy these relaxed constraints as well. By taking $\gamma$ to zero, we can get as close as needed to the original target powers $P_\ell$. Afterwards, we can make the average probability of error as small to be as desired by choosing $\epsilon$. Finally, since the rate expressions are continuous functions of the powers, we can operate as close as we would like to any rate tuple in the original rate region using fixed dithers.

\bibliographystyle{ieeetr}

\end{document}